\documentclass[acmsmall]{acmart}
\usepackage{xcolor}
\usepackage{cleveref} 
\usepackage{graphicx}

\definecolor{blue}{RGB}{58, 54, 255}
\AtBeginDocument{%
  }

\setcopyright{none}
\acmDOI{10.1145/3757650}
\begin{document}

\title[\textsc{Remini}: Chatbot-Mediated Mutual Reminiscence]{\textsc{Remini}: Leveraging Chatbot-Mediated Mutual Reminiscence for Promoting Positive Affect and Feeling of Connectedness among Loved Ones}

\author{Zhuoqun Jiang}
\email{zhuoqun_jiang@mymail.sutd.edu.sg}
\orcid{0000-0002-4296-0996}
\affiliation{%
  \institution{Singapore University of Technology and Design}
  \country{Singapore}
}
\authornote{\textbf{Preprint Notice:} This paper has been accepted for publication in the \textit{Proceedings of the ACM on Human-Computer Interaction} (PACM HCI), Volume 9, Issue 7, November 2025 (CSCW 2025).}

\author{ShunYi Yeo}
\affiliation{%
  \institution{Singapore University of Technology and Design}
  \country{Singapore}}
\email{yeoshunyi.sutd@gmail.com}
\orcid{0000-0002-8067-3402}

\author{Wei Xuan Donovan Seow}
\affiliation{%
  \institution{Singapore University of Technology and Design}
  \country{Singapore}}
\email{seowdonovan@gmail.com}
\orcid{0009-0009-6405-3631}

\author{Simon Perrault}
\affiliation{%
  \institution{Singapore University of Technology and Design}
  \country{Singapore}}
\email{perrault.simon@gmail.com}
\orcid{0000-0002-3105-9350}

\renewcommand{\shortauthors}{Jiang et al.}
\begin{abstract}
  Mutual reminiscence, defined as revisiting shared positive memories through reciprocal self-disclosure, strengthens emotional bonds, enhances well-being, and deepens intimacy. However, most technology-mediated reminiscence tools emphasize individual reflection or one-way storytelling, which overlooks the dynamic, interactive dialogue essential for meaningful mutual reminiscence. To address this limitation, we introduce \textsc{Remini}, a chatbot designed to support reciprocal self-disclosure between close partners such as couples, friends, or family members. Grounded in the Social Functions of Autobiographical Memory (SFAM) framework, \textsc{Remini} uses conversational AI to guide emotionally rich exchanges through five narrative phases: rapport building, memory narration, elaboration, reflection, and summary. In a mixed-method, both between- and within-subjects study (N = 48, 24 dyads), we compare \textsc{Remini} to a baseline chatbot that offers minimal memory-trigger prompts. Our findings show that structured guidance from \textsc{Remini} significantly improves positive affect, feeling of connection, and engagement. It also fosters more detailed narrative co-construction and greater reciprocal self-disclosure. Participant feedback highlights the practical value, perceived benefits, and design considerations of chatbot-mediated reminiscence. We contribute empirically grounded design implications for conversational agents that strengthen human connection through mutual reminiscence.
\end{abstract}



\maketitle

\section{Introduction}
\label{sec:introduction}

Sharing personal memories with others often evokes joy, laughter, feelings of connection, and mutual understanding. Consider how enjoyable it is to swap family stories around the dinner table or to share a funny anecdote from years ago with a friend. This act of revisiting and reflecting on one’s past, referred to as \textbf{reminiscence}, can be carried out either alone or with others~\cite{bohlmeijer2007effects}. Research shows that reminiscing is more than just nostalgic indulgence; it is a valuable activity for bolstering psychological well-being. When we recall and interpret memories in solitude, we often enhance our self-esteem, experience a sense of belonging, and find meaning in life~\cite{webster1993construction, westerhof2014celebrating, routledge2013nostalgia}. When conducted socially, social reminiscence stimulates engaging conversations, strengthens interpersonal ties, and elevates positive emotional states~\cite{o2017reminiscence, cappeliez2007functions}. Crucially, when memories are revisited by individuals who initially experienced these events together, a process termed \textbf{mutual reminiscence}, they often lead to deeper emotional resonance and increased intimacy within relationships~\cite{bietti2010sharing, pasupathi2002conversational, beike2017sharing}.

A key mechanism in mutual reminiscence lies in \textbf{reciprocal self-disclosure}, which is characterized by the mutual exchange of emotions, intimate thoughts, and narrative details between participants~\cite{reis1988intimacy, laurenceau1998intimacy, alea2003you}. Reciprocal self-disclosure not only facilitates deeper emotional bonding but also contributes significantly to trust, intimacy, and relationship sustainability. This relational dynamics at the heart of mutual reminiscence resonates with the broader goal of CSCW: designing technologies that foster genuine feelings of connection~\cite{stepanova2022strategies}. The \textbf{Social Functions of Autobiographical Memory (SFAM)} framework outlines three interrelated dimensions into how reciprocal self-disclosure, particularly its depth and emotional richness, shapes the adaptive outcomes of mutual reminiscence as individuals co-construct emotionally vivid narratives through reciprocal emotional sharing.

In the field of HCI and CSCW, technology-mediated reminiscence have primarily focused on individual and social reminiscence. Prior studies have examined a range of memory triggering technologies---such as photos or hybrid artifacts---and demonstrated positive outcomes in individual reflective contexts~\cite{chris2016aquantified, thomas2016reminiscence, yoo2024remembering, baumann2024mnemosyne, chen2023exploring, gibson2023gifting}. In social reminiscence, especially among intergenerational families and older adults, these technologies have similarly been used to encourage storytelling and strengthen group bonds~\cite{li2023exploring, shen2024legacysphere, kang2021momentmeld, baker2021school}. However, existing designs often emphasize memory cues alone, which may suffice for individual or social reminiscence centered on personal reflection or memory sharing, but fall short in adequately supporting the reciprocal and emotionally engaged dialogue essential to mutual reminiscence. Moreover, many studies assume a uni-directional ``narrator–audience'' format from a storyteller to a passive audience~\cite{li2023exploring}, often neglecting how reciprocal self-disclosure unfolds dynamically, whereby memories are co-constructed through two-way interaction. This raises open questions about how to support richer, reciprocal mutual reminiscence among diverse close relationships that goes beyond specialized populations~\cite{jeffrey2007reminiscence, hallford2016brief, pasupathi2003age, kelly2016brief}.

Recent advances in conversational agents and chatbots present a promising solution for addressing these challenges. In particular, chatbots' abilities for reciprocal dialogue exchange, empathetic feedback, and context-sensitive prompting, naturally support the verbal self-disclosure and conversational depth inherent to mutual reminiscence~\cite{park2021wrote, seo2024chacha, lee2020ihearyou}. For example, chatbots have been used to facilitate relationship-building among strangers~\cite{shin2023introbot} and in mediating long-distance relationships~\cite{zheng2021pocketbot}. In the context of reminiscence, chatbots tend to focus on dialogue-based memory triggers for older adults in individual settings~\cite{jin2024exploring, zhai2024exploring}. This points to a missed opportunity: \textit{chatbot-based reminiscence research often overlooks one of the core functions of memory sharing---strengthening interpersonal bonds through reciprocal self-disclosure.}

To address these limitations and support a \textbf{mutual reminiscence} process, we draw on the SFAM framework and leverage the interactive capabilities of chatbots. We introduce \textsc{Remini}, a chatbot that guides close partners (e.g., couples, friends, or family members) through the collaborative narration of shared, positive memories. \textsc{Remini} was designed by aligning and adapting the three interrelated dimensions of the SFAM---including narrative detail, emotional expression, and listener responsiveness---with further reference drawn from Hallford’s positive reminiscence induction manual~\cite{hallford2022remembering}. \textsc{Remini} uses shortened, informal prompts, adapted for everyday contexts, to help partners retell and co-construct meaningful past events. Powered by large language models (LLMs), \textsc{Remini} provides real-time, context-aware guidance through multiple structured phases, encouraging storytelling and emotional disclosure. Our study aims to address two primary research questions: \textbf{RQ1:} How does a reminiscence companion chatbot's structured guidance influence the depth of shared memories and the psychosocial benefits derived from reminiscence? \textbf{RQ2:} How do close partners perceive and experience chatbot-mediated mutual reminiscence?

We conducted a mixed-method and mixed-design study ($N=48$) to evaluate the feasibility and impact of \textsc{Remini}. To isolate the effect of structured facilitation, we compared \textsc{Remini} with a simpler, baseline chatbot that provides only simple memory triggers. Our results show that both \textsc{Remini} and the baseline chatbot improved positive affect and social bonding, but \textsc{Remini} yielded significantly stronger effects. Participants who used \textsc{Remini} reported significantly higher perceived emotional synchrony and partner responsiveness. Chat log analyses also revealed that \textsc{Remini} led to deeper, more detailed memory narratives, heightened emotional expression, and more self-disclosure between partners. Finally, interviews offered insights into \textsc{Remini}’s feasibility, benefits, and potential pitfalls.

We make the following contributions to the HCI and CSCW communities:
\begin{enumerate}
    \item Design and Implementation: We propose a novel chatbot system (\textsc{Remini}) that adapts SFAM’s principles for guiding mutual reminiscence among paired loved ones through structured, context-sensitive prompts.
    
    \item Empirical Evaluation: Through a comparative study, we empirically demonstrate \textsc{Remini}’s effectiveness in fostering positive affect, richer memory narratives, emotional exchange, and interpersonal bonding among close partners.

    \item Broader Implications: We illuminate design considerations for future chatbot-mediated interventions in polyadic settings that aim to enhance human–human connections, going beyond memory triggers to support deeper, more reciprocal interactions. 
\end{enumerate}

\begin{figure*}[t]
  \centering
  \includegraphics[width=\linewidth]{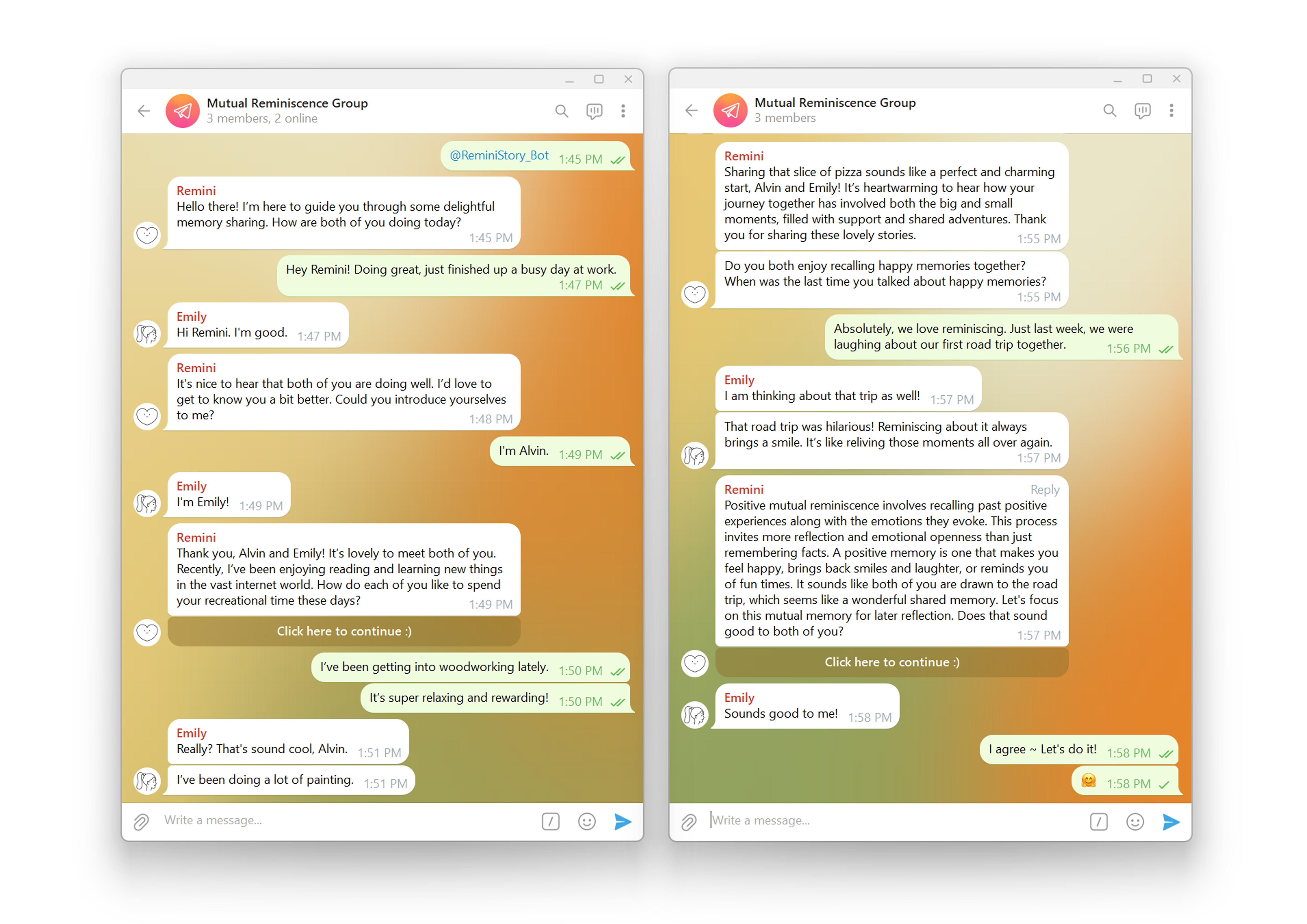}
  \caption{Interface Demonstration and Sample Conversation Snippets with the Chatbot \textsc{Remini} with two users - Alvin and Emily. Note: This screenshot displays Alvin’s view. All user names are fictitious and responses are for illustration purpose. The “Click here to continue” button prompts Remini to generate the next message.}
  \Description{The figure illustrates the user interface of the chatbot, \textsc{Remini}, engaged in facilitating a mutual reminiscence conversation within a group chat setting. Displayed are two snippets of an example conversation between a user, Alvin, his friend Emily, and the chatbot, \textsc{Remini}. The conversation highlights the chatbot's role in guiding participants through sharing and reflecting on positive personal memories. In the first snippet, \textsc{Remini} initiates the conversation by encouraging personal introductions and asking about their hobbies. In the second snippet, the chatbot prompts Alvin and Emily to share and reflect on past happy experiences, thereby fostering a connection and enhancing the reminiscence process. Note: All user interactions and names shown are fictional and for demonstration purposes only.}
  \label{fig:Interface}
\end{figure*}

\section{Related Work}

We review prior research on the emotional and interpersonal value of mutual reminiscence, theoretical frameworks that explain its mechanism, and technologies that support memory sharing. We also identify key gaps in technology-mediated reminiscence, particularly around reciprocity and co-construction, that motivate our exploration of polyadic chatbots for socially grounded reminiscence.

\subsection{Mutual Reminiscence: Co-constructing Shared Memories}
\label{sec_narrative_reminiscence}

Reminiscence, a form of autobiographical memory, involves recalling personal past events beyond mere episodic details (i.e., when, what, and where) to encompass internal reflection, emotional resonance, and personal meaning-making~\cite{bluck1998reminiscence,fivush2011making,mcdams2001thepsychology}. While different reminiscence types exist, such as obsessive (focused on ruminative negative outcomes), integrative (aimed at enhancing identity coherence and individual reflection), narrative (emphasized on storytelling), and instrumental (oriented towards problem-solving), they differ markedly in their psychological impact~\cite{o2017reminiscence,westerhof2014celebrating,pennebaker1999forming,routledge2013nostalgia, bryant2005using, conway2000theconstruction}. Among them, integrative reminiscence is associated with psychological benefits and has informed therapeutic interventions (e.g., cognitive-reminiscence therapy~\cite{watt2000integrative}).

In narrative reminiscence, individuals shift from private mental reflection to the public act of storytelling. It refers to storytelling about personal memories with others, transforming internal reflections into communicative narratives that foster positive emotional engagement, entertain, and deepen relationships~\cite{pennebaker1999forming, smorti2011autobiographical, cappeliez2007functions}. Within narrative reminiscence, mutual reminiscence is defined as collaboratively recalling and co-constructing memories among those who originally experienced the events together~\cite{wolf2024howremembering}. Mutual reminiscence is characterized by interactive turn-taking, shared perspectives, and collaborative meaning-making during recollection~\cite{beike2017sharing}. Its primal role remains fostering specific interpersonal and emotional outcomes, including increased emotional closeness, stronger relational bonds, and enhanced positive mood~\cite{seemann2019reminiscing, kellas2005family, fivush1996remembering, philippe2008functions}.

A key mechanism underlying these positive interpersonal effects is affective self-disclosure~\cite{alea2003you}, the reciprocal sharing of emotions, reflective thoughts, and intimate experiences, which contributes significantly to trust and intimacy within relationships~\cite{reis1988intimacy,laurenceau1998intimacy}. Empirical evidence consistently supports mutual reminiscence as beneficial in relational contexts, such as friendships, couples, and families, demonstrating its ability to enhance participants’ motivation to engage in future shared experiences and foster sustained well-being~\cite{emma2024reminiscence,kellas2005family, kellas2010narrating}.

Despite significant attention in HCI and CSCW research directed toward individual and general social narrative reminiscence, mutual reminiscence remains relatively underexplored. This gap underscores opportunities to design for mutual reminiscence’s unique co-constructive, emotionally resonant, and collaborative nature.

In summary, mutual reminiscence promotes positive affect, intimacy, and relational closeness predominantly through fostering reciprocal affective self-disclosure. These defined terms and established benefits will inform subsequent sections.

\subsection{Understanding Mutual Reminiscence through the SFAM Framework}
\label{sec: SFAM Framework}
A key mechanism underlying mutual reminiscence is affective self-disclosure~\cite{reis1988intimacy, laurenceau1998intimacy, alea2003you}. Existing HCI/CSCW systems have taken a more indirect approach to support this, often relying on biofeedback sharing such as wearable pendants that share breathing rhythms~\cite{frey2018breeze}, smartwatches that transmit biosignals~\cite{liu2021significantotter}, and rings that transmit heartbeats across distance~\cite{werner2008unitedpulse}. Other approaches use symbolic or contextual cues like emotional icons~\cite{kowalski2013cubble}, tangible tokens~\cite{kaye2005communicating}, or location sharing~\cite{bales2011coupleVIBE} to foster awareness and emotional closeness. While these approaches can make affective self-disclosure visible, they typically represent emotional states abstractly, risking misinterpretation and often neglecting reciprocity, which is essential for intimacy~\cite{stepanova2022strategies}.

In contrast, mutual reminiscence centers on direct verbal self-disclosure through responsive dialogue, which more naturally support deepening intimacy~\cite{reis1988intimacy}. Systems supporting direct conversational exchanges, such as PocketBot’s~\cite{zheng2021pocketbot} ``20 questions'' feature for prompting deep conversations among long-distance couples, have gained user appreciation and fostered meaningful connections. The principles of Social Penetration Theory further reinforces the importance of reciprocal verbal exchanges to progressively deepening intimacy~\cite{altman1973social}, aligning well with the interactive and emotionally engaging narration characteristic of mutual reminiscence.

To understand the importance of rich, reciprocal verbal exchanges in fostering intimacy in mutual reminiscence, we employ the Social Functions of Autobiographical Memory (SFAM) framework~\cite{alea2003you}, with particular emphasis on its role in supporting intimacy development. This framework provides a structured theoretical lens to understand why and how mutual reminiscence produces beneficial outcomes in social interactions.

The SFAM framework posits three central dimensions that influence the social and psychological impacts of reminiscence:

\begin{enumerate}
    \item \textit{Memory Narration Characteristics}: How detailed, vivid, and emotionally charged a narrated memory is significantly affects its interpersonal and emotional resonance. Richer emotional detail typically enhances listener engagement, narrator well-being, and relational closeness~\cite{larsen2014whatisit,bluck2001predicting}.
    \item \textit{Traits of Memory Sharers}: The personal characteristics of reminiscence participants, including age, gender, personality traits, relationship length, familiarity, and similarity, significantly shape the quality and emotional depth of their shared narratives~\cite{webster2013adult,davis1999gender,mcadams1997stories, karney2000memorybias}.
    \item \textit{Listener Responsiveness}: Listener engagement through active participation, empathic listening, and responsiveness crucially facilitates deeper self-disclosures and more emotionally resonant storytelling, thus reinforcing interpersonal intimacy~\cite{bavelas2000listeners,gable2004you}.
\end{enumerate}

Taken together, these dimensions suggest pathways to facilitate technology-mediated mutual reminiscence:

\begin{itemize}
    \item Facilitate interactions among familiar, positively bonded relationships (e.g., friends, family, couples) to leverage inherent relational trust.
    \item Encourage narrators to articulate rich emotional content and vivid narrative details, increasing listener engagement and mutual emotional resonance.
    \item Promote reciprocal conversational participation and active listener responsiveness, supporting deeper intimacy and genuine emotional exchange.
\end{itemize}

In summary, the SFAM framework enables a structured understanding of mutual reminiscence's emotional and interpersonal benefits. We employed this framework in our study both to identify key outcome measures (Section~\ref{sec: survey DVs}) and to directly inform the design strategies incorporated into \textsc{Remini} (Section~\ref{sec:remini}).

\subsection{Technology-Mediated Reminiscence}
\label{sec: prior work on mutual reminiscence}

\subsubsection{Individual and Social Reminiscence in Prior Work}

Research in HCI and CSCW has explored reminiscence extensively, initially emphasizing integrative reminiscence (see Section~\ref{sec_narrative_reminiscence}), which facilitates individual reflection using meaningful digital possessions, such as hybrid artifacts~\cite{gray2020towards}, personal photographs~\cite{chen2023photoclock, mcgookin2019reveal}, emails~\cite{hangal2011muse}, music~\cite{kim2022slide2remember}, and contextual metadata~\cite{white2023memory, odom2020exploring, isaacs2013echoes}.

More recently, attention has shifted toward social reminiscence, emphasizing how digital artifacts can support interpersonal storytelling and strengthen social bonds~\cite{li2019story, apted2006tabletop, petrelli2010fmradio, neumann2017memory, uriu2009caraclock, dib2010sonic, houben2019foregrounding}. For instance, MomentMeld~\cite{kang2021momentmeld} facilitates intergenerational reminiscence conversation through photographic interactions in family settings. By algorithmically generating semantically related juxtapositions of photos, what the authors coined as mutually stimulatory mementos, the system embeds moments of reminiscence into everyday routines, prompting meaningful conversations across generations.

Other projects have explored diverse modalities for social reminiscence. The School Days project~\cite{baker2021school} leverages virtual reality (VR) to support group reminiscence among older adults. The system integrates a `historically-inspired' schoolroom environment, digitized personal artifacts, and open-ended prompts to stimulate social reminiscence. In a similar vein, Li et al.~\cite{li2023exploring} explored how augmented reality (AR) can enhance photo-based intergenerational storytelling between grandparents and grandchildren. They found that AR enriched social reminiscence by evoking stronger emotions, augment storytelling by introducing narrative structures, and engaging listeners through multimedia augmentation. Magic Thing~\cite{liagat2022withahint} uses participatory design methods to support cross-cultural, multilingual storytelling within immigrant families, illustrating how reminiscence tools can be tailored to diverse communicative and cultural needs.

These studies collectively demonstrate the capability of technology to facilitate narrative sharing and emotional engagement, primarily focusing on initial cues and one-directional storytelling, underscoring the need for reciprocal interactions central to mutual reminiscence.
 
\subsubsection{Mutual Reminiscence: Extending the Scope of Reminiscence Technologies}

Our work makes four key contributions to HCI/CSCW research in technology-mediated reminiscence. First, despite extensive exploration of individual and social reminiscence, the subdomain of mutual reminiscence, where individuals collaboratively recall and co-construct shared memories, remains underexplored. Uhde et al.~\cite{alarith2022time} observed that people’s most cherished memories frequently center around relationships, love, friendships, and shared experiences. This evidence highlights the potential interpersonal benefits of reminiscing mutually, where the co-constructive nature of memory narration actively strengthens relational bonds.

Second, existing systems have predominantly targeted intergenerational or familial groups. Psychological studies indicate, however, that reminiscence benefits extend beyond specific age demographics~\cite{jeffrey2007reminiscence, pasupathi2003age}, suggesting a broader applicability for these technologies that remains underleveraged.

Third, many existing designs focus primarily on generating initial memory cues to spark storytelling but offer little structural support for the unfolding process of reminiscence itself~\cite{kang2021momentmeld, baker2021school}. Theoretical framework of SFAM suggests that the adaptive outcomes of reminiscence are strongly connected to the vividness and emotional richness of memory narration as well as the interactional dynamics between participants (see Section~\ref{sec: SFAM Framework}). However, it is important to recognize that narrating emotions and past experiences can be intricate, as individuals vary in their ability to articulate their stories and feelings~\cite{einam2024shedding, fioretti2015emotional, peter1990emotional}. Moreover, individuals differ in the amount of positive memories recalled and in the frequency of using their memories for social and mood-enhancement functions~\cite{wolf2015differences, wolf2021thevalence}. These variations highlight the importance of designing adaptive systems that can support individual expressive capacities while still facilitating meaningful reminiscence. Thus, our work explores how technology can scaffold this ongoing process, facilitating elaboration, responsiveness, and emotional connection rather than merely initiating it.

Finally, prior systems often follow a one-directional storyteller and listener model, which overlooks the dynamic and reciprocal nature of mutual reminiscence which involves interactive turn-taking, emotional responsiveness, and collaborative meaning-making. The need for reciprocity in reminiscence technologies has been echoed in prior work: Magic Thing~\cite{liagat2022withahint} advocates for flexible narrative structures, diverse narrative exchanges, and fluid participant roles as future directions, while Li et al.~\cite{li2023exploring} call for more reciprocal interactions after finding that grandparents predominantly narrated while grandchildren remained passive listeners.

In response to these gaps, our work moves beyond static prompts and one-sided narration to design for structured reciprocity in memory sharing across diverse relational contexts. Through guided co-construction and interactive storytelling, we foreground the collaborative and emotionally resonant processes at the heart of mutual reminiscence.

\subsection{Chatbot-Mediated Reminiscence}

Chatbots, also known as conversational agents, are software systems designed to engage users through natural language interactions~\cite{allouch2021conversational, rapp2021thehumanside}. In the HCI and CSCW communities, interest in chatbots has grown rapidly, particularly for socially-oriented applications~\cite{klopfenstein2017therise}, driven in part by recent advancements in generative AI~\cite{zhang2023completesurveygenerativeai}. Chatbots uniquely offer interactive dialogue and natural language engagement, making them particularly promising for mediating reciprocal and emotionally rich mutual reminiscence.

\subsubsection{Dyadic Chatbots for Individual Reminiscence}

To date, most chatbot research has focused on \textit{dyadic} interactions, defined as one-on-one exchanges between a single human and the system~\cite{zheng2022ux}. Prior studies consistently showed that chatbots can promote expressivity, particularly in the form of self-disclosure~\cite{cai2023listen, kim2024mindfuldiary} emotional articulation~\cite{inkster2018anempathydriven, fitzpatrick2017delivering, tsai2019employing}, and personal storytelling~\cite{lee2019caring, wu2024mindshift, kumar2022exploringdesignpromptsapplying, kumar2023mindfulness}. For example, Lee et al.~\cite{lee2020ihearyou} developed a chatbot capable of active self-disclosure during casual conversations, prompting users to share deeper and more personal narratives in return. Diarybot~\cite{park2021wrote} supports users in transforming free-form journaling into guided dialogue, helping structure their thoughts into coherent stories. More recently, ChaCha~\cite{seo2024chacha}, an LLM-based chatbot for children, encourages emotional expression and storytelling through natural, free-form conversations. These expressive capabilities have also given rise to companion-like chatbots (e.g., Replika\footnote{https://replika.ai}), which simulate human-like interactions and creating emotional bonds with users. Research on Human–Chatbot Relationships (HCR) reveals that people may form meaningful emotional bonds with these systems, ranging from friendships to romantic or even familial ties~\cite{skjuve2021mychatbot, pentina2023exploring, ramadan2021alexafying}. While such relationships can provide emotional support~\cite{pentina2023exploring}, they also raise concerns about dysfunctional attachment and potential harm to real-world relationships~\cite{laestadius2024toohuman}.

These examples demonstrate that dyadic chatbots can foster key elements of narrative reminiscence, including self-disclosure and narrative depth. Despite the potential of LLM-based chatbots, their application in reminiscence settings remains limited, with a few exceptions such as~\cite{zhai2024exploring, xu2024memory}. Jin et al.~\cite{jin2024exploring}, for instance, leverage LLMs to support music-driven reminiscence with older adults by generating contextually relevant conversations and imagery. Yet, these initiatives primarily focus on individual or general social reminiscence with chatbots or unfamiliar partners, rather than mutual reminiscence among trusted individuals sharing past experiences. In our work, we thus move beyond dyadic designs to support multi-user interactions through polyadic chatbots that offer a promising alternative: one that retains the expressive benefits of chatbots while minimizing risks of over-dependence and enabling richer, socially grounded reminiscence experiences.

\subsubsection{Polyadic Chatbots for Social Reminiscence}

While dyadic chatbots focus on one-on-one interactions, polyadic chatbots are designed to support conversations involving multiple human participants, making them especially well-suited for fostering narrative reminiscence in social contexts. In this light, research on polyadic chatbots is gaining momentum across domains such as collaborative learning~\cite{gregory2012towards}, group facilitation~\cite{kim2021moderator, wagner2022comparing, niksirat2023onthepotential, dejong2024assessing}, online community moderation~\cite{seering2019beyond, govers2024aidriven}, and workplace collaboration~\cite{toxtli2018understanding, li2022improving}. These chatbots have been shown to support group tasks by enhancing information retrieval~\cite{avula2018searchbots}, enabling second-language learners to participate in conversation~\cite{guo2019using}, and facilitating equitable participation and time management in group discussions~\cite{kim2020botinthebunch}.

Beyond task coordination, polyadic chatbots have also contributed to \textit{relationship formation} and \textit{social cohesion}. For example, Shin et al.~\cite{shin2023introbot} demonstrated that chatbot-mediated introductions between unfamiliar teammates increased group trust, cohesion, and ideation. Chatbots have also been used to foster belonging in communities by scaffolding relationship-building activities~\cite{low2021socia}. In more intimate settings~\cite{hassenzahl2012all, balaji2024designing}, PocketBot~\cite{zheng2021pocketbot} supports long-distance couples by offering prompts for deep conversations, humor, and conflict resolution, encouraging more emotionally meaningful exchanges.

Taken together, polyadic chatbots have shown promise in shaping group dynamics, facilitating responsive and balanced communication, and enhancing interpersonal closeness—qualities central to mutual reminiscence~\cite{reis1988intimacy, bavelas2000listeners}. Yet, the role of LLM-powered polyadic chatbots in supporting reciprocal and co-constructed reminiscence remains an open question. In this work, we explore how polyadic chatbots can facilitate \textit{reciprocal mutual reminiscence} by integrating conversational AI with principles of narrative reminiscence. Unlike prior work such as PocketBot, which relies on rule-based scripts and predefined prompts, our approach leverages the generative capabilities of LLMs to enable more fluid, free-form conversations, particularly those involving emotionally rich topics. By supporting reciprocal exchanges and co-construction of narratives between loved ones, our system aims to support the social and emotional value of reminiscing together.

\section{REMINI}
\label{sec:remini}

To address the underexplored design space of mutual reminiscence, we developed \textsc{Remini}, a chatbot designed to facilitate mutual reminiscence among loved ones by encouraging positive memory sharing and emotional reflection. Built on Telegram which is one of the most widely used messaging platforms globally, \textsc{Remini} integrates seamlessly into group chats as if it were a regular user. Powered by a large language model (GPT-4o), \textsc{Remini} facilitates meaningful and dynamic conversations. 

\subsection{Design Rationale}
Drawing directly from the SFAM framework’s emphasis on emotionally rich narration, responsive listening, and relational context (see Section~\ref{sec: SFAM Framework}), as well as mutual reminiscence’s core function of fostering intimacy and positive affect through shared storytelling (see Section~\ref{sec_narrative_reminiscence}), our design rationale centers on three strategies:

\subsubsection{Support emotional expression through facilitating positive memory sharing}

Reminiscing typically emphasizes positive memories, enhancing social bonding and positive emotions. Due to a `pleasantness bias'~\cite{rasmussen2009emotional, wolf2021thevalence}, positive memories are recalled more frequently, boosting mood and social connection~\cite{mclean2008whyrecall, wolf2019themood}. Thus, \textsc{Remini} fosters a positive environment, gently engaging users with casual topics like recreational activities before prompting joyful memory sharing, guided by hints such as \texttt{``A positive memory is one that makes you feel happy, brings a smile to your face, or reminds you of fun times.''} Its empathetic, non-judgmental conversation style further enhances user well-being and interaction quality~\cite{skjuve2021mychatbot, ta2020user, laestadius2024toohuman}. For example, after users detailed their memories, \textsc{Remini} responds warmly with affirmations like \texttt{``It's wonderful to hear about all these delightful details! The silence challenge, the hunger, the delicious ramen, the good-looking runners—it all adds such richness to the memory.''} (G12)

\subsubsection{Provide contextual prompts for in-depth reminiscence} 

Autobiographical memories are characterized by varying degrees of vividness and detail, often richer when they carry significant emotional weight, leading to intense re-experiencing of past events~\cite{larsen2014whatisit, wright1992construction, holland2010emotion}. Such vivid memories, like recalling the joy of a concert with loved ones, are not only more memorable but also facilitate deeper emotional disclosures and intimacy~\cite{laurenceau1998intimacy}. To leverage this, \textsc{Remini} is designed to enhance memory narration by prompting users to add details and emotions to their recollections. For instance, after participants shared a specific memory such as taking their children to an aquarium, \textsc{Remini} asked, \texttt{``What specific sights or exhibits stood out to you at the aquarium? Were there any specific moments with your child that melted your heart?''} and  \texttt{``How did each of you felt during these particular moments at the aquarium?''} (G2) These prompts help users relive their memories vividly, strengthening emotional bonds and intimacy.

\subsubsection{Promote mutual reminiscence through reciprocal self-disclosure between loved ones}

Mutual reminiscence serves crucial social functions, including enhancing intimacy and fostering empathy between individuals~\cite{alea2007illkeepyouinmind, theodore2014relations, pohl2014autobiographical}. These functions are influenced by the dynamics of memory sharing, such as responsiveness and engagement, which are integral to the quality of social interactions~\cite{laurenceau1998intimacy}. In addition, reciprocal self-disclosure, particularly through turn-taking, is known to be effective for fostering intimacy and positive emotions~\cite{susan2013takingturns}. To promote balanced participation and reciprocal self-disclosure, \textsc{Remini} is designed to ensure that both users equally contribute and respond to each question before moving to the next prompt. It also encourages users to share their perspectives on shared memories and engage deeply in discussions about their experiences, aiming to enhance emotional connectedness. For example, \textsc{Remini} would prompt \texttt{``Do you feel more emotionally connected during this reminiscence conversation? Please share your thoughts and feelings with each other.''}

\subsection{Phases of \textsc{Remini} for In-depth Reminiscence}
\label{label: phases}

We designed \textsc{Remini}'s conversation system as a state machine~\cite{winograd1986alanguage}, where the system's tasks are divided into multiple phases. Dividing the overall prompts for large language models (LLMs) into different stages—each containing specialized prompts with dedicated goals—has been proven effective in previous studies applying LLMs in human-chatbot interaction to promote mental well-being~\cite{kim2024mindfuldiary, seo2024chacha}. 

The \textsc{Remini} dialogue framework is primarily grounded in the SFAM, which emphasizes narrative detail, emotional expression, and responsive listening~\cite{alea2003you, philippe2008functions}. As a practical reference for structuring prompts, we also drew on Hallford’s positive reminiscence induction~\cite{hallford2022remembering}, a component of cognitive-reminiscence therapy designed to promote healthy psychological outcomes through recalling and exploring positive past experiences~\cite{hallford2024guided, watt2000integrative}. Though initially structured as three weekly sessions focusing on relationships, coping, and meaning-making~\cite{mf2023remembering, hallford2024randomized} in individual or group therapy settings, Hallford’s approach sufficiently aligns with SFAM’s principles for improving reminiscence quality. Thus, we condensed it into a single, relaxed mutual reminiscence session for loved ones in daily life, guiding participants to elicit, narrate, and reflect on uplifting memories, thereby enhancing positive affect and interpersonal connection. Detailed descriptions and GPT prompts are provided in subsequent sections and Appendix~\ref{app:remini}.

\subsubsection{Rapport Building}

The \textit{Rapport Building} phase serves as an ice-breaker, creating a warm, friendly environment to help users feel comfortable discussing personal memories. After initiating the conversation, \textsc{Remini} greets users: \texttt{``Hello! I'm Remini, your reminiscence companion. How are you both doing today?''} It then invites each user to share recent hobbies, recreational activities, and fun facts about each other, while also disclosing information about itself to encourage reciprocal self-disclosure~\cite{lee2020ihearyou}, \texttt{``Even though I’m a chatbot, I absolutely love `traveling' through the stories you share with me.''}(G8). Finally, \textsc{Remini} shifts its focus to their relationship, prompting users to describe how they met and how their connection has evolved, preparing them for the subsequent \textit{Memory Narration} phase.

\subsubsection{Memory Narration}

In this phase, \textsc{Remini} guides users in narrating memorable, positive shared experiences. Initially, it poses transitional questions to introduce reminiscence (e.g., \texttt{``Are you both enjoying recalling happy memories together? When was the last time you discussed happy memories?''}), then provides examples for inspiration (e.g., \texttt{It could be a special trip, an unforgettable party, or a hobby you both enjoyed.''}). \textsc{Remini} then invites users to vividly write down a chosen memory in detail, visualizing the setting, emotions, and events as if reliving it. Users read each other's narratives, discuss, and share reflections before moving on to the \textit{Elaboration} phase.

\subsubsection{Elaboration}

The \textit{Elaboration} phase enhances users' autobiographical memories by prompting them to add factual details and emotions to their previous narrations. In this phase, \textsc{Remini} first identifies each user's most cherished moment, then uses targeted prompts to increase the vividness of their memories, such as asking, \texttt{``How did you feel by the lake in Hakone?''} These prompts are inspired by Gardner et al.'s methods for detailing autobiographical memory through various aspects like context and emotions~\cite{gardner2012quantitative}. A specific prompt is provided to \textsc{Remini} for this purpose (see Appendix~\ref{app:remini}) . For example: \texttt{``Briefly introduce why adding details may promote reminiscing, and provide at least three contextual prompts to inspire both users to describe details about that moment, such as people, places, things, times, and contexts.''} Following this, the phase shifts to \textit{Reflection}.

\subsubsection{Reflection}

Prior research on chatbot-mediated expressive writing has demonstrated that follow-up questions encourage deeper reflection, enhancing insights into participants' narratives~\cite{park2021wrote}. Once the details of the memory are established, the \textit{Reflection} phase guides users through deeper reminiscence to promote more self-disclosure on thoughts and feelings. This includes reflecting on the meanings of their shared memories, feelings of emotional connectedness during reminiscence, and perceptions about themselves and others within the context of their relationship.

\subsubsection{Summary}

In the last phase, \textsc{Remini} briefly summarizes the entire conversation and encourages users to review their reminiscence. For example, \texttt{``Today you both shared profoundly meaningful memories and reflections. You revisited the cherished moments of your first hug and kiss, emphasizing how these experiences dissolved fears and deepened your connectedness.''} (G3) Users are encouraged to express gratitude and share any final thoughts, reinforcing gratitude's role in enhancing relationships~\cite{algoe2008beyond}. \textsc{Remini} then concludes the session with a farewell.

\subsection{Baseline Chatbot}

We developed a baseline chatbot offering only memory triggers to systematically compare the impact of \textsc{Remini}’s structured facilitation on mutual reminiscence. First, prior work has already demonstrated that chatbots can effectively support individual reminiscence~\cite{zhai2024exploring, xu2024memory}; our goal is to extend this to the domain of mutual reminiscence, where co-construction and reciprocity are central. Second, using a no-chatbot condition would have complicated direct comparisons, as participants would rely on various unmediated approaches (e.g., free-form conversation, self-guided reflection). Third, comparing \textsc{Remini} to a simpler chatbot, not just to the absence of a chatbot, situates our findings within the broader landscape of memory-based conversational systems, allowing us to isolate and examine how deeper facilitation techniques enhance emotional expression, reciprocal self-disclosure, and social bonding.

Thus, the baseline chatbot retains only the \textit{Rapport Building} phase (first phase), along with a simplified \textit{Memory Narration} phase (refer to Appendix~\ref{app:baseline} for details on the prompt engineering for the baseline). We include the Rapport Building phase so that the chatbot remains ``visible'' as a participant in the group conversation~\cite{zheng2022ux}, but it does not provide deeper structured guidance. After the initial prompts, participants are free to reminisce with no constraints on response length or content. The chatbot remains mostly passive, offering only brief acknowledgments (e.g., ``I really enjoyed listening to your stories. It’s clear you treasure these shared moments.'') and occasionally asking if they wish to continue.

\begin{figure*}[t]
  \centering
  \includegraphics[width=\linewidth]{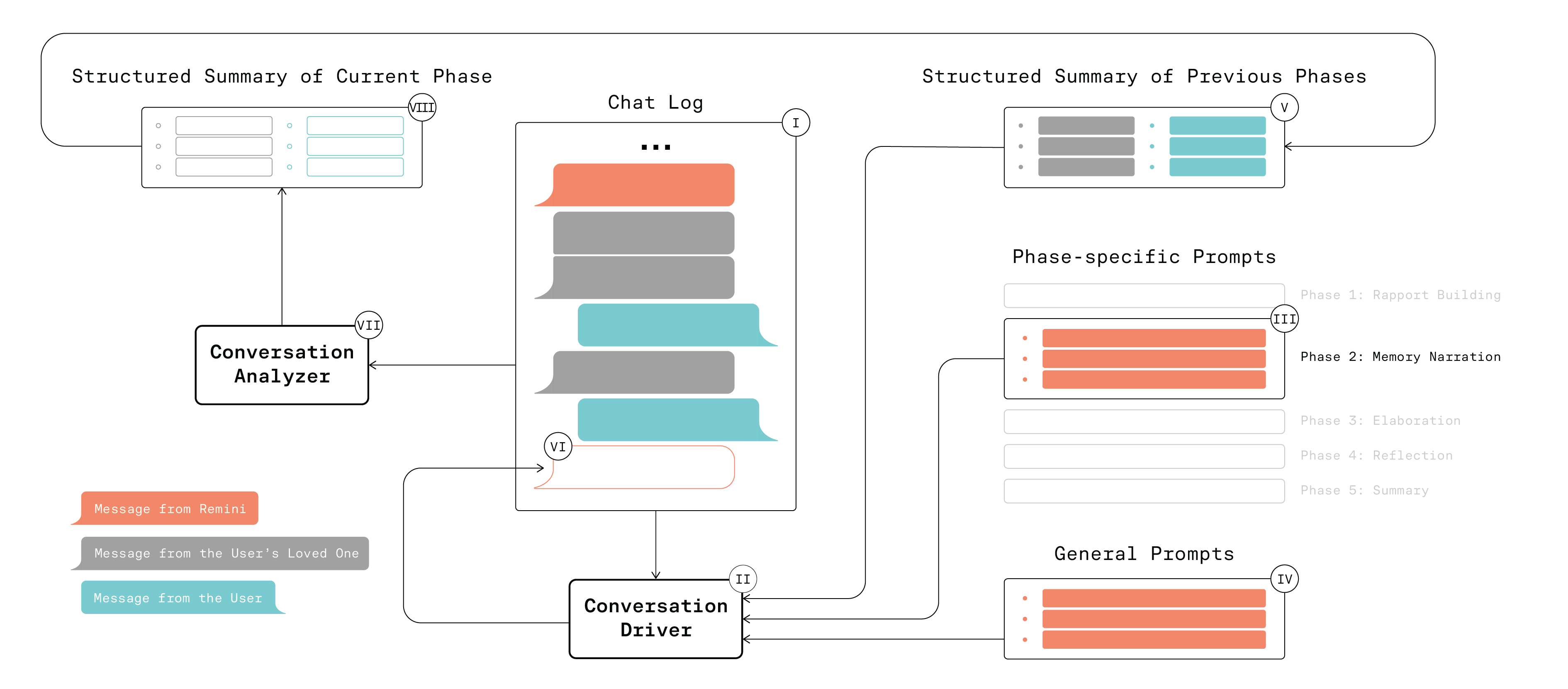}
  \caption{Example workflow of \textsc{Remini}'s Conversation System in Phase 2 of Memory Narration: After being mentioned in the group chat, (1) the conversation driver receives input consisting of combined prompts from the current chat log (I), phase-specific prompts for the current phase (III), general prompts applicable to all phases (IV), and a structured summary containing key information from both users' previous interactions (V). (2) The conversation driver then generates a response message (VI) and sends it back to the Telegram group chat. (3) Subsequently, the chat logs are forwarded to the conversation analyzer (VII), which generates a structured summary for the current phase (VIII).}
  \Description{The diagram describes the workflow of \textsc{Remini}'s Conversation System in Phase 2, highlighting how the system manages dialogue in a group chat setting. Initially, the conversation driver collects inputs from the current chat log, structured summaries of previous interactions, phase-specific prompts, and general prompts applicable across all phases. It then synthesizes these inputs to generate and send a tailored response back to the group. Concurrently, the conversation analyzer processes the chat logs to create a structured summary of the ongoing phase, ensuring that the conversation remains contextually relevant and progressively builds upon previous exchanges.}
  \label{fig:Interaction}
\end{figure*}

\subsection{Implementation and Prompt Engineering}

\subsubsection{Implementation}

According to Li et al. on AI-based conversational agents for well-being~\cite{li2023systematic}, mobile applications and instant messaging platforms offer benefits such as broader reach, ease of use, and convenience. We thus selected Telegram as the platform for \textsc{Remini} due to its robust support for chatbots, widespread use in our target region, and its group chat functionality, which facilitates seamless integration of \textsc{Remini}. The chatbot was developed using Python and Telegram's Bot API\footnote{https://core.telegram.org/api}, enabling efficient message exchanges between users and \textsc{Remini}. For the underlying LLMs, we employed \textit{GPT-4o} via OpenAI's API\footnote{https://platform.openai.com/docs/overview}, specifically the \texttt{gpt-4o-2024-05-13} model for its superior speed and comparable performance to \textit{GPT-4}. This choice was crucial for reducing response times and enhancing the conversational flow, making the user experience smoother and more engaging.

\subsubsection{Interaction with \textsc{Remini} in Group Chat}

\textsc{Remini} is deployed in a group chat on Telegram. However, chatbots in polyadic settings may be perceived as intrusive or distracting~\cite{toxtli2018understanding}, suggesting that they should ideally be ``ignorable'' when human interaction is the primary focus~\cite{zheng2022ux}. To avoid this and respect human interaction, \textsc{Remini} responds only when directly mentioned via the ``@ReminiStory\_Bot'' function. This approach differs from typical instant-reply chatbots, prioritizing user convenience and control. Additionally, a ``Continue'' button under each \textsc{Remini} message allows users to progress the conversation at their pace. Utilizing interactive buttons has proven to enhance user engagement and clarify chatbot functions~\cite{isabel2022understanding}. This setup ensures that users can interact freely and maintain control over the conversation flow.

This interaction design serves two main purposes:

\begin{itemize}
\item {\textit{User Autonomy}}: Users have the freedom to engage in extended discussions without unsolicited interruptions from \textsc{Remini}.
\item {\textit{Controlled Progression}}: Users have the initiative to advance the conversation, fostering a more natural and comfortable communication environment.
\end{itemize}

\subsubsection{Architecture of \textsc{Remini}}

We integrated two LLM-driven modules, the Conversation Driver and the Conversation Analyzer, to power \textsc{Remini}. Aware of the issues that overly complex or verbose prompts pose for language models' reliability and accuracy~\cite{perez2021truefewshot, brown2020language}, we used phase-specific tailored prompts to enhance model performance, as supported by previous studies~\cite{reynolds2021prompt}. The conversation is structured into five distinct phases, detailed in section~\ref{label: phases}, with users identified by user ID. Figure \ref{fig:Interaction} shows the system design activated whenever \textsc{Remini} is mentioned in the chat.

\subsubsection{Conversation Driver}

The Conversation Driver is responsible for receiving instructions from the system and messages from the users as input, generating appropriate responses to send back to the group chat, and determining when to transit to the next phase of the conversation.

The instructions from the system are divided into four specific segments:

\begin{itemize}
\item {\textit{Conversation History of the Current Phase}}: Records of the dialogue that has occurred in the current phase.
\item {\textit{Phase-Specific Prompts}}: Task lists designed according to \textsc{Remini}'s objectives in each phase.
\item {\textit{General Prompts}}: Instructions that apply to all phases, defining overall guidelines such as \textsc{Remini}'s role, tone, language rules, and a task overview.
\item {\textit{Conversation Summary}}: A summary generated by the Conversation Analyzer, containing key information related to each task completed by the Conversation Driver in the previous phases.
\end{itemize}

These segments prompt \textsc{Remini} to generate its next message in the group chat, consisting of two parts: responding to users' previous inputs and posing the next question to advance the conversation based on phase objectives. If a question is unanswered or a user doesn't respond, \textsc{Remini} may re-prompt or invite that user to reply. It only re-prompts if there's no response but will proceed if users choose not to answer, respecting their autonomy. Once both users have responded to the final question of a phase, the Conversation Driver signals ``phase done'' to transit to the next phase.

\subsubsection{Conversation Analyzer}

Upon completing a phase, the Conversation Analyzer generates a summary using three elements: the conversation history of that phase, general prompts outlining the phase's role and objectives, and phase-specific prompts that guide the summary creation. It processes this information to produce a concise summary of the completed phase. Summaries from previous phases are compiled and provided to the Conversation Driver, ensuring continuity and context-awareness for subsequent phases.

\section{User Study}

\subsection{Recruitment and Participants}
\label{sec:Covariates}
Participants were recruited via posters and emails distributed across our university campus, as well as through a Telegram group chat specifically designed to connect local researchers with potential study volunteers. The selection criteria included: (a) participants must be paired with a friend (with a friendship lasting over six months), a close family member, or a significant other; (b) willingness to share personal mutual memories with their paired partner; (c) proficiency in both written and spoken English; (d) regular use of Telegram as a daily social messaging platform; (e) be aged 18 or above.

We successfully recruited 48 participants, consisting of 15 males, 32 females, and 1 non-binary individual, with ages ranging from 18 to 33 ($Median= 22$). After registering for the experiment, participants were asked to complete a survey to gather their demographic information as shown in Table~\ref{tab:demo}. Participants were matched across two groups (one per design), based on age, gender, relationship type, relationship length, and personality traits to control for demographic differences that may influence mutual reminiscence as highlighted by SFAM. To account for personality differences, we used the Ten-Item Personality Inventory (TIPI) to measure the Big-Five personality traits~\cite{samuel2003averybrif}. Participants were compensated according to our local IRB guidelines. In the analysis, we included  age, gender, relationship type, relationship length, and personality traits as covariates.

\begin{table*}[t]
  \small
  \centering
  \caption{Demographics Information of Participants. Groups 1 to 12 were assigned to the \textsc{Remini} group, Groups 13 to 24 to the \emph{Baseline} group. The length of relationship (Rel. Length) is measured in years. Note: NB=Non-Binary.}
  \label{tab:demo}
  \resizebox{\textwidth}{!}{%
  \begin{tabular}{cccccc|cccccc}
    \toprule
    Group ID & ID & Gender & Rel. Type & Rel. Length & Age & Group ID & ID & Gender & Rel. Type & Rel. Length & Age\\
    \midrule
    G1 & P1 & Male & Lovers & 2 & 25 & G13 & P25 & Female & Siblings & 24 & 24\\
    G1 & P2 & Female & Lovers & 2 & 24 & G13 & P26 & Female & Siblings & 24 & 24\\
    G2 & P3 & Female & Lovers & 8 & 30 & G14 & P27 & Female & Lovers & 7 & 29\\
    G2 & P4 & Male & Lovers & 8 & 33 & G14 & P28 & Male & Lovers & 7 & 28\\
    G3 & P5 & Female & Lovers & 2 & 27 & G15 & P29 & Male & Lovers & 6 & 27\\
    G3 & P6 & Male & Lovers & 2 & 25 & G15 & P30 & Female & Lovers & 6 & 25\\
    G4 & P7 & Female & Lovers & 1 & 21 & G16 & P31 & Male & Lovers & 2 & 21\\
    G4 & P8 & Male & Lovers & 1 & 21 & G16 & P32 & Female & Lovers & 2 & 21\\
    G5 & P9 & Male & Lovers & 3 & 24 & G17 & P33 & Female & Lovers & 0.7 & 27\\
    G5 & P10 & Female & Lovers & 3 & 23 & G17 & P34 & Male & Lovers & 0.7 & 27\\
    G6 & P11 & Female & Friends & 0.8 & 21 & G18 & P35 & Female & Friends & 1 & 20\\
    G6 & P12 & Female & Friends & 0.8 & 20 & G18 & P36 & Female & Friends & 1 & 19\\
    G7 & P13 & Female & Friends & 12 & 25 & G19 & P37 & Male & Friends & 1 & 28\\
    G7 & P14 & Female & Friends & 12 & 22 & G19 & P38 & Male & Friends & 1 & 29\\
    G8 & P15 & Female & Friends & 9 & 21 & G20 & P39 & Female & Friends & 2 & 18\\
    G8 & P16 & Female & Friends & 9 & 21 & G20 & P40 & NB & Friends & 2 & 19\\
    G9 & P17 & Female & Friends & 6 & 21 & G21 & P41 & Female & Friends & 10 & 27\\
    G9 & P18 & Female & Friends & 6 & 21 & G21 & P42 & Female & Friends & 10 & 27\\
    G10 & P19 & Female & Friends & 0.7 & 21 & G22 & P43 & Female & Friends & 11 & 21\\
    G10 & P20 & Female & Friends & 0.7 & 22 & G22 & P44 & Male & Friends & 11 & 21\\
    G11 & P21 & Male & Friends & 1 & 22 & G23 & P45 & Female & Friends & 6 & 22\\
    G11 & P22 & Male & Friends & 1 & 21 & G23 & P46 & Female & Friends & 6 & 23\\
    G12 & P23 & Female & Friends & 1 & 25 & G24 & P47 & Female & Friends & 1 & 21\\
    G12 & P24 & Female & Friends & 1 & 30 & G24 & P48 & Male & Friends & 1 & 24\\
  \bottomrule
\end{tabular}%
}
\end{table*}

\subsection{Independent Variables (IV)}
The goal of our experiment is to investigate mutual reminiscence through \textsc{Remini}. As such, we have two independent variables. The first one \emph{Chatbot}, is between-subject and has two levels: \{\textsc{Remini}, Baseline Chatbot\}, and the second one is the \emph{Phase}, which is within-subject and has two levels: \{Pre-interaction, Post-interaction\}. The \emph{Phase} IV is used to measure attitude change due to the experimental setup.

\subsection{Procedure}

Participants joined remotely in pairs via Zoom from different locations using personal computers to avoid off-platform communication. The experimenter outlined the study's purpose, procedures, participants' rights, including optional withdrawal and data deletion, and that chat logs would be anonymized for analysis, before obtaining consent. Participants completed a pre-interaction survey and were assigned to one of two groups without disclosure of their group assignment or its rationale. Each pair entered a Telegram group chat with the experimenter and a chatbot (\textsc{Remini} or \emph{Baseline}), and the experimenter exited before interactions began to ensure privacy. Participants could skip chatbot questions without explanation. After interacting, they completed additional surveys and participated in open-ended interviews. Average session durations were approximately 54 minutes (SD = 15.5) for the \emph{Baseline} group and 107 minutes (SD = 47.7) for the \textsc{Remini} group. The study received approval from the university's IRB (No. IRB-24-660).

\subsection{Dependent Variables}
Our choice of dependent variables was guided by the SFAM framework (Section~\ref{sec: SFAM Framework}) and mutual reminiscence’s core function of fostering positive affect and relational closeness (Section~\ref{sec_narrative_reminiscence}). SFAM highlights memory detail, emotional richness, and listener responsiveness as key factors in meaningful memory sharing. To capture these dimensions, we measured dependent variables using (a) chat log data and (b) post-conversation surveys, focusing on both emotional expression and social bonding throughout the reminiscence process.

\subsubsection{DVs extracted from Chat Logs}
\paragraph{Quantitative.} 
All participant conversations were recorded with consent, and we extracted:
\begin{itemize}
    \item \textbf{Reminiscence duration}: the total time spent in shared reminiscence.
    \item  \textbf{Word counts},  \textbf{message counts}, and  \textbf{words per message}: computed using LIWC22~\cite{boyd2022thedevelopment, yla2010thepsychological}, reflecting established chatbot interaction evaluation metrics on user engagement~\cite{hill2015real, li2023systematic, lee2020ihearyou, kim2024mindfuldiary}. Richer narratives and more frequent exchanges indicate higher engagement, a core goal of our \textsc{Remini} system.
\end{itemize}

\paragraph{Qualitative.} 
Two independent coders (one and four years of experience in content analysis) coded all chat logs using ATLAS.ti~\cite{AtlasTi2023}. Their coding focused on:
\begin{itemize}
    \item \textbf{Number and diversity of details}: aligning with Gardner et al.’s quantitative guidelines for autobiographical memory~\cite{gardner2012quantitative}. The inter-rater reliability for presence/absence of details was 
$\alpha = 0.803$, exceeding the recommended 0.70 threshold~\cite{viera2005understanding, mchugh2012interrater}.
    \item \textbf{Self-disclosure (SD)}: we adapted coding rubrics originally used to measure SD in online forums~\cite{barak2007degree}. The coding rubrics, explained with definitions and examples for each category, are shown in \Cref{tab:sd}. SD contributes significantly to interpersonal relationships and may promote the development of liking, understanding, and intimacy~\cite{kreiner2019selfdisclosure}. We categorized disclosures into informational, thoughts, and emotional, evaluating each interaction phase as a unit. As the last phase on \textit{Summary} was very brief (mostly farewells), its data were combined with the \textit{Reflection} phase. Likewise, the coders achieved a high inter-rater reliability (Cohen’s $\kappa\geq0.77$ for all three SD categories), above the typical 0.70 threshold~\cite{viera2005understanding,mchugh2012interrater}.
\end{itemize}

Illustrative excerpts from participants’ reminiscence dialogues are provided in Appendix~\ref{app:snippet}.

\begin{table*}[t]
  \small
  \centering
  \renewcommand{\arraystretch}{1.5}
  \caption{Definition of Different Depth and Categories of Self-disclosure with Sample Participant's Messages.}
  \label{tab:sd}
  \resizebox{\textwidth}{!}{%
  \begin{tabular}{l p{4cm} p{4cm} p{4cm}}
  
    \hline
    
     & \textbf{Informational} & \textbf{Thoughts} & \textbf{Feelings} \\
     
    \hline
    
    \textbf{Level 1} & No information provided or provides basic, surface-level information without going into detail or personal significance. & No thoughts given or states simple thoughts without deeper exploration or personal insight. & No emotions expressed or mentions emotions briefly without elaboration or exploration of their significance. \\
    
    Example & \textit{``We have known each other for 3 years.''} (P30) & \textit{``I think we've always felt emotionally connected.''} (P14) & \textit{``It makes me feel a bit stressful.''} (P40) \\

    \hline
    
    \textbf{Level 2} & Shares specific details that are more personalized but still not very detailed. & Shares more involved thoughts, indicating some level of introspection. & Provides more descriptive accounts of their emotions, giving a clearer picture of their emotional state. \\

    Example & \textit{``So we first met actually at the airport. I was just arriving to [COUNTRY] and she was sending her parents off to home. That was a very awkward but funny encounter.''} (P24) & \textit{``I think it's rare that we could spend so much time together and do the things we love for an extended period of time (and we didn't have to part once the day ends), hence that was why this memory meant so much to me.''} (P10) & \textit{``Spending time exploring [COUNTRY] at a chill pace with her and [NAME] while enjoying things such as taking pictures wearing beautiful hanbok, eating interesting foods and playing with chonky cats was very relaxing :3''} (P12) \\

    \hline
    
    \textbf{Level 3} & Reveals a huge amount of personalized details, contextual information. & Expresses profound cognitive reflections that often involve self-awareness and understanding of the memory's impact on their life. & Shares a deep emotional account, describing complex feelings and may include reflections. \\

    Example & \textit{``We were with each other. We were looking at the lake from the torii gate in Hakone. The wind was strong with a great smell of sea salt, there were many nature paths with rocky trails, it was a peaceful place with a peaceful weather. I felt happy to be there with him.''} (P1) & \textit{``I learnt that during tough times, we should recall on those happy memories and moments we shared together, try to become each other's motivation to move forward, and not take it as hindrance to stop us from progressing, or to cause more trouble in our daily lives.''} (P13) & \textit{``For me, the first hug was a pivotal moment in extinguishing any fear or nervousness. In fact, it boosted my comfort and trust with him as he was a sweet and cuddly person after all, behind his tough, hard and bulky exterior.''} (P5) \\
    
    \hline
    
  \end{tabular}%
}
\end{table*}

\subsubsection{DVs measured by Surveys}
\label{sec: survey DVs}

To evaluate the interpersonal and emotional impact of our mutual reminiscence intervention, we selected five survey-based dependent variables aligned with the SFAM framework and the core functions of mutual reminiscence.

\paragraph{Positive affect (PA)} Positive affect reflects the emotional uplift that mutual reminiscence aims to promote~\cite{cappeliez2007functions}. To measure positive affect, we utilize the Positive Affect subscale of the well-validated Positive and Negative Affect Schedule (PANAS)~\cite{watson1988development}, often employed in technology interventions targeting reminiscence or well-being~\cite{wagener2022mood, alarith2022time, wagener2023selvreflect, schlagowski2023wish}. Participants rated 10 items on a 5-point Likert scale (1 = very slightly or not at all; 5 = extremely), yielding a total PA score ranging from 10 to 50. Higher scores indicate stronger positive affect. PA was measured both \textbf{pre- and post-interaction} to assess changes in emotional state.

\paragraph{Perceived emotional synchrony (PES)} Mutual reminiscence not only fosters individual emotional benefits but also cultivates shared emotional experiences. Emotional alignment or synchrony is one of the main strategies for fostering genuine feeling of connection in technology-mediated systems~\cite{stepanova2022strategies}. Perceived Emotional Synchrony refers to experiencing a shared positive affect (e.g., joy, elation) within a collective setting~\cite{wlodarczyk2020perceived}. To measure this, we adapted a shortened 6-item version of the PES scale from Dario et al.~\cite{paez2015psychosocial} (to reduce participant burden and focus on subscales most pertinent to reminiscence), measured on a 7-point Likert scale (scores 6–42). PES was collected \textbf{post-interaction}.

\paragraph{Perceived relationship quality (PRQ)} Building and reinforcing close relationships is a central function of mutual reminiscence in the SFAM framework~\cite{cappeliez2007functions, alea2003you}. To measure relationship quality, we used a shortened version of Fletcher et al.’s Perceived Relationship Quality Components (PRQC) scale~\cite{garth2000themeasurement}, which has been applied in computer-mediated communication studies~\cite{halpern2017texting}. This adaptation includes six items covering commitment and intimacy, each rated on a 7-point Likert scale (score range 6–42). We measured PRQ \textbf{pre- and post-interaction}. The scale showed strong internal consistency (MacDonald's $\omega = .96$).

\paragraph{Inclusion of other in the self (IOS)} As a complement to PRQ, we employed the IOS scale~\cite{aron1992inclusion}, a single-item measure widely used to gauge closeness in computer-mediated communication research~\cite{an2024emowear, zhang2023icebreaking}. It presents seven Venn diagram-like images depicting two circles (representing two individuals) at varying levels of overlap (1–7). Participants select the diagram that best reflects their perceived closeness. IOS was measured \textbf{pre- and post-interaction}.

\paragraph{Perceived Partner Responsiveness (PPR)} Listener responsiveness is a cornerstone of the SFAM framework and a key enabler of intimacy. Feeling understood, validated, and cared for fosters deeper connections during mutual reminiscence~\cite{Beyzanur2023responsiveness}. Prior system has used this scale in multiparty interactions, including couples~\cite{utami2020collaborative}. We adopted a 4-item short form of the Perceived Partner Responsiveness Scale (PPRS), each item rated on a 7-point Likert scale for a total range of 4–28. This scale showed good reliability (McDonald’s $\omega = .91$) and was measured only \textbf{post-interaction}.

\subsubsection{Interviews}

At the conclusion of the study, we conducted open-ended interviews to gather qualitative data on participants' experiences of mutual reminiscence conversations with the chatbots. We then explored whether the chatbot facilitated meaningful conversations by inquiring if it positively influenced their emotional state, aided in recalling details and emotions, or prompted deeper reflection on their memories. Participants also compared these interactions to sharing memories with their partners in daily life. For the \textsc{Remini} group, we additionally sought suggestions for improving the system and asked if they would consider using \textsc{Remini} regularly for recording or reminiscing about their memories. All interviews were audio-recorded, anonymized, transcribed verbatim, and subjected to thematic analysis as described by Braun and Clarke~\cite{virginia2006using}. 

\subsection{Statistical Analysis}
To analyze our data, we first checked the normality of the data distribution using the \textbf{Shapiro-Wilk test}. If the data were normally distributed, we used a parametric approach, and applied either \textbf{t-tests} (for dependent variables measured for a single independent variable with two levels), or \textbf{ANCOVA} (for dependent variables measured for two IVs). For ANCOVA, we included the covariates described in section~\ref{sec:Covariates}. For non-parametric tests, we used either the \textbf{Aligned Rank Transform (ART) ANOVA}~\cite{wobbrock2011aligned} (for DVs measured for two IVs), or \textbf{Mann-Whitney U} tests (for DVs measured for a between-subject IV), or \textbf{Friedman} tests (for DVs measured repeatedly within subjects), or \textbf{Wilcoxon Signed-Rank} tests (for DVs measured for a within-subject IV).

\section{Findings}

\subsection{Quantitative Findings}

In this section, we present the quantitative findings that primarily address \textbf{RQ1} to evaluate whether Remini’s guided approach elicits meaningful psychosocial and relational outcomes compared to a minimally guided baseline chatbot. 

\subsubsection{\textsc{Remini} Facilitates Greater Positive Affect}
\begin{figure}[t]
  \centering
  \includegraphics[width=\linewidth]{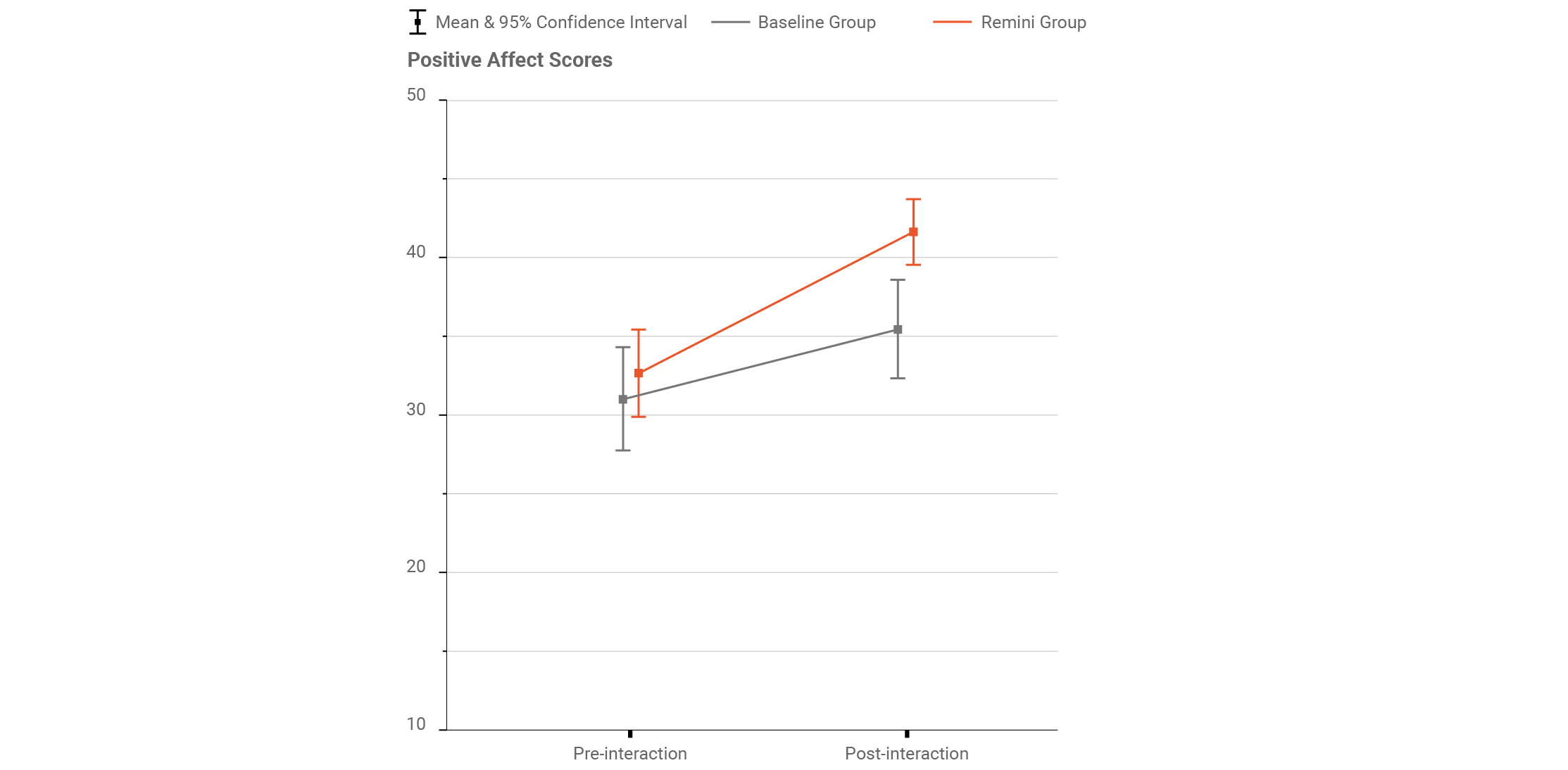}
  \caption{Line plot comparing pre- and post-study positive affect scores between the \emph{Baseline} and \textsc{Remini} groups.}
  \Description{This graph presents a comparison of Positive Affect Scores between the \emph{Baseline} and \textsc{Remini} groups, before and after interaction sessions. Boxplots illustrate the median scores and interquartile ranges for each group at both time points. The lines connecting the pre- and post-interaction medians highlight the trend of scores within each group and shows the 95\% confidence intervals and the mean scores. The \textsc{Remini} group shows a noticeable larger increase in scores from pre- to post-interaction, which is statistically significant as indicated by the interaction effect of Time * Condition (F(1,37) = 6.33, p = 0.016*).} 
  \label{fig:PA}
\end{figure}

The data for Positive Affect (PA) was normally distributed. An ANCOVA showed no significant effect of any covariates measured, nor of any individual independent variable (all $p>.05$). We did however find a significant interaction ($F_{1,37}=6.33$, $p=.016$), which suggests that the average PA scores increase from \emph{pre-} to \emph{post-interaction} and was more substantial in the \textsc{Remini} group compared to the \emph{Baseline} group. For the \textsc{Remini} group, average PA scores increased from 32.67 ($SD=6.88$) to 41.63 ($SD=5.21$). In the \emph{Baseline} group, scores rose from 31.04 ($SD=8.16$) to 35.46 ($SD=7.80$). Figure \ref{fig:PA} displays a line plot comparing the average PA scores between the two groups, both during pre- and post-interaction.

\subsubsection{Reminiscence Boosts Relationship Quality and Closeness }
\begin{figure*}[t]
  \centering
  \includegraphics[width=0.8\linewidth]{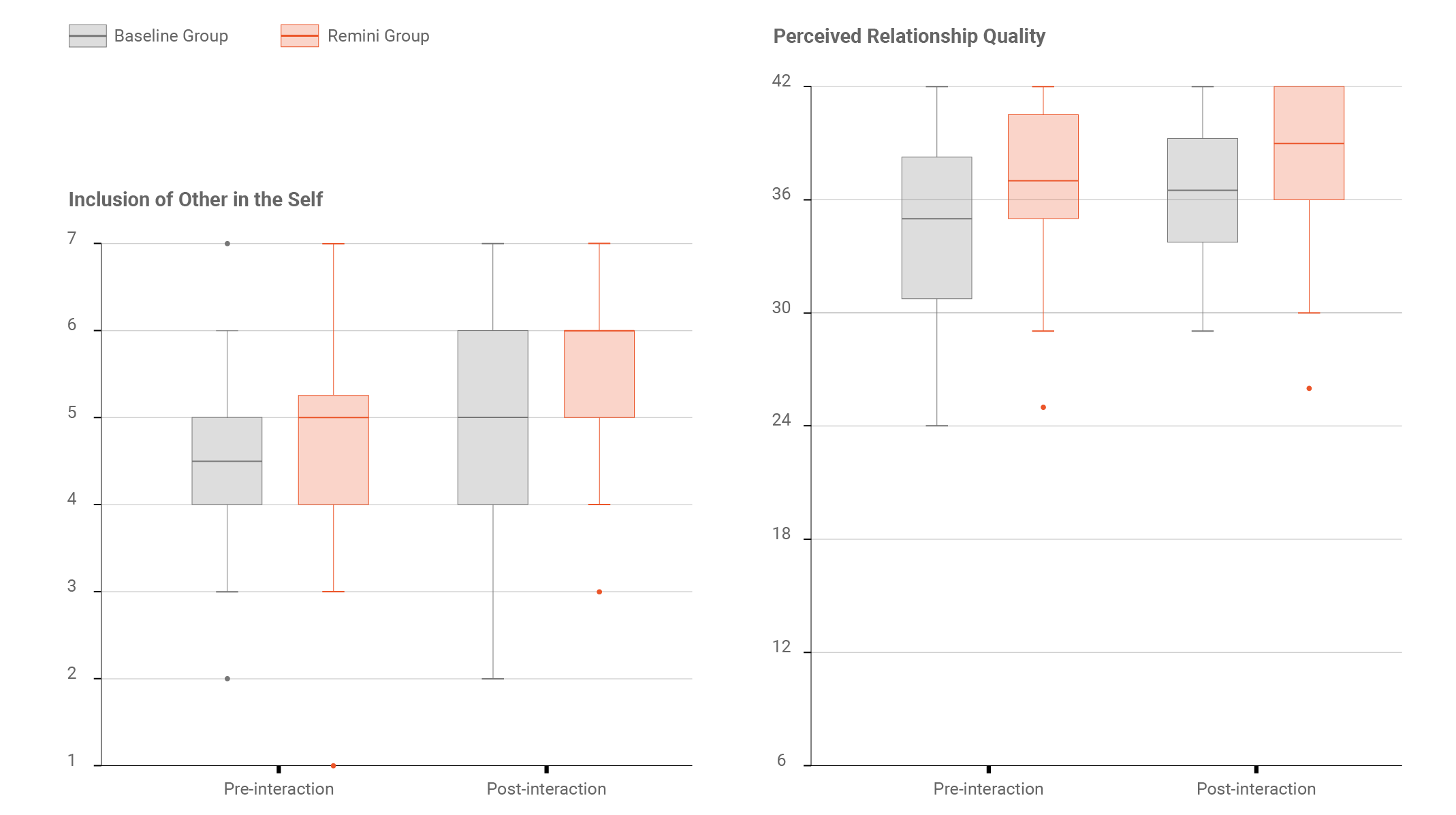}
  \caption{Boxplots comparing pre- and post-study Inclusion of Other In Self scores and Perceived Partner Responsiveness scores between the \emph{Baseline} and \textsc{Remini} groups.}
  \Description{These graphs compare changes in 'Inclusion of Other in the Self' (IOS) and 'Perceived Partner Responsiveness' (PPR) between the \emph{Baseline} and \textsc{Remini} groups before and after interventions. The left graph shows an increase in IOS scores for the \emph{Baseline} and \textsc{Remini} group post-intervention. The right graph illustrates both groups experiencing a rise in PPR scores. Each boxplot includes median lines, interquartile ranges, and outliers.}
  \label{fig:IOS}
\end{figure*}

\paragraph{Perceived Relationship Quality (PRQ)} The PRQ data were not normally distributed; thus, we conducted an ART ANOVA, suitable for non-parametric data. Results showed a significant main effect of \emph{Chatbot Design} ($F_{1,92}=7.26$, $p<.01$) as well as an effect of \emph{Phase}, with the PRQ increasing between the beginning and the end of the session ($F_{1,92}=4.84$, $p=.03$). Median PRQ scores for the \textsc{Remini} group increased from 37 ($IQR (35, 41)$) to 39 ($IQR (36, 42)$), and for the \emph{Baseline} group from 35 ($IQR (30.5, 38.5)$) to 36.5 ($IQR (33.5, 39.5)$). No significant interactions were found. Figure \ref{fig:IOS} (right) displays the comparative PRQ boxplots for both groups and times.

\paragraph{Inclusion of Other in Self (IOS)} An ART ANOVA showed a significant main effect of \emph{Phase} on IOS, ($F_{1,92}=10.19$, $p<.01$) with IOS being higher in the Post-experimental measurement. Median IOS scores for the \textsc{Remini}  group increased from 5 ($IQR (4, 5.5)$) to 6 ($IQR (5, 6)$),  and for the \emph{Baseline} group, from 4.5 ($IQR (4, 5)$) to 5 ($IQR (4, 6)$). No additional effects or interactions were detected. Figure \ref{fig:IOS} (left) shows the boxplots comparing IOS between the two groups and across pre- and post-interaction periods.

\subsubsection{\textsc{Remini} Promotes Higher Perceived Emotional Synchrony}
The data for Perceived Emotional Synchrony (PES) was normally distributed. A t-test showed a significant main effect of our \emph{Chatbot Design} ($t_{46}=2.69$, $p=.01$), with participants using \textsc{Remini} achieving higher scores ($M=33.67$, $SD=5.31$) compared to the \emph{Baseline} group condition ($M=28.83$, $SD=7.01$). Figure \ref{fig:PES} (left) shows the boxplots that compares PES between the two groups.

\subsubsection{\textsc{Remini} Facilitates Higher Perceived Partner Responsiveness}
\begin{figure*}[t]
  \centering
  \includegraphics[width=0.8\linewidth]{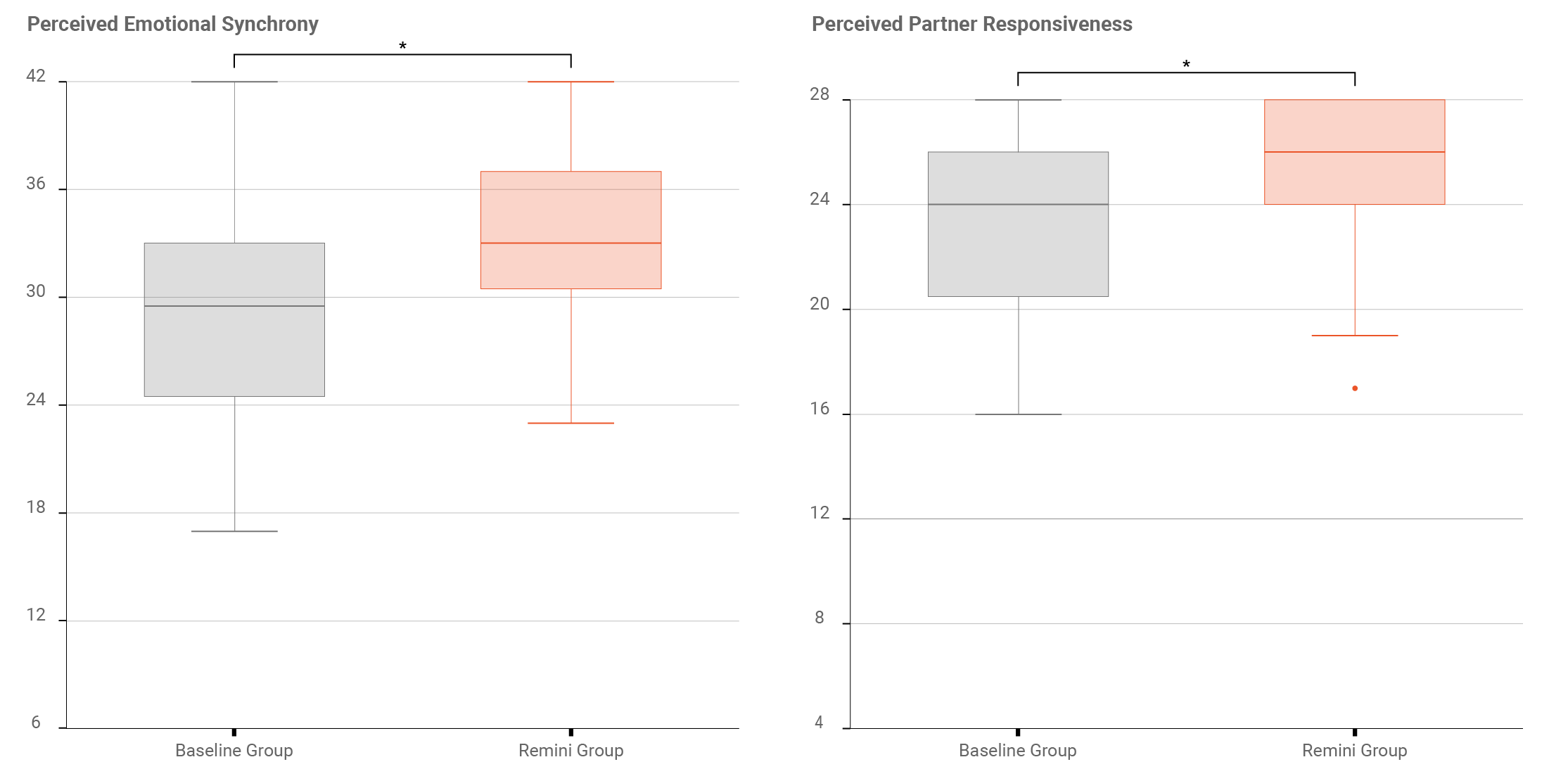}
  \caption{Boxplots comparing pre- and post-interaction scores for Perceived Emotional Synchrony and Perceived Partner Responsiveness across the \emph{Baseline} and \textsc{Remini} groups.}
  \Description{This graph presents boxplots that illustrate the distribution of scores for Perceived Emotional Support (PES) and Perceived Partner Responsiveness (PPR) before and after interaction sessions. The boxplots are grouped by intervention type — Baseline and \textsc{Remini}. Each boxplot displays the median, interquartile ranges, and outliers.}
  \label{fig:PES}
\end{figure*}

A Mann-Whitney U Test with a right-tailed hypothesis showed a statistically significant effect in terms of Perceived Partner Responsiveness (PPR) ($U=378.5$, $p=.029$). Participants utilizing \textsc{Remini} achieved higher scores ($Med=26.0$, $IQR (24.0, 28.0)$) compared to those in the \emph{Baseline} group ($Med=24.0$, $IQR= (20.5, 26.0)$). Figure~\ref{fig:PES} (right) displays the boxplots comparing PPR scores between the two groups.

\subsubsection{\textsc{Remini} Promotes Higher Conversation Engagement}
\begin{figure*}[t]
  \centering
  \includegraphics[width=0.8\linewidth]{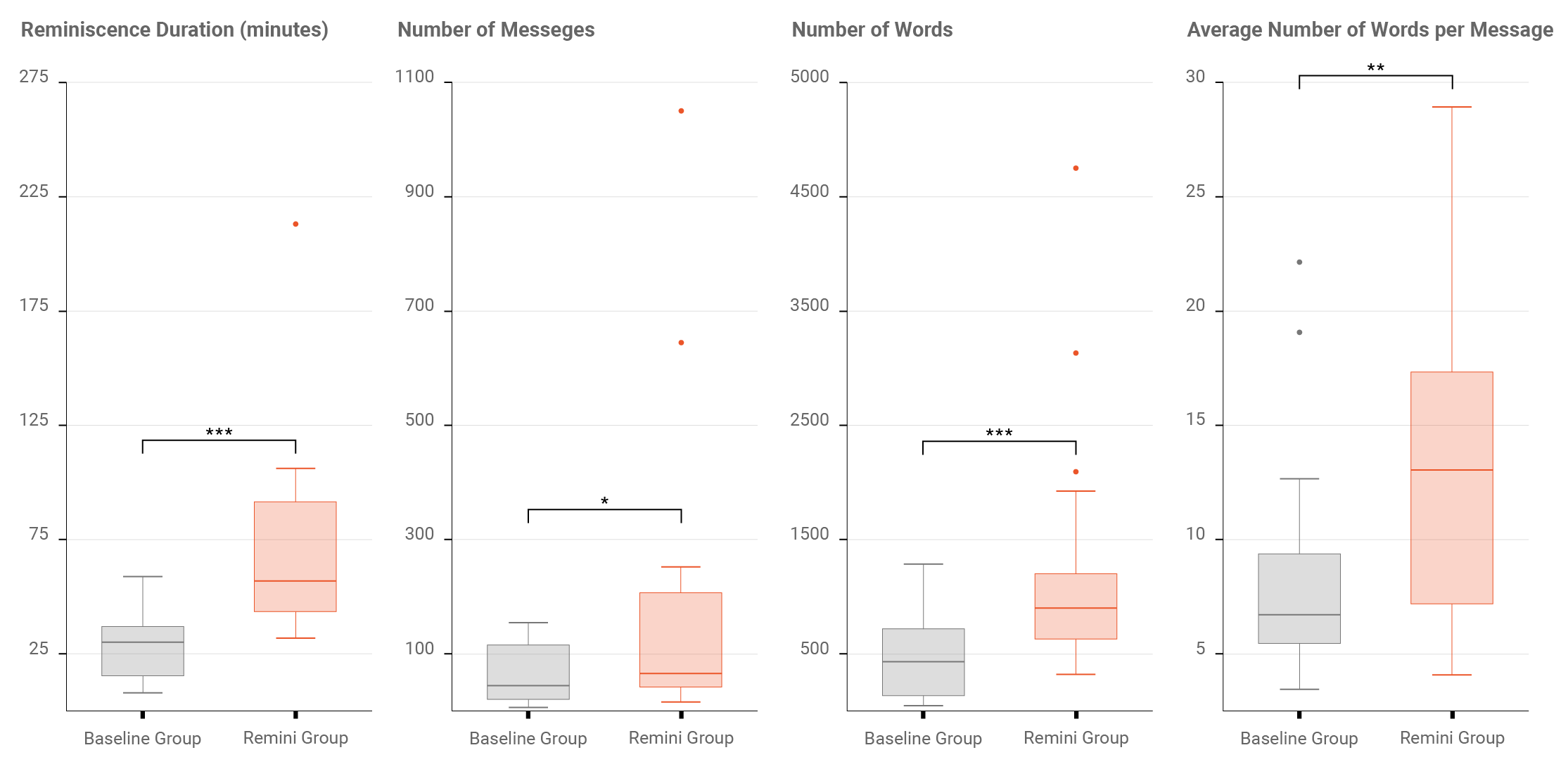}
  \caption{Boxplots comparing communication metrics between the \emph{Baseline} and \textsc{Remini} groups: analysis of reminiscence duration, number of messages, number of words, and average words per message.}
  \Description{These boxplots compare communication metrics—Reminiscence Duration, Number of Messages, Number of Words, and Average Number of Words per Message—between the \emph{Baseline} and \textsc{Remini} groups, both before and after interaction sessions. Across all metrics, the \textsc{Remini} group demonstrates significantly higher engagement. Specifically, the \textsc{Remini} group shows a markedly longer Reminiscence Duration and a higher Number of Words, both with statistical significance (p < .001***). Additionally, the \textsc{Remini} group sent more messages and used more words per message, indicating a deeper level of interaction compared to the \emph{Baseline Chatbot} group.}
  \label{fig:WORD}
\end{figure*}

A Mann-Whitney U right-tailed hypothesis revealed significant differences in the conversation duration ($U=528$, $p<.001$), number of messages ($U=382.5$, $p=.003$), total number of words ($U=479$, $p<.001$) and average number of words per message ($U=402$, $p=.009$). For conversation duration, participants in the \textsc{Remini} group had higher median ($Med=57$, $IQR (43.5, 91.5)$), compared to the \emph{Baseline} group ($Med=30$, $IQR (15.5, 37)$). Similarly, the \textsc{Remini} group exchanged more messages ($Med=63.5$, $IQR (41, 207)$) than the \emph{Baseline} group ($Med=43$, $IQR= (19.5, 115.5)$). Total word counts were also higher in the \textsc{Remini} group ($Med=891$, $IQR (626, 1195.5)$), compared to the \emph{Baseline} group ($Med=431.5$, $IQR (137, 717.5)$). Additionally, the average number of words per message was greater in the \textsc{Remini} group ($Med=13.05$, $IQR (7.2, 17.35)$), than the \emph{Baseline} group ($Med=6.7$, $IQR (5.45, 9.4)$). Figure \ref{fig:WORD} displays boxplots comparing all these factors between the two groups.

\subsubsection{\textsc{Remini} Enables More Vivid Reminiscence Conversation}
\begin{figure*}[t]
  \centering
  \includegraphics[width=0.8\linewidth]{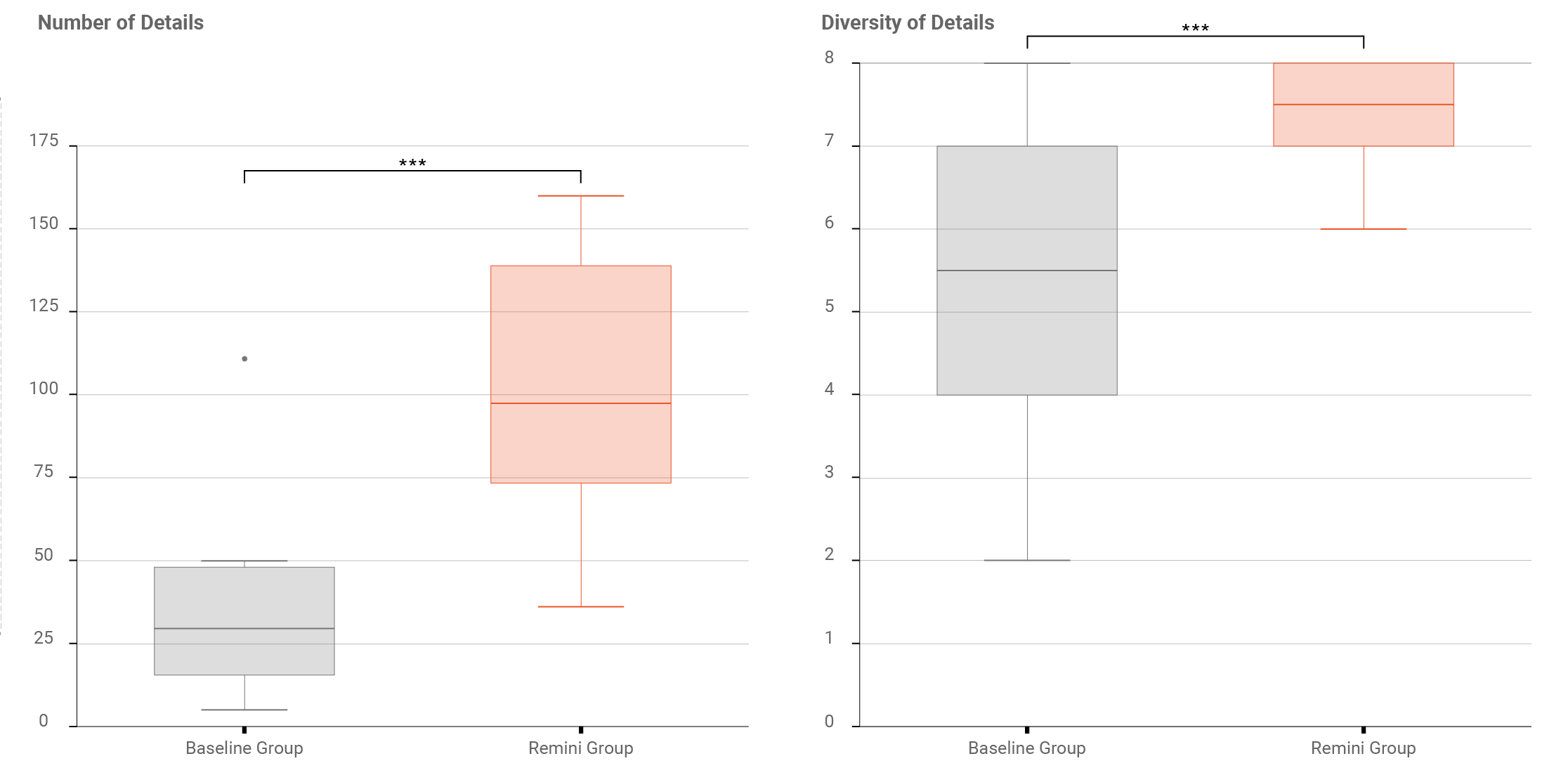}
  \caption{Boxplots comparing the number and diversity of details between the \emph{Baseline} and \textsc{Remini} groups.}
  \Description{These boxplots display the number and diversity of details recalled by the \emph{Baseline} and \textsc{Remini} groups. The \textsc{Remini} group reports significantly higher numbers and more diverse details compared to the \emph{Baseline} group, as indicated by the statistical significance marked with asterisks.}
  \label{fig:MD}
\end{figure*}

A Mann-Whitney U Test, conducted with a right-tailed hypothesis, indicated a significant effect of \emph{Chatbot Design} on the number of details ($U=532$, $p<.001$) and on the diversity of details ($U=462$, $p<.001$). Participants in the \textsc{Remini} group reported a higher number of details ($Med=97.5$, $IQR (73.5, 139.0)$), compared to the \emph{Baseline} group ($Med=29.5$, $IQR= (15.5, 48.0)$). Similarly, the diversity of details was greater for \textsc{Remini} ($Med=7.5$, $IQR (7.0, 8.0)$), versus the \emph{Baseline} group ($Med=5.5$, $IQR (4.0, 7.0)$). Figure \ref{fig:MD} displays boxplots comparing both the number and diversity of details between the two groups.

\subsubsection{\textsc{Remini}'s Different phases of In-depth Reminiscence Promotes Different Types of Self-Disclosure}

This analysis focuses on the different Phases implemented within \textsc{Remini}. We discuss self-disclosure as detailed in Table~\ref{tab:sd}. For reference (see section~\ref{label: phases}), Phase 1 is Rapport Building, Phase 2 is Memory Narration, Phase 3 is Elaboration, Phase 4 combines Reflection and Summary. 

A Friedman Test conducted across four phases of reminiscence indicated significant differences in all categories: Informational ($\chi^2(3) = 54.636$, $p<.001$), Thoughts ($\chi^2(3) = 26.209$, $p<.001$), Feelings ($\chi^2(3) = 25.787$, $p<.001$). Dunn's post-hoc tests with Bonferroni correction identified significant pairwise differences.

\paragraph{Informational Self-disclosure} Significant differences were observed between Phases 1 and 2, 1 and 3, 2 and 4, and 3 and 4 (all $p<.001$). Median scores across the phases for Informational self-disclosure were $Med=1$, $IQR (1, 2)$ for Phase 1; $Med=3$, $IQR (3, 3)$ for Phase 2; $Med=3$, $IQR (2, 3)$ for Phase 3; and $Med=1$, $IQR (1, 1)$ for Phase 4.

\paragraph{Thoughts Self-disclosure} Significant differences were noted between Phases 1 and 4 ($p<.001$), Phase 2 and 4 ($p=.04$), and Phases 3 and 4 ($p=.002$). Median scores for Thoughts self-disclosure were $Med=1$, $IQR (1, 1.5)$ for Phase 1; $Med=2$, $IQR (1, 2)$ for Phase 2; $Med=1.5$, $IQR (1, 2)$ for Phase 3; and $Med=2.5$, $IQR (2, 3)$ for Phase 4.

\paragraph{Feelings Self-disclosure} Significant differences were found between Phases 1 and 2, 1 and 3, and 1 and 4 (all $p<.001$). Median scores for Feelings self-disclosure were $Med=1$, $IQR (1, 1)$ for Phase 1; $Med=2$, $IQR (2, 2.5)$ for Phase 2; $Med=2$, $IQR (2, 3)$ for Phase 3; and $Med=2.5$, $IQR (1, 3)$ for Phase 4.

\begin{figure}[t]
  \centering
  \includegraphics[width=\linewidth]{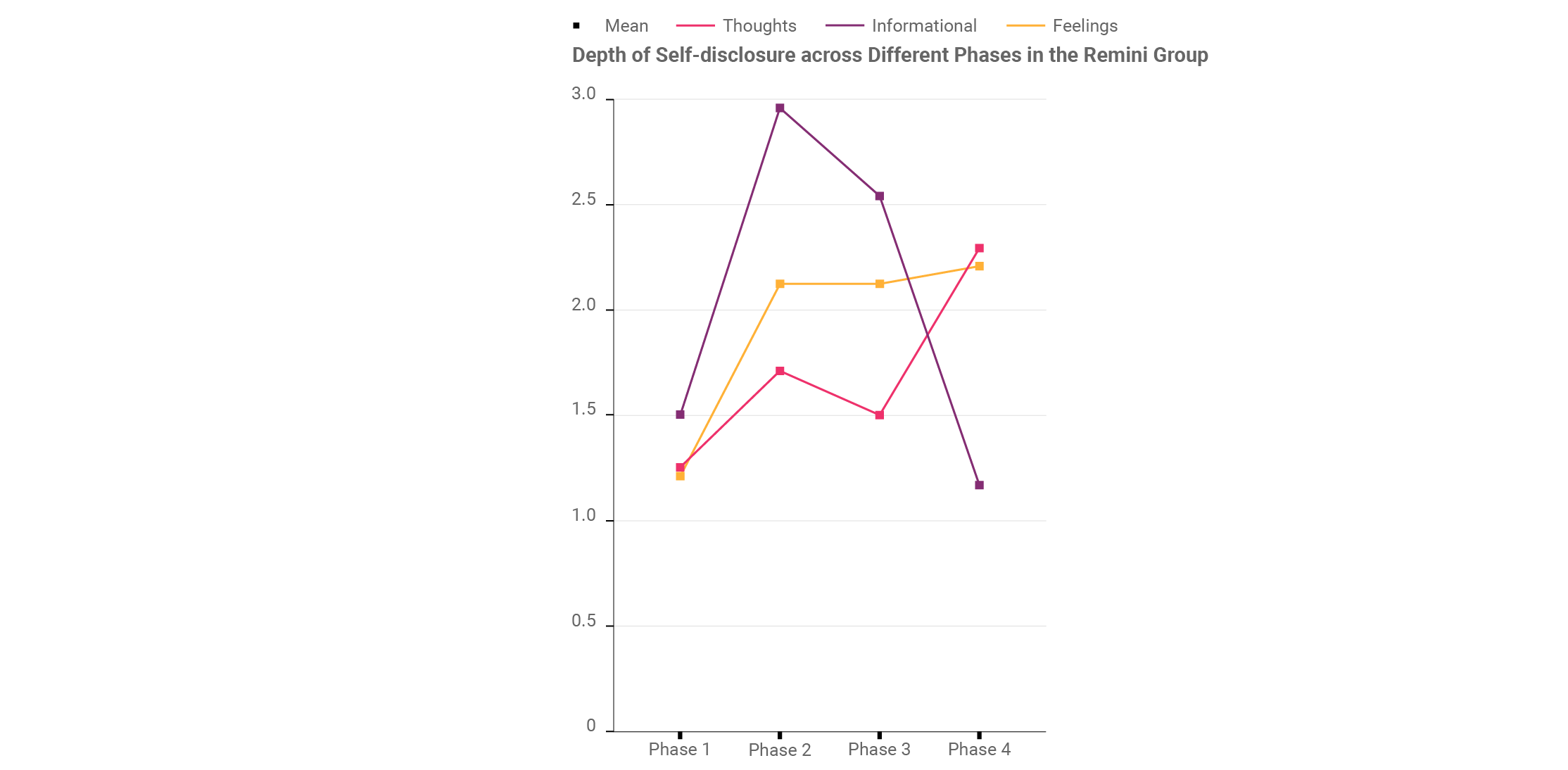}
  \caption{Line plots comparing the average self-disclosure levels in \textsc{Remini} groups across various phases, segmented by Informational, Thoughts, and Emotional categories.}
  \Description{This line plot displays average self-disclosure levels in \textsc{Remini} groups across four phases, segmented into Informational, Thoughts, and Feelings categories. Notably, the group peaks in Phase 2 for Informational, and Phase 4 for Thoughts. Scores of feelings for Phases 2, 3, 4 are similar but higher than Phase 1.}
  \label{fig:SD}
\end{figure}

Figure \ref{fig:SD} illustrates the average levels of self-disclosure across all three categories—informational, thoughts, and feelings—across different phases.

In summary, the structured design of \textsc{Remini} consistently facilitated significantly greater positive affect, deeper emotional synchrony, higher conversation engagement, and more vivid reminiscence recall compared to the minimally guided baseline chatbot. Taken together, these results offer strong support for RQ1, demonstrating how \textsc{Remini}’s structured facilitation contributes to deeper shared memories and robust psychosocial benefits.

\subsection{Qualitative Findings}

We now present the qualitative findings addressing \textbf{RQ2}, which build on the quantitative results showing how varying levels of chatbot guidance influence reminiscence outcomes. Our analysis of interview data reveals distinct experiences between the \textsc{Remini} and \emph{Baseline} groups, yet also identifies shared experiences among participants. In the sections below, we offer deeper insights into participants' mutual reminiscence interactions.

\subsubsection{Facilitating In-Depth Reminiscence}

Participants across both \textsc{Remini} and \emph{Baseline} groups noted that the chatbot’s prompts helped them recall significant yet long-unspoken memories. Several acknowledged that, without these prompts, certain moments would likely remain unmentioned in daily life, remarking, \textit{``We managed to think about memories that … we don’t actively think about on a daily basis''} (P42), and memories that, \textit{``might have slipped through the cracks otherwise''} (P35) or \textit{``we hadn’t thought about in a long time''} (P27).

Beyond simply eliciting memories, \textsc{Remini}’s use of detailed, context-rich prompts offered a more structured recall process that enhanced both detail and depth. Participants felt the prompts prompted them to, \textit{``go very detailed, like fix on your memory space to relive this moment''} (P3), and praised the chatbot for facilitating rich recollections: \textit{``…the bot really facilitated us in recalling exactly what we did and then stuff that we might have forgotten.''} (P12) The follow-up questions contribute to the detailed narration, such as, \textit{``The bot would start by asking us to think about a memory… then ask follow-up questions… which were useful in making me think more deeply.''} (P13) However, some users in the \textsc{Remini} group found that the structured guidance could feel limiting: \textit{``…it’s very structured, [but] on the other hand, it might limit me a little…''} (P13). Conversely, some in the \emph{Baseline} group appreciated the less directive prompts as they fostered a freer flow, while others still wished for more probing questions, \textit{``Asking more questions… would help dig out more details.''} (P25) At the same time, participants appreciated the \textsc{Remini}’s responsive and contextual responses. P10 felt \textsc{Remini} \textit{`` really weaves in and abstracts out the things you mentioned and breezes quite well into its response, questions, and prompts to us.''} Participants also highlighted its ability to sustain the discussion, noting, \textit{``Even if you feel you’re running out of things to say, you can ask the chatbot for more prompts.''} (P22) In closing, the chatbot’s summaries effectively synthesized these discussions, providing \textit{``a more structured conversation about our favorite memories''} (P13) than might arise in spontaneous conversation alone. 

\subsubsection{Experiencing Feeling of Connection through Collaborative Narration and Reciprocal Self-disclosure}

\textsc{Remini}’s structured approach fostered collaboration: participants were prompted to first narrate individually, then read each other's narration, and further elaborate collaborativly. This process created a sense of teamwork or embarking on a ``quest'' (P5) of narrating memory step-by-step during which participants supplement each other's narration as noted by P15 \textit{``When one of us forgets some details and then the other person brings it up.''} Eventually, this collaboration created narrations that cannot be achieved if the narrators and listeners roles are fixed, as remarked by P18 \textit{``feels like the memory is more rounded now, very whole, because there are two different sides to it.''} 

This collaborative process further enabled them to explore each other's distinct perspectives which is often missed in daily conversations. As P19 observed, \textit{``I didn't know how she interpreted the specific memory, and then knowing her interpretation made me think of that memory through her eyes.''} Many participants shared similar reflection, \textit{``…in daily reminiscence, maybe we are less willing to talk about things from our own perspectives…''} (P9) and \textit{``when we talked about past memories, it was just one-sided, so I never really hear what the memory looked like from her perspective.''} As a consequence, participants noted that this discovery of partner's perspectives cultivated engagement in reminiscence, such as P31, who said, \textit{``I was a bit excited throughout the process because I was waiting for her answer to see her views.''} And also new insights of their partners, as P15 observed, \textit{``it does make me feel like, ‘Oh, actually she paid attention and took the effort to remember it.’''} This experience even sparked an eagerness to create new memories, with P10 commenting, \textit{``It makes me look forward to making more memories to reminisce about,''} and P9 adding, \textit{``I think I want to try this bot with a family member… to see or hear from their perspectives.''}

This form of reminiscence attach importance to not only the shared memory itself but also individual viewpoints facilitate participants to voice out their own feelings, an element often overlooked in daily conversation. One participant noted, \textit{``Normally when we talk, we just focus on the events, but not so much on how we felt. But this bot really brought those emotions to life.''} (P21) Other than expressing themselves, participants also validated each other’s feelings, as noted by P15: \textit{``When we were reminiscing, we sort of shared similar feelings… which I think validated my own feelings.''} This two-way interaction reflects the high responsiveness between participants, indicate attentiveness and comprehension, manifest the process of reciprocal self-disclosure, which reinforced their intimacy, with P5 noting, \textit{``This exercise reminded us why we are putting the effort into our relationship… it’s very meaningful.''} and P14, \textit{``It allows us to reflect back on the happy memories, and it makes us appreciate all this little joy in life more.''}

\subsubsection{Influence on Conversation Dynamics}

Participants initially felt uneasy talking in the presence of the chatbot but grew increasingly at ease. As one participant explained, \textit{``At first we were quite awkward with it… but slowly it became more and more easygoing.''} (P16) This shift to a more relaxed environment was aided by the chatbot’s non-judgmental stance and supportive nature, which some likened to \textit{``a mutual friend, like a person in real life''} (P32). Furthermore, knowing that a ``third party'' was listening, motivated participants to highlight more positive aspects of their memories: \textit{`` We might have encountered some challenges during the travel, but I don't want to share those with a third party, even if it's a chatbot. So, I'll focus on the positive things, and that made me happy.''} (P44)

Participants described the chatbot in various ways, characterizing it as a therapist, moderator, and facilitator due to the introspective questions it posed, which encouraged self-disclosure about emotions and thoughts seldom discussed in daily life. It was also seen as an observer, with several noting that it \textit{``doesn’t really interrupt you''} (P17) beyond \textit{``providing the initial prompt''}, which led P35 to forget about its presence. Similarly, P36 mentioned that the interaction felt natural, as they \textit{``talked like they normally do''}. But some wished it would be more proactive and human-like, feeling it should \textit{``sometimes interject… rather than us needing to continue ourselves.''} (P21) However, some participants found its non-human nature off-putting,: \textit{``It feels more awkward because the question being asked is based on algorithms instead of natural thought being put into it.''} (P31) Some also worried about it being too human-like, which could be unsettling: \textit{``If the language becomes a bit more human-like, I would feel a bit freaked out.''} (P19) 

Because the chatbot was seen as an ``outsider'' with no prior knowledge, participants felt compelled to share more backstory and context than in typical one-on-one interactions. This led to richer, more comprehensive storytelling. As P4 explained, \textit{``We were treating the chatbot as someone who doesn’t know anything, so we go from the very basics…''}, and P41 added, \textit{``If it’s just the two of us talking, then I only have to mention keywords.''}

Another effect of having a third party was its potential to diffuse tension and encourage patience. P3 noted that the chatbot’s presence could help \textit{``calm down your emotions, make you more neutral,''} preventing discussions from turning into scolding. They elaborated, \textit{``You treat the person like a beginner… instead of assuming the other party already knows,''} thereby reducing frustration. However, this same presence could also formalize the conversation, with one participant stating, \textit{``Because of the presence of the bot, the conversation… was more formal than usual.''} (P30) And P29 further explained: \textit{`` In our conversation, sometimes we'll  use nicknames or we'll send  emojis or pictures instead of text. So I'm not too sure whether the chatbot can take in those inputs.''} Others suggested for more natural language use that mirrors participants' own styles (P37, P29).

\section{Discussion}

This section examines the multifaceted impact of our chatbot on mutual reminiscence. We first assess how structured, context-aware guidance enhances narrative construction by eliciting richer details. Next, we explore collaborative memory sharing through reciprocal self-disclosure that deepens interpersonal bonds. Finally, we analyze the chatbot’s role in intimate conversations, addressing its social mediation alongside challenges of AI anthropomorphism and privacy.

\subsection{Narrative Construction through Guided Reminiscence}

Our findings indicate that a chatbot providing structured guidance and contextual prompts significantly enriches shared memories compared to broad prompts alone. The \textsc{Remini} group experienced longer, more engaged conversations, as shown by increased duration, turn-taking, and word count, and elicited richer, more diverse details. These results build on prior work showing the value of curated cues and conversation starters for social reminiscence~\cite{baker2021school, kang2021momentmeld}, but further highlight that \emph{structured} guidance yields more coherent, engaged narrative-building than generic triggers like \emph{Baseline} chatbot provided. Participants appreciated how \textsc{Remini} helped them recall or discover details they might have overlooked, pointing to appreciation for vivid, immersive chatbot-mediated reminiscence, consistent with prior work~\cite{zhai2024exploring}.

Results from our findings suggest that blending structured guidance with contextual prompts greatly improves storytelling in chatbot-mediated reminiscence. For instance, in intergenerational contexts where improvised narratives often lack organization~\cite{papacharissi2016affective, liem2020structure}, a chatbot can actively facilitate conversation flow, prompt recollections, and summarize stories to enrich overall narrative construction. Additionally, future work could integrate this form of guidance with digital memorabilia or VR-based memory cues to foster deeper and more immersive reminiscence experiences~\cite{baker2021school}. While Baker et al.~\cite{baker2021school} relied on human facilitators for both technical and emotional support, our findings suggest that a thoughtfully designed chatbot with clear guidance like \textsc{Remini} could assume these roles, offering a scalable alternative.

\subsection{Collaborative Narration and Reciprocal Self-disclosure}

\subsubsection{Moving from One-directional Storytelling to Collaborative Memory Reconstruction}

While prior work has shown that technology-mediated reminiscence can be enjoyable in social settings, it often emphasizes one-directional storytelling, where a single individual narrates to passive listeners who did not share the original experience. For instance, designs used in intergenerational storytelling~\cite{liagat2022withahint, shen2024legacysphere, li2023exploring} and group reminiscence among older adults~\cite{baker2021school} frequently adopt a one-directional format, where roles of narrator and listener are fixed. This approach leaves an important gap: it limits opportunities for deeper, reciprocal engagement and mutual contribution.

This pattern persists not only in group-based designs but also in chatbot-mediated reminiscence systems, which typically engage users individually. Studies have found that users prefer sharing personal memories with trusted individuals, such as close friends or family~\cite{jin2024exploring}, highlighting the limitations of individual chatbot interactions for capturing the intimacy and co-experiential richness of human connection~\cite{zhai2024exploring, xu2024memory}.

Building on these insights, our study shifts the focus toward two-way interaction and genuine human-to-human connection. We observed that participants reported having collaborative experiences when using \textsc{Remini} to recall shared memories. Rather than adhering to a fixed storyteller-audience model, \textsc{Remini} engages each person in articulating and interpreting the same experience through their unique lens. Participants reviewed and responded to each other’s perspectives, ultimately building more comprehensive accounts of their shared memories. Many described this process as akin to teamwork, where simultaneous sharing and mutual contributions enabled a richer reminiscence experience that fixed narrator-listener roles or individual reminiscence with chatbot simply cannot provide. These findings suggest that future reminiscence systems should prioritize co-construction to better support emotionally resonant social experiences.

\subsubsection{Enhancing Intimacy Through Reciprocal Sharing of Personal Perspectives}

The \textsc{Remini} group’s collaborative narrative format promoted more emotional expressions and fostered deeper reflective conversations compared to the \emph{Baseline} group. Participants in the \textsc{Remini} condition were encouraged to co-construct narratives, enabling them to articulate their personal thoughts and feelings while simultaneously gaining insights into their partner’s perspective on the same event. For example, several participants reported valuing the opportunity to view a cherished memory “through their partner’s eyes” (P19) and expressed being “more willing to talk about things from their own perspectives” (P9). This process not only stimulated the sharing of emotions but also created an interactive environment in which partners could provide responsive feedback, thereby facilitating synchronized positive emotions directly connected to their shared experiences.

These findings align with the Social Function of Autobiographical Memory framework~\cite{alea2003you}, which emphasizes memory sharing as a pathway to intimacy. They also reflect Stepanova et al.’s~\cite{stepanova2022strategies} emphasis on reciprocal self-disclosure in technology-mediated systems. Unlike earlier systems that relied on implicit cues (e.g., biofeedback, symbolic tokens), \textsc{Remini} supports active, reciprocal verbal disclosure—a mode of interaction that better mirrors natural conversations and enables emotional attunement in real time. In addition, the work of Baumann et al.~\cite{baumann2024mnemosyne} suggests that for populations like older adults, the sense of connectedness achieved through direct interaction with loved ones may be more critical than reminiscence through digital cues alone. These studies collectively underscore the value of reciprocal self-disclosure as a robust strategy for fostering intimacy through technology-mediated conversations.

These results point to the value of conversational scaffolding that promotes turn-taking and mutual reflection, suggesting broader applications for chatbots designed to mediate emotionally rich social interactions~\cite{zheng2022ux}. By helping participants engage with each other’s perspectives, \textsc{Remini} may also support processes like relational meaning-making or conflict resolution, particularly in relationships where emotional narratives are contested or under-communicated~\cite{zheng2021pocketbot}. Reciprocal prompts also appeared to support more balanced participation, which could be especially beneficial in relationships where one partner is more expressive than the other~\cite{ruan2020can}. Some participants also suggested that the structured nature of chatbot prompts helped them articulate thoughts they might not have otherwise shared, pointing to the role of chatbots in creating emotionally safe spaces \cite{pentina2023exploring}.

While the \textsc{Remini} group exhibited higher levels of positive affect, emotional synchrony, self-disclosure, and perceived partner responsiveness, there was no significant short-term improvement in overall relationship quality. This highlights a potential limitation of single-session interventions and suggests that more substantial relational benefits may emerge through repeated or ongoing use. Future research should explore how sustained chatbot-facilitated reminiscence can support long-term relational growth.

\subsection{Negotiating Chatbot Presence in Intimate Conversation}

\subsubsection{Social Presence and Mediating Influence}

Drawing on theories of AI Anthropomorphism and the Computers Are Social Actors (CASA) framework~\cite{epley2007seeing, nass1994computers}, people often attribute human qualities to artificial agents that display social cues. In \textsc{Remini}, users frequently described the chatbot as demonstrating empathy and responsiveness (\textit{``actively listens to our conversation and remembers everything''} —P43), evoking a sense of social presence akin to \textit{``a mutual friend''} (P32). This aligns with studies showing that people can quickly develop rapport with chatbots, at times disclosing personal information sooner than they would with human partners~\cite{skjuve2021mychatbot}. Several participants noted how the conversation progressed from a brief ``warm-up'' phase (e.g., exchanging daily pleasantries) into more intimate reflections on shared memories. One participant remarked, \textit{``it felt a little weird knowing there’s a bot there, but we got more comfortable and started sending stickers as we would to each other''} (P9). However, some participants highlighted the unnaturalness of chatbot responses, criticizing overly formal phrasing that resembled structured Q\&A rather than casual conversation. This aligns with prior AIMC research indicating that verbose and unnatural responses or languages with excessive emotional intensity in AI outputs can hinder interpersonal perceptions like social attraction~\cite{fu2024from}. Future work could provide customizable options, such as sliders to adjust politeness, formality, and emotional expressiveness, helping users better align chatbot interactions with their authentic voice.

At the same time, \textsc{Remini} meaningfully influenced conversational dynamics in ways that shaped how people disclosed personal details. Because the chatbot was seen as an ``outsider'' with no prior knowledge, participants felt compelled to share more backstory and context than in typical one-on-one interactions. For some, this enriched storytelling, as P4 explained: \textit{``We were treating the chatbot as someone who doesn’t know anything, so we go from the very basics…''}, and P41 added, \textit{``If it’s just the two of us talking, then I only have to mention keywords.''} However, this additional contextual effort might occasionally burden participants, potentially creating conversational friction or discomfort. Future work should further examine when automating mutual reminiscence enhances conversational depth and when it unintentionally disrupts conversational authenticity or relational dynamics. Participants also noted the chatbot’s affective influence. Some observed that \textsc{Remini}’s consistent positive stance helped diffuse minor tensions which align with prior works~\cite{zheng2022ux}, while a few found the conversation grew more ``formal'' (P24) than usual. Interestingly, many noted that the chatbot could also fade into the background; once participants became deeply engaged with each other, the chatbot felt ``ignorable'' or easy to overlook. These varied responses highlight how a chatbot can serve as both a supportive mediator and, at times, an invisible bystander in intimate exchanges. Future works should investigate the duality in chatbot perception, examining how users simultaneously experience these agents as empathetic conversational partners and as detached, utilitarian tools, and assess how this tension influences disclosure practices and relational dynamics.

\subsubsection{Ambivalence and Tensions around AI Anthropomorphism}
Despite appreciating \textsc{Remini}’s facilitative role and wishing for more human-like behaviors from the chatbot, not all participants felt comfortable treating the chatbot as a human-like partner. One participant explained, \textit{``It feels more awkward because the question being asked is based on algorithms, not real thought''} (P31). This underscores a critique of AI Anthropomorphism: it can exaggerate AI capabilities by attributing human-like traits to systems that do not actually possess them~\cite{placani2024anthropomorphism}, and thereby fuel mistaken beliefs about AI’s cognitive or moral status~\cite{stuart2021guilty}. While \textsc{Remini} uses empathetic language to foster comfort and disclosure, it does not claim genuine moral character or emotional capacity. The interaction design further reinforces this, as users interact with \textsc{Remini} within a Telegram group and trigger it with bot-specific commands, complete with playful reminders of its non-human nature (e.g., “I have been enjoying wandering my virtual garden!”). 

Still, the tension remains. On one hand, deeper anthropomorphism can enhance user engagement, especially in companionship-oriented AI~\cite{pentina2023exploring, skjuve2021mychatbot}. On the other hand, research on highly human-like chatbots, such as Replika, shows that mimicking human conversation without genuine empathy risks fostering overdependence and further emotional distress when users eventually confront the AI’s limitations~\cite{laestadius2024toohuman}. While both Replika and \textsc{Remini} encourage emotional disclosure, \textsc{Remini} is explicitly designed to facilitate reminiscence among human partners, focusing on related topics rather than creating a human–chatbot bond, which helps reduce the risk of users becoming overly attached to the chatbot. Even so, the awkwardness described by P31 underscores broader concerns about conflating friendly facilitation with genuine human thought~\cite{namvarpour2024uncovering}. Moving forward, clearly communicating the chatbot’s role and nature can help calibrate user expectations , avoid projecting unwarranted human qualities onto the system, and promote responsible design choices~\cite{placani2024anthropomorphism}.

\subsubsection{Privacy Considerations}

Participants expressed concerns about revealing personal information—including references to family and friends—when using a reminiscence chatbot in group settings. Although our study anonymized all data and used it solely for research, real-world polyadic conversational agents raise additional challenges for managing privacy and social boundaries. According to Communication Privacy Management Theory (CPM), once private information is shared within a group (including an AI participant), each member becomes a co-owner who must negotiate privacy rules to prevent boundary turbulence~\cite{petronio20113breif}. For designers of polyadic conversational agents in memory sharing related settings, this implies a need for transparent, user-defined controls over how personal data is accessed, stored, and shared, such as improving the visibility and usability of privacy settings~\cite{im2023less}.

Another issue that emerged was participants’ desire for a more proactive chatbot that ``pops up'' rather than only responding when called upon. While increased proactivity could enhance social presence, it also carries risks of expectancy violations such as bringing feelings of being monitored. Expectancy Violations Theory suggests that users’ evaluations—and the emotional or converstional validation they experience—are shaped by the gap between their expectations and the chatbot’s behavior~\cite{minjin2024whena}. In our study, the chatbot processed conversations only when participants explicitly invoked it, granting them partial control over shared content. More autonomous and human-like systems could trigger both positive and negative expectancy violations. To mitigate potential negative effects, future designs should ensure users maintain clear agency over the chatbot’s interventions—such as through a ``privacy mode'' that temporarily suspends participation~\cite{zheng2021pocketbot}. Addressing these design tensions early is key to safeguarding trust while harnessing the benefits of AI-mediated intimate conversations.

\section{Limitations and Future Work}

This study has several limitations. First, our participant pool consisted of young adults from a single geographic region, which may introduce cultural biases and a unique sociolect shaping both engagement with the chatbot and perceptions of mutual reminiscence~\cite{guan2022does}. Although age-related differences in autobiographical memory could also influence user behavior, prior research indicates that people increasingly rely on past memories to maintain intimacy as they grow older~\cite{webster2013adult}, suggesting that our findings may nonetheless be informative for other demographics.

Second, we only compared a structured chatbot (\textsc{Remini}) to a minimal-guidance chatbot (baseline). As a result, we cannot fully isolate the impact of using a chatbot itself from the effect of structured guidance. Future work could explore this by comparing chatbot-based interventions to non-chatbot or entirely offline conditions, thus clarifying the medium’s distinct influence on mutual reminiscence.

Looking ahead, we plan to address these limitations by broadening our participant base to include diverse age groups, cultural backgrounds, and social contexts, thereby strengthening the generalizability of our findings. Additionally, future iterations of \textsc{Remini} may incorporate more natural and long-term forms of interaction, such as handling spoken dialogue or integrating emotion-aware responses. We envision leveraging cutting-edge AI to seamlessly incorporate voice messages, enabling richer emotional expression during both everyday and significant reminiscence moments. Through these advancements, we aim to enhance well-being by ensuring that our conversational agents not only facilitate—but also deepen—the preservation and sharing of meaningful memories.

\section{Conclusion}

This study developed and implemented a LLM-powered \textsc{Remini} chatbot to facilitate mutual reminiscence among loved ones for the enhancement of positive affect and feeling of connection. We identified three key outcomes: the novel design of \textsc{Remini}, insights into user interactions, and the implications of using chatbots for deeper reminiscence in polyadic settings. User study with 48 participants revealed that \textsc{Remini}, as opposed to a baseline condition, significantly improved positive affect, emotional synchrony, and self-disclosure. Both versions, however, improved relationship quality and social bonding. Our findings demonstrate the chatbot’s potential to enrich reminiscence conversations and suggest important design implications for future interpersonal connection technologies.

\bibliographystyle{ACM-Reference-Format}
\bibliography{sample-base}


\begin{thebibliography}{179}


\ifx \showCODEN    \undefined \def \showCODEN     #1{\unskip}     \fi
\ifx \showISBNx    \undefined \def \showISBNx     #1{\unskip}     \fi
\ifx \showISBNxiii \undefined \def \showISBNxiii  #1{\unskip}     \fi
\ifx \showISSN     \undefined \def \showISSN      #1{\unskip}     \fi
\ifx \showLCCN     \undefined \def \showLCCN      #1{\unskip}     \fi
\ifx \shownote     \undefined \def \shownote      #1{#1}          \fi
\ifx \showarticletitle \undefined \def \showarticletitle #1{#1}   \fi
\ifx \showURL      \undefined \def \showURL       {\relax}        \fi
\providecommand\bibfield[2]{#2}
\providecommand\bibinfo[2]{#2}
\providecommand\natexlab[1]{#1}
\providecommand\showeprint[2][]{arXiv:#2}

\bibitem[Alea and Bluck(2003)]%
        {alea2003you}
\bibfield{author}{\bibinfo{person}{Nicole Alea} {and} \bibinfo{person}{Susan Bluck}.} \bibinfo{year}{2003}\natexlab{}.
\newblock \showarticletitle{Why are you telling me that? A conceptual model of the social function of autobiographical memory}.
\newblock \bibinfo{journal}{\emph{Memory}} \bibinfo{volume}{11}, \bibinfo{number}{2} (\bibinfo{year}{2003}), \bibinfo{pages}{165--178}.
\newblock
\href{https://doi.org/10.1080/741938207}{doi:\nolinkurl{10.1080/741938207}}


\bibitem[Alea and Bluck(2007)]%
        {alea2007illkeepyouinmind}
\bibfield{author}{\bibinfo{person}{Nicole Alea} {and} \bibinfo{person}{Susan Bluck}.} \bibinfo{year}{2007}\natexlab{}.
\newblock \showarticletitle{I'll keep you in mind: the intimacy function of autobiographical memory}.
\newblock \bibinfo{journal}{\emph{Applied Cognitive Psychology}} \bibinfo{volume}{21}, \bibinfo{number}{8} (\bibinfo{year}{2007}), \bibinfo{pages}{1091--1111}.
\newblock
\href{https://doi.org/10.1002/acp.1316}{doi:\nolinkurl{10.1002/acp.1316}}


\bibitem[Algoe et~al\mbox{.}(2008)]%
        {algoe2008beyond}
\bibfield{author}{\bibinfo{person}{Sara~B Algoe}, \bibinfo{person}{Jonathan Haidt}, {and} \bibinfo{person}{Shelly~L Gable}.} \bibinfo{year}{2008}\natexlab{}.
\newblock \showarticletitle{Beyond reciprocity: gratitude and relationships in everyday life}.
\newblock \bibinfo{journal}{\emph{Emotion}} \bibinfo{volume}{8}, \bibinfo{number}{3} (\bibinfo{year}{2008}), \bibinfo{pages}{425--429}.
\newblock
\href{https://doi.org/10.1037/1528-3542.8.3.425}{doi:\nolinkurl{10.1037/1528-3542.8.3.425}}


\bibitem[Allouch et~al\mbox{.}(2021)]%
        {allouch2021conversational}
\bibfield{author}{\bibinfo{person}{Merav Allouch}, \bibinfo{person}{Amos Azaria}, {and} \bibinfo{person}{Rina Azoulay}.} \bibinfo{year}{2021}\natexlab{}.
\newblock \showarticletitle{Conversational Agents: Goals, Technologies, Vision and Challenges}.
\newblock \bibinfo{journal}{\emph{Sensors}} \bibinfo{volume}{21}, \bibinfo{number}{24} (\bibinfo{year}{2021}), \bibinfo{pages}{8448}.
\newblock
\href{https://doi.org/10.3390/s21248448}{doi:\nolinkurl{10.3390/s21248448}}


\bibitem[Altman and Taylor(1973)]%
        {altman1973social}
\bibfield{author}{\bibinfo{person}{Irwin Altman} {and} \bibinfo{person}{Dalmas~A Taylor}.} \bibinfo{year}{1973}\natexlab{}.
\newblock \bibinfo{booktitle}{\emph{Social penetration: The development of interpersonal relationships}}.
\newblock \bibinfo{publisher}{Holt, Rinehart \& Winston}, \bibinfo{address}{New York}.
\newblock


\bibitem[An et~al\mbox{.}(2024)]%
        {an2024emowear}
\bibfield{author}{\bibinfo{person}{Pengcheng An}, \bibinfo{person}{Jiawen~Stefanie Zhu}, \bibinfo{person}{Zibo Zhang}, \bibinfo{person}{Yifei Yin}, \bibinfo{person}{Qingyuan Ma}, \bibinfo{person}{Che Yan}, \bibinfo{person}{Linghao Du}, {and} \bibinfo{person}{Jian Zhao}.} \bibinfo{year}{2024}\natexlab{}.
\newblock \showarticletitle{EmoWear: Exploring Emotional Teasers for Voice Message Interaction on Smartwatches}. In \bibinfo{booktitle}{\emph{Proceedings of the 2024 CHI Conference on Human Factors in Computing Systems}} (Honolulu, HI, USA) \emph{(\bibinfo{series}{CHI '24})}. \bibinfo{publisher}{Association for Computing Machinery}, \bibinfo{address}{New York, NY, USA}, Article \bibinfo{articleno}{279}, \bibinfo{numpages}{16}~pages.
\newblock
\showISBNx{9798400703300}
\href{https://doi.org/10.1145/3613904.3642101}{doi:\nolinkurl{10.1145/3613904.3642101}}


\bibitem[Apted et~al\mbox{.}(2006)]%
        {apted2006tabletop}
\bibfield{author}{\bibinfo{person}{Trent Apted}, \bibinfo{person}{Judy Kay}, {and} \bibinfo{person}{Aaron Quigley}.} \bibinfo{year}{2006}\natexlab{}.
\newblock \showarticletitle{Tabletop sharing of digital photographs for the elderly}. In \bibinfo{booktitle}{\emph{Proceedings of the SIGCHI Conference on Human Factors in Computing Systems}} (Montr\'{e}al, Qu\'{e}bec, Canada) \emph{(\bibinfo{series}{CHI '06})}. \bibinfo{publisher}{Association for Computing Machinery}, \bibinfo{address}{New York, NY, USA}, \bibinfo{pages}{781–790}.
\newblock
\showISBNx{1595933727}
\href{https://doi.org/10.1145/1124772.1124887}{doi:\nolinkurl{10.1145/1124772.1124887}}


\bibitem[Arican-Dinc and Gable(2023)]%
        {Beyzanur2023responsiveness}
\bibfield{author}{\bibinfo{person}{Beyzanur Arican-Dinc} {and} \bibinfo{person}{Shelly~L. Gable}.} \bibinfo{year}{2023}\natexlab{}.
\newblock \showarticletitle{Responsiveness in romantic partners’ interactions}.
\newblock \bibinfo{journal}{\emph{Current Opinion in Psychology}}  \bibinfo{volume}{53} (\bibinfo{year}{2023}), \bibinfo{pages}{101652}.
\newblock
\href{https://doi.org/10.1016/j.copsyc.2023.101652}{doi:\nolinkurl{10.1016/j.copsyc.2023.101652}}


\bibitem[Aron and Smollan(1992)]%
        {aron1992inclusion}
\bibfield{author}{\bibinfo{person}{Elaine N.~Aron Aron, Arthur} {and} \bibinfo{person}{Danny Smollan}.} \bibinfo{year}{1992}\natexlab{}.
\newblock \showarticletitle{Inclusion of Other in the Self Scale and the structure of interpersonal closeness}.
\newblock \bibinfo{journal}{\emph{Journal of Personality and Social Psychology}} \bibinfo{volume}{63}, \bibinfo{number}{4} (\bibinfo{year}{1992}), \bibinfo{pages}{596--612}.
\newblock
\href{https://doi.org/10.1037/0022-3514.63.4.596}{doi:\nolinkurl{10.1037/0022-3514.63.4.596}}


\bibitem[{ATLAS.ti Scientific Software Development GmbH}(2023)]%
        {AtlasTi2023}
\bibfield{author}{\bibinfo{person}{{ATLAS.ti Scientific Software Development GmbH}}.} \bibinfo{year}{2023}\natexlab{}.
\newblock \bibinfo{title}{ATLAS.ti Mac}.
\newblock \bibinfo{howpublished}{Macintosh software}.
\newblock
\newblock
\shownote{Available software for qualitative data analysis}.


\bibitem[Avula et~al\mbox{.}(2018)]%
        {avula2018searchbots}
\bibfield{author}{\bibinfo{person}{Sandeep Avula}, \bibinfo{person}{Gordon Chadwick}, \bibinfo{person}{Jaime Arguello}, {and} \bibinfo{person}{Robert Capra}.} \bibinfo{year}{2018}\natexlab{}.
\newblock \showarticletitle{SearchBots: User Engagement with ChatBots during Collaborative Search}. In \bibinfo{booktitle}{\emph{Proceedings of the 2018 Conference on Human Information Interaction \& Retrieval}} (New Brunswick, NJ, USA) \emph{(\bibinfo{series}{CHIIR '18})}. \bibinfo{publisher}{Association for Computing Machinery}, \bibinfo{address}{New York, NY, USA}, \bibinfo{pages}{52–61}.
\newblock
\showISBNx{9781450349253}
\href{https://doi.org/10.1145/3176349.3176380}{doi:\nolinkurl{10.1145/3176349.3176380}}


\bibitem[Azy and Orit(2007)]%
        {barak2007degree}
\bibfield{author}{\bibinfo{person}{Barak Azy} {and} \bibinfo{person}{Gluck-Ofri Orit}.} \bibinfo{year}{2007}\natexlab{}.
\newblock \showarticletitle{Degree and reciprocity of self-disclosure in online forums}.
\newblock \bibinfo{journal}{\emph{CyberPsychology \& Behavior}} \bibinfo{volume}{10}, \bibinfo{number}{3} (\bibinfo{year}{2007}), \bibinfo{pages}{407--417}.
\newblock
\href{https://doi.org/10.1089/cpb.2006.9938}{doi:\nolinkurl{10.1089/cpb.2006.9938}}


\bibitem[Baker et~al\mbox{.}(2021)]%
        {baker2021school}
\bibfield{author}{\bibinfo{person}{Steven Baker}, \bibinfo{person}{Ryan~M. Kelly}, \bibinfo{person}{Jenny Waycott}, \bibinfo{person}{Romina Carrasco}, \bibinfo{person}{Roger Bell}, \bibinfo{person}{Zaher Joukhadar}, \bibinfo{person}{Thuong Hoang}, \bibinfo{person}{Elizabeth Ozanne}, {and} \bibinfo{person}{Frank Vetere}.} \bibinfo{year}{2021}\natexlab{}.
\newblock \showarticletitle{School's Back: Scaffolding Reminiscence in Social Virtual Reality with Older Adults}.
\newblock \bibinfo{journal}{\emph{Proc. ACM Hum.-Comput. Interact.}} \bibinfo{volume}{4}, \bibinfo{number}{CSCW3}, Article \bibinfo{articleno}{267} (\bibinfo{date}{Jan.} \bibinfo{year}{2021}), \bibinfo{numpages}{25}~pages.
\newblock
\href{https://doi.org/10.1145/3434176}{doi:\nolinkurl{10.1145/3434176}}


\bibitem[Balaji et~al\mbox{.}(2024)]%
        {balaji2024designing}
\bibfield{author}{\bibinfo{person}{Divyaa Balaji}, \bibinfo{person}{Gert-Jan De~Bruijn}, \bibinfo{person}{Tibor Bosse}, \bibinfo{person}{Carolin Ischen}, \bibinfo{person}{Margot Van Der~Goot}, {and} \bibinfo{person}{Reinout Wiers}.} \bibinfo{year}{2024}\natexlab{}.
\newblock \showarticletitle{Designing a Couples-Based Conversational Agent to Promote Safe Sex in New, Young Couples: A User-Centred Design Approach}. In \bibinfo{booktitle}{\emph{Proceedings of the 6th ACM Conference on Conversational User Interfaces}} (Luxembourg, Luxembourg) \emph{(\bibinfo{series}{CUI '24})}. \bibinfo{publisher}{Association for Computing Machinery}, \bibinfo{address}{New York, NY, USA}, Article \bibinfo{articleno}{4}, \bibinfo{numpages}{11}~pages.
\newblock
\showISBNx{9798400705113}
\href{https://doi.org/10.1145/3640794.3665556}{doi:\nolinkurl{10.1145/3640794.3665556}}


\bibitem[Bales et~al\mbox{.}(2011)]%
        {bales2011coupleVIBE}
\bibfield{author}{\bibinfo{person}{Elizabeth Bales}, \bibinfo{person}{Kevin~A. Li}, {and} \bibinfo{person}{William Griwsold}.} \bibinfo{year}{2011}\natexlab{}.
\newblock \showarticletitle{CoupleVIBE: mobile implicit communication to improve awareness for (long-distance) couples}. In \bibinfo{booktitle}{\emph{Proceedings of the ACM 2011 Conference on Computer Supported Cooperative Work}} (Hangzhou, China) \emph{(\bibinfo{series}{CSCW '11})}. \bibinfo{publisher}{Association for Computing Machinery}, \bibinfo{address}{New York, NY, USA}, \bibinfo{pages}{65–74}.
\newblock
\showISBNx{9781450305563}
\href{https://doi.org/10.1145/1958824.1958835}{doi:\nolinkurl{10.1145/1958824.1958835}}


\bibitem[Baumann et~al\mbox{.}(2024)]%
        {baumann2024mnemosyne}
\bibfield{author}{\bibinfo{person}{Andrea Baumann}, \bibinfo{person}{Peter Shaw}, \bibinfo{person}{Ludwig Trotter}, \bibinfo{person}{Sarah Clinch}, {and} \bibinfo{person}{Nigel Davies}.} \bibinfo{year}{2024}\natexlab{}.
\newblock \showarticletitle{Mnemosyne - Supporting Reminiscence for Individuals with Dementia in Residential Care Settings}. In \bibinfo{booktitle}{\emph{Proceedings of the 2024 CHI Conference on Human Factors in Computing Systems}} (Honolulu, HI, USA) \emph{(\bibinfo{series}{CHI '24})}. \bibinfo{publisher}{Association for Computing Machinery}, \bibinfo{address}{New York, NY, USA}, Article \bibinfo{articleno}{217}, \bibinfo{numpages}{17}~pages.
\newblock
\showISBNx{9798400703300}
\href{https://doi.org/10.1145/3613904.3642783}{doi:\nolinkurl{10.1145/3613904.3642783}}


\bibitem[Bavelas et~al\mbox{.}(2000)]%
        {bavelas2000listeners}
\bibfield{author}{\bibinfo{person}{Janet~B Bavelas}, \bibinfo{person}{Linda Coates}, {and} \bibinfo{person}{Trudy Johnson}.} \bibinfo{year}{2000}\natexlab{}.
\newblock \showarticletitle{Listeners as co-narrators}.
\newblock \bibinfo{journal}{\emph{Journal of personality and social psychology}} \bibinfo{volume}{79}, \bibinfo{number}{6} (\bibinfo{year}{2000}), \bibinfo{pages}{941}.
\newblock
\href{https://doi.org/10.1037/0022-3514.79.6.941}{doi:\nolinkurl{10.1037/0022-3514.79.6.941}}


\bibitem[Beike et~al\mbox{.}(2017)]%
        {beike2017sharing}
\bibfield{author}{\bibinfo{person}{Denise~R. Beike}, \bibinfo{person}{Holly~E. Cole}, {and} \bibinfo{person}{Carmen~R. Merrick}.} \bibinfo{year}{2017}\natexlab{}.
\newblock \showarticletitle{Sharing specific “We” autobiographical memories in close relationships: the role of contact frequency}.
\newblock \bibinfo{journal}{\emph{Memory}} \bibinfo{volume}{25}, \bibinfo{number}{10} (\bibinfo{year}{2017}), \bibinfo{pages}{1425--1434}.
\newblock
\href{https://doi.org/10.1080/09658211.2017.1313990}{doi:\nolinkurl{10.1080/09658211.2017.1313990}}


\bibitem[Bietti(2010)]%
        {bietti2010sharing}
\bibfield{author}{\bibinfo{person}{Lucas~M Bietti}.} \bibinfo{year}{2010}\natexlab{}.
\newblock \showarticletitle{Sharing memories, family conversation and interaction}.
\newblock \bibinfo{journal}{\emph{Discourse \& Society}} \bibinfo{volume}{21}, \bibinfo{number}{5} (\bibinfo{year}{2010}), \bibinfo{pages}{499--523}.
\newblock
\href{https://doi.org/10.1177/0957926510373973}{doi:\nolinkurl{10.1177/0957926510373973}}


\bibitem[Bluck and Levine(1998)]%
        {bluck1998reminiscence}
\bibfield{author}{\bibinfo{person}{Susan Bluck} {and} \bibinfo{person}{Linda~J. Levine}.} \bibinfo{year}{1998}\natexlab{}.
\newblock \showarticletitle{Reminiscence as autobiographical memory: A catalyst for reminiscence theory development}.
\newblock \bibinfo{journal}{\emph{Aging \& Society}} \bibinfo{volume}{18}, \bibinfo{number}{1} (\bibinfo{date}{March} \bibinfo{year}{1998}), \bibinfo{pages}{185--208}.
\newblock
\href{https://doi.org/10.1017/S0144686X98006862}{doi:\nolinkurl{10.1017/S0144686X98006862}}


\bibitem[Bluck and Li(2001)]%
        {bluck2001predicting}
\bibfield{author}{\bibinfo{person}{Susan Bluck} {and} \bibinfo{person}{Karen~Z.H. Li}.} \bibinfo{year}{2001}\natexlab{}.
\newblock \showarticletitle{Predicting memory completeness and accuracy: emotion and exposure in repeated autobiographical recall}.
\newblock \bibinfo{journal}{\emph{Applied Cognitive Psychology}} \bibinfo{volume}{15}, \bibinfo{number}{2} (\bibinfo{year}{2001}).
\newblock
\href{https://doi.org/10.1002/1099-0720(200103/04)15:2<145::AID-ACP693>3.0.CO;2-T}{doi:\nolinkurl{10.1002/1099-0720(200103/04)15:2<145::AID-ACP693>3.0.CO;2-T}}


\bibitem[Bohlmeijer et~al\mbox{.}(2007)]%
        {bohlmeijer2007effects}
\bibfield{author}{\bibinfo{person}{Ernst Bohlmeijer}, \bibinfo{person}{Marte Roemer}, \bibinfo{person}{Pim Cuijpers}, {and} \bibinfo{person}{Filip Smit}.} \bibinfo{year}{2007}\natexlab{}.
\newblock \showarticletitle{The effects of reminiscence on psychological well-being in older adults: A meta-analysis}.
\newblock \bibinfo{journal}{\emph{Aging and Mental Health}} \bibinfo{volume}{11}, \bibinfo{number}{3} (\bibinfo{year}{2007}), \bibinfo{pages}{291--300}.
\newblock
\href{https://doi.org/10.1080/13607860600963547}{doi:\nolinkurl{10.1080/13607860600963547}}


\bibitem[Boyd et~al\mbox{.}(2022)]%
        {boyd2022thedevelopment}
\bibfield{author}{\bibinfo{person}{Ryan~L. Boyd}, \bibinfo{person}{Ashwini Ashokkumar}, \bibinfo{person}{Sarah Seraj}, {and} \bibinfo{person}{James~W. Pennebaker}.} \bibinfo{year}{2022}\natexlab{}.
\newblock \showarticletitle{The development and psychometric properties of LIWC-22}.
\newblock \bibinfo{journal}{\emph{Austin, TX: University of Texas at Austin}}  \bibinfo{volume}{10} (\bibinfo{year}{2022}).
\newblock


\bibitem[Braun and Clarke(2006)]%
        {virginia2006using}
\bibfield{author}{\bibinfo{person}{Virginia Braun} {and} \bibinfo{person}{Victoria Clarke}.} \bibinfo{year}{2006}\natexlab{}.
\newblock \showarticletitle{Using thematic analysis in psychology}.
\newblock \bibinfo{journal}{\emph{Qualitative Research in Psychology}} \bibinfo{volume}{3}, \bibinfo{number}{2} (\bibinfo{year}{2006}), \bibinfo{pages}{77--101}.
\newblock
\href{https://doi.org/10.1191/1478088706qp063oa}{doi:\nolinkurl{10.1191/1478088706qp063oa}}


\bibitem[Brown et~al\mbox{.}(2020)]%
        {brown2020language}
\bibfield{author}{\bibinfo{person}{Tom~B. Brown}, \bibinfo{person}{Benjamin Mann}, \bibinfo{person}{Nick Ryder}, \bibinfo{person}{Melanie Subbiah}, \bibinfo{person}{Jared Kaplan}, \bibinfo{person}{Prafulla Dhariwal}, \bibinfo{person}{Arvind Neelakantan}, \bibinfo{person}{Pranav Shyam}, \bibinfo{person}{Girish Sastry}, \bibinfo{person}{Amanda Askell}, \bibinfo{person}{Sandhini Agarwal}, \bibinfo{person}{Ariel Herbert-Voss}, \bibinfo{person}{Gretchen Krueger}, \bibinfo{person}{Tom Henighan}, \bibinfo{person}{Rewon Child}, \bibinfo{person}{Aditya Ramesh}, \bibinfo{person}{Daniel~M. Ziegler}, \bibinfo{person}{Jeffrey Wu}, \bibinfo{person}{Clemens Winter}, \bibinfo{person}{Christopher Hesse}, \bibinfo{person}{Mark Chen}, \bibinfo{person}{Eric Sigler}, \bibinfo{person}{Mateusz Litwin}, \bibinfo{person}{Scott Gray}, \bibinfo{person}{Benjamin Chess}, \bibinfo{person}{Jack Clark}, \bibinfo{person}{Christopher Berner}, \bibinfo{person}{Sam McCandlish}, \bibinfo{person}{Alec Radford}, \bibinfo{person}{Ilya Sutskever},
  {and} \bibinfo{person}{Dario Amodei}.} \bibinfo{year}{2020}\natexlab{}.
\newblock \showarticletitle{Language models are few-shot learners}. In \bibinfo{booktitle}{\emph{Proceedings of the 34th International Conference on Neural Information Processing Systems}} (Vancouver, BC, Canada) \emph{(\bibinfo{series}{NIPS '20})}. \bibinfo{publisher}{Curran Associates Inc.}, \bibinfo{address}{Red Hook, NY, USA}, Article \bibinfo{articleno}{159}, \bibinfo{numpages}{25}~pages.
\newblock
\showISBNx{9781713829546}


\bibitem[Bryant et~al\mbox{.}(2005)]%
        {bryant2005using}
\bibfield{author}{\bibinfo{person}{Fred~B Bryant}, \bibinfo{person}{Colette~M Smart}, {and} \bibinfo{person}{Scott~P King}.} \bibinfo{year}{2005}\natexlab{}.
\newblock \showarticletitle{Using the past to enhance the present: Boosting happiness through positive reminiscence}.
\newblock \bibinfo{journal}{\emph{Journal of Happiness Studies}}  \bibinfo{volume}{6} (\bibinfo{year}{2005}), \bibinfo{pages}{227--260}.
\newblock
\href{https://doi.org/10.1007/s10902-005-3889-4}{doi:\nolinkurl{10.1007/s10902-005-3889-4}}


\bibitem[Cai et~al\mbox{.}(2023)]%
        {cai2023listen}
\bibfield{author}{\bibinfo{person}{Wanling Cai}, \bibinfo{person}{Yucheng Jin}, \bibinfo{person}{Xianglin Zhao}, {and} \bibinfo{person}{Li Chen}.} \bibinfo{year}{2023}\natexlab{}.
\newblock \showarticletitle{“Listen to Music, Listen to Yourself”: Design of a Conversational Agent to Support Self-Awareness While Listening to Music}. In \bibinfo{booktitle}{\emph{Proceedings of the 2023 CHI Conference on Human Factors in Computing Systems}} (Hamburg, Germany) \emph{(\bibinfo{series}{CHI '23})}. \bibinfo{publisher}{Association for Computing Machinery}, \bibinfo{address}{New York, NY, USA}, Article \bibinfo{articleno}{119}, \bibinfo{numpages}{19}~pages.
\newblock
\showISBNx{9781450394215}
\href{https://doi.org/10.1145/3544548.3581427}{doi:\nolinkurl{10.1145/3544548.3581427}}


\bibitem[Cappeliez et~al\mbox{.}(2008)]%
        {philippe2008functions}
\bibfield{author}{\bibinfo{person}{Philippe Cappeliez}, \bibinfo{person}{Marilyn Guindon}, {and} \bibinfo{person}{Annie Robitaille}.} \bibinfo{year}{2008}\natexlab{}.
\newblock \showarticletitle{Functions of reminiscence and emotional regulation among older adults}.
\newblock \bibinfo{journal}{\emph{Journal of Aging Studies}} \bibinfo{volume}{22}, \bibinfo{number}{3} (\bibinfo{year}{2008}), \bibinfo{pages}{266--272}.
\newblock
\href{https://doi.org/10.1016/j.jaging.2007.06.003}{doi:\nolinkurl{10.1016/j.jaging.2007.06.003}}


\bibitem[Cappeliez et~al\mbox{.}(2007)]%
        {cappeliez2007functions}
\bibfield{author}{\bibinfo{person}{P. Cappeliez}, \bibinfo{person}{V. Rivard}, {and} \bibinfo{person}{S. Guindon}.} \bibinfo{year}{2007}\natexlab{}.
\newblock \showarticletitle{Functions of reminiscence in later life: proposition of a model and applications}.
\newblock \bibinfo{journal}{\emph{European Review of Applied Psychology}} \bibinfo{volume}{57}, \bibinfo{number}{3} (\bibinfo{year}{2007}), \bibinfo{pages}{151--156}.
\newblock
\href{https://doi.org/10.1016/j.erap.2005.02.014}{doi:\nolinkurl{10.1016/j.erap.2005.02.014}}


\bibitem[Chen et~al\mbox{.}(2023a)]%
        {chen2023photoclock}
\bibfield{author}{\bibinfo{person}{Amy Yo~Sue Chen}, \bibinfo{person}{William Odom}, \bibinfo{person}{Sol Kang}, {and} \bibinfo{person}{Carman Neustaedter}.} \bibinfo{year}{2023}\natexlab{a}.
\newblock \showarticletitle{PhotoClock: Reliving Memories in Digital Photos as the Clock Ticks in the Present Moment}. In \bibinfo{booktitle}{\emph{Proceedings of the 2023 ACM Designing Interactive Systems Conference}} (Pittsburgh, PA, USA) \emph{(\bibinfo{series}{DIS '23})}. \bibinfo{publisher}{Association for Computing Machinery}, \bibinfo{address}{New York, NY, USA}, \bibinfo{pages}{1015–1031}.
\newblock
\showISBNx{9781450398930}
\href{https://doi.org/10.1145/3563657.3596020}{doi:\nolinkurl{10.1145/3563657.3596020}}


\bibitem[Chen et~al\mbox{.}(2023b)]%
        {chen2023exploring}
\bibfield{author}{\bibinfo{person}{Amy Yo~Sue Chen}, \bibinfo{person}{William Odom}, \bibinfo{person}{Carman Neustaedter}, \bibinfo{person}{Ce Zhong}, {and} \bibinfo{person}{Henry Lin}.} \bibinfo{year}{2023}\natexlab{b}.
\newblock \showarticletitle{Exploring Memory-Oriented Interactions with Digital Photos In and Across Time: A Field Study of Chronoscope}. In \bibinfo{booktitle}{\emph{Proceedings of the 2023 CHI Conference on Human Factors in Computing Systems}} (Hamburg, Germany) \emph{(\bibinfo{series}{CHI '23})}. \bibinfo{publisher}{Association for Computing Machinery}, \bibinfo{address}{New York, NY, USA}, Article \bibinfo{articleno}{631}, \bibinfo{numpages}{20}~pages.
\newblock
\showISBNx{9781450394215}
\href{https://doi.org/10.1145/3544548.3581012}{doi:\nolinkurl{10.1145/3544548.3581012}}


\bibitem[Chris~Elsden and Durrant(2016)]%
        {chris2016aquantified}
\bibfield{author}{\bibinfo{person}{David S.~Kirk Chris~Elsden} {and} \bibinfo{person}{Abigail~C. Durrant}.} \bibinfo{year}{2016}\natexlab{}.
\newblock \showarticletitle{A Quantified Past: Toward Design for Remembering With Personal Informatics}.
\newblock \bibinfo{journal}{\emph{Human–Computer Interaction}} \bibinfo{volume}{31}, \bibinfo{number}{6} (\bibinfo{year}{2016}), \bibinfo{pages}{518--557}.
\newblock
\href{https://doi.org/10.1080/07370024.2015.1093422}{doi:\nolinkurl{10.1080/07370024.2015.1093422}}


\bibitem[Conway and Pleydell-Pearce(2000)]%
        {conway2000theconstruction}
\bibfield{author}{\bibinfo{person}{Martin~A. Conway} {and} \bibinfo{person}{Christopher~W. Pleydell-Pearce}.} \bibinfo{year}{2000}\natexlab{}.
\newblock \showarticletitle{The construction of autobiographical memories in the self-memory system}.
\newblock \bibinfo{journal}{\emph{Psychological Review}} \bibinfo{volume}{107}, \bibinfo{number}{2} (\bibinfo{year}{2000}), \bibinfo{pages}{261--288}.
\newblock
\href{https://doi.org/10.1037/0033-295X.107.2.261}{doi:\nolinkurl{10.1037/0033-295X.107.2.261}}


\bibitem[de~Jong et~al\mbox{.}(2024)]%
        {dejong2024assessing}
\bibfield{author}{\bibinfo{person}{Sander de Jong}, \bibinfo{person}{Joel Wester}, \bibinfo{person}{Tim Schrills}, \bibinfo{person}{Kristina S.~Secher}, \bibinfo{person}{Carla F.~Griggio}, {and} \bibinfo{person}{Niels van Berkel}.} \bibinfo{year}{2024}\natexlab{}.
\newblock \showarticletitle{Assessing Cognitive and Social Awareness among Group Members in AI-assisted Collaboration}. In \bibinfo{booktitle}{\emph{Proceedings of the International Conference on Mobile and Ubiquitous Multimedia}} \emph{(\bibinfo{series}{MUM '24})}. \bibinfo{publisher}{Association for Computing Machinery}, \bibinfo{address}{New York, NY, USA}, \bibinfo{pages}{338–350}.
\newblock
\showISBNx{9798400712838}
\href{https://doi.org/10.1145/3701571.3701582}{doi:\nolinkurl{10.1145/3701571.3701582}}


\bibitem[Dib et~al\mbox{.}(2010)]%
        {dib2010sonic}
\bibfield{author}{\bibinfo{person}{Lina Dib}, \bibinfo{person}{Daniela Petrelli}, {and} \bibinfo{person}{Steve Whittaker}.} \bibinfo{year}{2010}\natexlab{}.
\newblock \showarticletitle{Sonic souvenirs: exploring the paradoxes of recorded sound for family remembering}. In \bibinfo{booktitle}{\emph{Proceedings of the 2010 ACM Conference on Computer Supported Cooperative Work}} (Savannah, Georgia, USA) \emph{(\bibinfo{series}{CSCW '10})}. \bibinfo{publisher}{Association for Computing Machinery}, \bibinfo{address}{New York, NY, USA}, \bibinfo{pages}{391–400}.
\newblock
\showISBNx{9781605587950}
\href{https://doi.org/10.1145/1718918.1718985}{doi:\nolinkurl{10.1145/1718918.1718985}}


\bibitem[DJ et~al\mbox{.}(2024)]%
        {hallford2024randomized}
\bibfield{author}{\bibinfo{person}{Hallford DJ}, \bibinfo{person}{Gerring K}, \bibinfo{person}{Button M}, \bibinfo{person}{Campbell R}, {and} \bibinfo{person}{Sutton K}.} \bibinfo{year}{2024}\natexlab{}.
\newblock \showarticletitle{A Randomized Controlled Component Study of Cognitive Reminiscence Therapy for Psychological Resources and Mental Well-Being: Group-Based CRT Versus Self-Guided Homework Sheets}.
\newblock \bibinfo{journal}{\emph{Cognitive Therapy and Research}} (\bibinfo{year}{2024}), \bibinfo{pages}{1--16}.
\newblock
\href{https://doi.org/10.1007/s10608-024-10554-1}{doi:\nolinkurl{10.1007/s10608-024-10554-1}}


\bibitem[Dyke et~al\mbox{.}(2012)]%
        {gregory2012towards}
\bibfield{author}{\bibinfo{person}{Gregory Dyke}, \bibinfo{person}{David Adamson}, \bibinfo{person}{Iris Howley}, {and} \bibinfo{person}{Carolyn Penstein~Ros{\'e}}.} \bibinfo{year}{2012}\natexlab{}.
\newblock \showarticletitle{Towards Academically Productive Talk Supported by Conversational Agents}. In \bibinfo{booktitle}{\emph{Intelligent Tutoring Systems}}, \bibfield{editor}{\bibinfo{person}{Stefano~A. Cerri}, \bibinfo{person}{William~J. Clancey}, \bibinfo{person}{Giorgos Papadourakis}, {and} \bibinfo{person}{Kitty Panourgia}} (Eds.). \bibinfo{publisher}{Springer Berlin Heidelberg}, \bibinfo{address}{Berlin, Heidelberg}, \bibinfo{pages}{531--540}.
\newblock
\showISBNx{978-3-642-30950-2}


\bibitem[Einam et~al\mbox{.}(2024)]%
        {einam2024shedding}
\bibfield{author}{\bibinfo{person}{Haran Einam}, \bibinfo{person}{Mario Mikulincer}, {and} \bibinfo{person}{Ron Shachar}.} \bibinfo{year}{2024}\natexlab{}.
\newblock \showarticletitle{Shedding a light on the teller: on storytelling, meaning in life, and personal goals}.
\newblock \bibinfo{journal}{\emph{The Journal of Positive Psychology}} (\bibinfo{year}{2024}), \bibinfo{pages}{1--15}.
\newblock
\href{https://doi.org/10.1080/17439760.2024.2431684}{doi:\nolinkurl{10.1080/17439760.2024.2431684}}


\bibitem[Epley et~al\mbox{.}(2007)]%
        {epley2007seeing}
\bibfield{author}{\bibinfo{person}{Nicholas Epley}, \bibinfo{person}{Adam Waytz}, {and} \bibinfo{person}{John~T Cacioppo}.} \bibinfo{year}{2007}\natexlab{}.
\newblock \showarticletitle{On seeing human: a three-factor theory of anthropomorphism}.
\newblock \bibinfo{journal}{\emph{Psychological Review}} \bibinfo{volume}{114}, \bibinfo{number}{4} (\bibinfo{year}{2007}), \bibinfo{pages}{864--886}.
\newblock
\href{https://doi.org/10.1037/0033-295X.114.4.864}{doi:\nolinkurl{10.1037/0033-295X.114.4.864}}


\bibitem[Fioretti and Smorti(2015)]%
        {fioretti2015emotional}
\bibfield{author}{\bibinfo{person}{Chiara Fioretti} {and} \bibinfo{person}{Andrea Smorti}.} \bibinfo{year}{2015}\natexlab{}.
\newblock \showarticletitle{How emotional content of memories changes in narrating}.
\newblock \bibinfo{journal}{\emph{Narrative Inquiry}} \bibinfo{volume}{25}, \bibinfo{number}{1} (\bibinfo{year}{2015}), \bibinfo{pages}{37--56}.
\newblock
\href{https://doi.org/10.1075/ni.25.1.03fio}{doi:\nolinkurl{10.1075/ni.25.1.03fio}}


\bibitem[Fitzpatrick et~al\mbox{.}(2017)]%
        {fitzpatrick2017delivering}
\bibfield{author}{\bibinfo{person}{Kathleen~Kara Fitzpatrick}, \bibinfo{person}{Alison Darcy}, {and} \bibinfo{person}{Molly Vierhile}.} \bibinfo{year}{2017}\natexlab{}.
\newblock \showarticletitle{Delivering Cognitive Behavior Therapy to Young Adults With Symptoms of Depression and Anxiety Using a Fully Automated Conversational Agent (Woebot): A Randomized Controlled Trial}.
\newblock \bibinfo{journal}{\emph{JMIR Mental Health}} \bibinfo{volume}{4}, \bibinfo{number}{2} (\bibinfo{year}{2017}), \bibinfo{pages}{e7785}.
\newblock
\href{https://doi.org/10.2196/mental.7785}{doi:\nolinkurl{10.2196/mental.7785}}


\bibitem[Fivush et~al\mbox{.}(2011)]%
        {fivush2011making}
\bibfield{author}{\bibinfo{person}{Robyn Fivush}, \bibinfo{person}{Tilmann Habermas}, \bibinfo{person}{Theodore~EA Waters}, {and} \bibinfo{person}{Widaad Zaman}.} \bibinfo{year}{2011}\natexlab{}.
\newblock \showarticletitle{The making of autobiographical memory: Intersections of culture, narratives and identity}.
\newblock \bibinfo{journal}{\emph{International Journal of Psychology}} \bibinfo{volume}{46}, \bibinfo{number}{5} (\bibinfo{year}{2011}), \bibinfo{pages}{321--345}.
\newblock
\href{https://doi.org/10.1080/00207594.2011.596541}{doi:\nolinkurl{10.1080/00207594.2011.596541}}


\bibitem[Fivush et~al\mbox{.}(1996)]%
        {fivush1996remembering}
\bibfield{author}{\bibinfo{person}{Robyn Fivush}, \bibinfo{person}{Catherine Haden}, \bibinfo{person}{Elaine Reese}, {et~al\mbox{.}}} \bibinfo{year}{1996}\natexlab{}.
\newblock \showarticletitle{Remembering, recounting, and reminiscing: The development of autobiographical memory in social context}.
\newblock In \bibinfo{booktitle}{\emph{Remembering our past: Studies in autobiographical memory}}. \bibinfo{publisher}{Cambridge University Press}, \bibinfo{address}{Cambridge}, \bibinfo{pages}{341--359}.
\newblock
\href{https://doi.org/10.1017/CBO9780511527913.014}{doi:\nolinkurl{10.1017/CBO9780511527913.014}}


\bibitem[Fletcher et~al\mbox{.}(2000)]%
        {garth2000themeasurement}
\bibfield{author}{\bibinfo{person}{Garth J.~O. Fletcher}, \bibinfo{person}{Jeffry~A. Simpson}, {and} \bibinfo{person}{Geoff Thomas}.} \bibinfo{year}{2000}\natexlab{}.
\newblock \showarticletitle{The Measurement of Perceived Relationship Quality Components: A Confirmatory Factor Analytic Approach}.
\newblock \bibinfo{journal}{\emph{Personality and Social Psychology Bulletin}} \bibinfo{volume}{26}, \bibinfo{number}{3} (\bibinfo{year}{2000}), \bibinfo{pages}{340--354}.
\newblock
\href{https://doi.org/10.1177/0146167200265007}{doi:\nolinkurl{10.1177/0146167200265007}}


\bibitem[Frey et~al\mbox{.}(2018)]%
        {frey2018breeze}
\bibfield{author}{\bibinfo{person}{J\'{e}r\'{e}my Frey}, \bibinfo{person}{May Grabli}, \bibinfo{person}{Ronit Slyper}, {and} \bibinfo{person}{Jessica~R. Cauchard}.} \bibinfo{year}{2018}\natexlab{}.
\newblock \showarticletitle{Breeze: Sharing Biofeedback through Wearable Technologies}. In \bibinfo{booktitle}{\emph{Proceedings of the 2018 CHI Conference on Human Factors in Computing Systems}} (Montreal QC, Canada) \emph{(\bibinfo{series}{CHI '18})}. \bibinfo{publisher}{Association for Computing Machinery}, \bibinfo{address}{New York, NY, USA}, \bibinfo{pages}{1–12}.
\newblock
\showISBNx{9781450356206}
\href{https://doi.org/10.1145/3173574.3174219}{doi:\nolinkurl{10.1145/3173574.3174219}}


\bibitem[Fu et~al\mbox{.}(2024)]%
        {fu2024from}
\bibfield{author}{\bibinfo{person}{Yue Fu}, \bibinfo{person}{Sami Foell}, \bibinfo{person}{Xuhai Xu}, {and} \bibinfo{person}{Alexis Hiniker}.} \bibinfo{year}{2024}\natexlab{}.
\newblock \showarticletitle{From Text to Self: Users’ Perception of AIMC Tools on Interpersonal Communication and Self}. In \bibinfo{booktitle}{\emph{Proceedings of the 2024 CHI Conference on Human Factors in Computing Systems}} (Honolulu, HI, USA) \emph{(\bibinfo{series}{CHI '24})}. \bibinfo{publisher}{Association for Computing Machinery}, \bibinfo{address}{New York, NY, USA}, Article \bibinfo{articleno}{977}, \bibinfo{numpages}{17}~pages.
\newblock
\showISBNx{9798400703300}
\href{https://doi.org/10.1145/3613904.3641955}{doi:\nolinkurl{10.1145/3613904.3641955}}


\bibitem[Gardner et~al\mbox{.}(2012)]%
        {gardner2012quantitative}
\bibfield{author}{\bibinfo{person}{Robert~S. Gardner}, \bibinfo{person}{Adam~T. Vogel}, \bibinfo{person}{Matteo Mainetti}, {and} \bibinfo{person}{Giorgio~A. Ascoli}.} \bibinfo{year}{2012}\natexlab{}.
\newblock \showarticletitle{Quantitative measurements of autobiographical memory content}.
\newblock \bibinfo{journal}{\emph{PLoS ONE}} \bibinfo{volume}{7}, \bibinfo{number}{9} (\bibinfo{year}{2012}), \bibinfo{pages}{e44809}.
\newblock
\href{https://doi.org/10.1371/journal.pone.0044809}{doi:\nolinkurl{10.1371/journal.pone.0044809}}


\bibitem[Gibson et~al\mbox{.}(2023)]%
        {gibson2023gifting}
\bibfield{author}{\bibinfo{person}{Rebecca Gibson}, \bibinfo{person}{Boriana Koleva}, \bibinfo{person}{Martin Flintham}, {and} \bibinfo{person}{Heidi Winklhofer}.} \bibinfo{year}{2023}\natexlab{}.
\newblock \showarticletitle{Gifting the Past in the Present: An Exploration of Evoking Nostalgia through Hybrid Gifts}. In \bibinfo{booktitle}{\emph{Proceedings of the 2023 ACM Designing Interactive Systems Conference}} (Pittsburgh, PA, USA) \emph{(\bibinfo{series}{DIS '23})}. \bibinfo{publisher}{Association for Computing Machinery}, \bibinfo{address}{New York, NY, USA}, \bibinfo{pages}{1047–1059}.
\newblock
\showISBNx{9781450398930}
\href{https://doi.org/10.1145/3563657.3596103}{doi:\nolinkurl{10.1145/3563657.3596103}}


\bibitem[Gosling et~al\mbox{.}(2003)]%
        {samuel2003averybrif}
\bibfield{author}{\bibinfo{person}{Samuel~D Gosling}, \bibinfo{person}{Peter~J Rentfrow}, {and} \bibinfo{person}{William~B Swann}.} \bibinfo{year}{2003}\natexlab{}.
\newblock \showarticletitle{A very brief measure of the Big-Five personality domains}.
\newblock \bibinfo{journal}{\emph{Journal of Research in Personality}} \bibinfo{volume}{37}, \bibinfo{number}{6} (\bibinfo{year}{2003}), \bibinfo{pages}{504--528}.
\newblock
\href{https://doi.org/10.1016/S0092-6566(03)00046-1}{doi:\nolinkurl{10.1016/S0092-6566(03)00046-1}}


\bibitem[Govers et~al\mbox{.}(2024)]%
        {govers2024aidriven}
\bibfield{author}{\bibinfo{person}{Jarod Govers}, \bibinfo{person}{Eduardo Velloso}, \bibinfo{person}{Vassilis Kostakos}, {and} \bibinfo{person}{Jorge Goncalves}.} \bibinfo{year}{2024}\natexlab{}.
\newblock \showarticletitle{AI-Driven Mediation Strategies for Audience Depolarisation in Online Debates}. In \bibinfo{booktitle}{\emph{Proceedings of the 2024 CHI Conference on Human Factors in Computing Systems}} (Honolulu, HI, USA) \emph{(\bibinfo{series}{CHI '24})}. \bibinfo{publisher}{Association for Computing Machinery}, \bibinfo{address}{New York, NY, USA}, Article \bibinfo{articleno}{803}, \bibinfo{numpages}{18}~pages.
\newblock
\showISBNx{9798400703300}
\href{https://doi.org/10.1145/3613904.3642322}{doi:\nolinkurl{10.1145/3613904.3642322}}


\bibitem[Gray et~al\mbox{.}(2020)]%
        {gray2020towards}
\bibfield{author}{\bibinfo{person}{Stuart Gray}, \bibinfo{person}{Rachel Hahn}, \bibinfo{person}{Kirsten Cater}, \bibinfo{person}{Debbie Watson}, \bibinfo{person}{Keir Williams}, \bibinfo{person}{Tom Metcalfe}, {and} \bibinfo{person}{Chloe Meineck}.} \bibinfo{year}{2020}\natexlab{}.
\newblock \showarticletitle{Towards A Design For Life: Redesigning For Reminiscence With Looked After Children}. In \bibinfo{booktitle}{\emph{Proceedings of the 2020 CHI Conference on Human Factors in Computing Systems}} (Honolulu, HI, USA) \emph{(\bibinfo{series}{CHI '20})}. \bibinfo{publisher}{Association for Computing Machinery}, \bibinfo{address}{New York, NY, USA}, \bibinfo{pages}{1–14}.
\newblock
\showISBNx{9781450367080}
\href{https://doi.org/10.1145/3313831.3376824}{doi:\nolinkurl{10.1145/3313831.3376824}}


\bibitem[Guan and Wang(2022)]%
        {guan2022does}
\bibfield{author}{\bibinfo{person}{Li Guan} {and} \bibinfo{person}{Qi Wang}.} \bibinfo{year}{2022}\natexlab{}.
\newblock \showarticletitle{Does sharing memories make us feel closer? The roles of memory type and culture}.
\newblock \bibinfo{journal}{\emph{Journal of Cross-Cultural Psychology}} \bibinfo{volume}{53}, \bibinfo{number}{3-4} (\bibinfo{year}{2022}), \bibinfo{pages}{344--361}.
\newblock
\href{https://doi.org/10.1177/00220221211072809}{doi:\nolinkurl{10.1177/00220221211072809}}


\bibitem[Guo and Inoue(2019)]%
        {guo2019using}
\bibfield{author}{\bibinfo{person}{Zixuan Guo} {and} \bibinfo{person}{Tomoo Inoue}.} \bibinfo{year}{2019}\natexlab{}.
\newblock \showarticletitle{Using a Conversational Agent to Facilitate Non-native Speaker's Active Participation in Conversation}. In \bibinfo{booktitle}{\emph{Extended Abstracts of the 2019 CHI Conference on Human Factors in Computing Systems}} (Glasgow, Scotland Uk) \emph{(\bibinfo{series}{CHI EA '19})}. \bibinfo{publisher}{Association for Computing Machinery}, \bibinfo{address}{New York, NY, USA}, \bibinfo{pages}{1–6}.
\newblock
\showISBNx{9781450359719}
\href{https://doi.org/10.1145/3290607.3313075}{doi:\nolinkurl{10.1145/3290607.3313075}}


\bibitem[Hallford et~al\mbox{.}(2024)]%
        {hallford2024guided}
\bibfield{author}{\bibinfo{person}{David Hallford}, \bibinfo{person}{Meg Woolfit}, \bibinfo{person}{Alicia Follett}, \bibinfo{person}{Elizabeth Jones}, \bibinfo{person}{Ollie Harrison}, {and} \bibinfo{person}{David Austin}.} \bibinfo{year}{2024}\natexlab{}.
\newblock \showarticletitle{Guided recall of positive autobiographical memories increases anticipated pleasure and psychological resources, and reduces depressive symptoms: a replication and extension of a randomised controlled trial of brief positive cognitive-reminiscence therapy}.
\newblock \bibinfo{journal}{\emph{Memory}} \bibinfo{volume}{32}, \bibinfo{number}{4} (\bibinfo{year}{2024}), \bibinfo{pages}{465--475}.
\newblock
\href{https://doi.org/10.1080/09658211.2024.2333510}{doi:\nolinkurl{10.1080/09658211.2024.2333510}}


\bibitem[Hallford et~al\mbox{.}(2022)]%
        {hallford2022remembering}
\bibfield{author}{\bibinfo{person}{David~John Hallford}, \bibinfo{person}{Sarah Hardgrove}, \bibinfo{person}{Meghna Sanam}, \bibinfo{person}{Stefany Oliveira}, \bibinfo{person}{Megan Pilon}, {and} \bibinfo{person}{Tyler Duran}.} \bibinfo{year}{2022}\natexlab{}.
\newblock \showarticletitle{Remembering for resilience: Brief cognitive-reminiscence therapy improves psychological resources and mental well-being in young adults}.
\newblock \bibinfo{journal}{\emph{Applied Psychology: Health and Well-Being}} \bibinfo{volume}{14}, \bibinfo{number}{3} (\bibinfo{year}{2022}), \bibinfo{pages}{1004--1021}.
\newblock
\href{https://doi.org/10.1111/aphw.12364}{doi:\nolinkurl{10.1111/aphw.12364}}


\bibitem[Hallford and Mellor(2016)]%
        {hallford2016brief}
\bibfield{author}{\bibinfo{person}{David~John Hallford} {and} \bibinfo{person}{David Mellor}.} \bibinfo{year}{2016}\natexlab{}.
\newblock \showarticletitle{Brief reminiscence activities improve state well-being and self-concept in young adults: A randomised controlled experiment}.
\newblock \bibinfo{journal}{\emph{Memory}} \bibinfo{volume}{24}, \bibinfo{number}{10} (\bibinfo{year}{2016}), \bibinfo{pages}{1311--1320}.
\newblock
\href{https://doi.org/10.1080/09658211.2015.1103875}{doi:\nolinkurl{10.1080/09658211.2015.1103875}}


\bibitem[Halpern and Katz(2017)]%
        {halpern2017texting}
\bibfield{author}{\bibinfo{person}{Daniel Halpern} {and} \bibinfo{person}{James~E. Katz}.} \bibinfo{year}{2017}\natexlab{}.
\newblock \showarticletitle{Texting's consequences for romantic relationships: A cross-lagged analysis highlights its risks}.
\newblock \bibinfo{journal}{\emph{Computers in Human Behavior}}  \bibinfo{volume}{71} (\bibinfo{year}{2017}), \bibinfo{pages}{386--394}.
\newblock
\href{https://doi.org/10.1016/j.chb.2017.01.051}{doi:\nolinkurl{10.1016/j.chb.2017.01.051}}


\bibitem[Han et~al\mbox{.}(2023)]%
        {li2023systematic}
\bibfield{author}{\bibinfo{person}{Li Han}, \bibinfo{person}{Renwen Zhang}, \bibinfo{person}{Yi-Chieh Lee}, \bibinfo{person}{Robert~E. Kraut}, {and} \bibinfo{person}{David~C. Mohr}.} \bibinfo{year}{2023}\natexlab{}.
\newblock \showarticletitle{Systematic review and meta-analysis of AI-based conversational agents for promoting mental health and well-being}.
\newblock \bibinfo{journal}{\emph{NPJ Digital Medicine}} \bibinfo{volume}{6}, \bibinfo{number}{1} (\bibinfo{year}{2023}), \bibinfo{pages}{236}.
\newblock
\href{https://doi.org/10.1038/s41746-023-00979-5}{doi:\nolinkurl{10.1038/s41746-023-00979-5}}


\bibitem[Hangal et~al\mbox{.}(2011)]%
        {hangal2011muse}
\bibfield{author}{\bibinfo{person}{Sudheendra Hangal}, \bibinfo{person}{Monica~S. Lam}, {and} \bibinfo{person}{Jeffrey Heer}.} \bibinfo{year}{2011}\natexlab{}.
\newblock \showarticletitle{MUSE: reviving memories using email archives}. In \bibinfo{booktitle}{\emph{Proceedings of the 24th Annual ACM Symposium on User Interface Software and Technology}} (Santa Barbara, California, USA) \emph{(\bibinfo{series}{UIST '11})}. \bibinfo{publisher}{Association for Computing Machinery}, \bibinfo{address}{New York, NY, USA}, \bibinfo{pages}{75–84}.
\newblock
\showISBNx{9781450307161}
\href{https://doi.org/10.1145/2047196.2047206}{doi:\nolinkurl{10.1145/2047196.2047206}}


\bibitem[Hassenzahl et~al\mbox{.}(2012)]%
        {hassenzahl2012all}
\bibfield{author}{\bibinfo{person}{Marc Hassenzahl}, \bibinfo{person}{Stephanie Heidecker}, \bibinfo{person}{Kai Eckoldt}, \bibinfo{person}{Sarah Diefenbach}, {and} \bibinfo{person}{Uwe Hillmann}.} \bibinfo{year}{2012}\natexlab{}.
\newblock \showarticletitle{All You Need is Love: Current Strategies of Mediating Intimate Relationships through Technology}.
\newblock \bibinfo{journal}{\emph{ACM Trans. Comput.-Hum. Interact.}} \bibinfo{volume}{19}, \bibinfo{number}{4}, Article \bibinfo{articleno}{30} (\bibinfo{date}{Dec.} \bibinfo{year}{2012}), \bibinfo{numpages}{19}~pages.
\newblock
\showISSN{1073-0516}
\href{https://doi.org/10.1145/2395131.2395137}{doi:\nolinkurl{10.1145/2395131.2395137}}


\bibitem[Haugeland et~al\mbox{.}(2022)]%
        {isabel2022understanding}
\bibfield{author}{\bibinfo{person}{Isabel Kathleen~Fornell Haugeland}, \bibinfo{person}{Asbjørn Følstad}, \bibinfo{person}{Cameron Taylor}, {and} \bibinfo{person}{Cato~Alexander Bjørkli}.} \bibinfo{year}{2022}\natexlab{}.
\newblock \showarticletitle{Understanding the user experience of customer service chatbots: An experimental study of chatbot interaction design}.
\newblock \bibinfo{journal}{\emph{International Journal of Human-Computer Studies}}  \bibinfo{volume}{161} (\bibinfo{year}{2022}), \bibinfo{pages}{102788}.
\newblock
\href{https://doi.org/10.1016/j.ijhcs.2022.102788}{doi:\nolinkurl{10.1016/j.ijhcs.2022.102788}}


\bibitem[Hill et~al\mbox{.}(2015)]%
        {hill2015real}
\bibfield{author}{\bibinfo{person}{Jennifer Hill}, \bibinfo{person}{W. {Randolph Ford}}, {and} \bibinfo{person}{Ingrid~G. Farreras}.} \bibinfo{year}{2015}\natexlab{}.
\newblock \showarticletitle{Real conversations with artificial intelligence: A comparison between human–human online conversations and human–chatbot conversations}.
\newblock \bibinfo{journal}{\emph{Computers in Human Behavior}}  \bibinfo{volume}{49} (\bibinfo{year}{2015}), \bibinfo{pages}{245--250}.
\newblock
\href{https://doi.org/10.1016/j.chb.2015.02.026}{doi:\nolinkurl{10.1016/j.chb.2015.02.026}}


\bibitem[Holland and Kensinger(2010)]%
        {holland2010emotion}
\bibfield{author}{\bibinfo{person}{Alisha~C Holland} {and} \bibinfo{person}{Elizabeth~A Kensinger}.} \bibinfo{year}{2010}\natexlab{}.
\newblock \showarticletitle{Emotion and autobiographical memory}.
\newblock \bibinfo{journal}{\emph{Physics of Life Reviews}} \bibinfo{volume}{7}, \bibinfo{number}{1} (\bibinfo{year}{2010}), \bibinfo{pages}{88--131}.
\newblock
\href{https://doi.org/10.1016/j.plrev.2010.01.006}{doi:\nolinkurl{10.1016/j.plrev.2010.01.006}}


\bibitem[Houben et~al\mbox{.}(2019)]%
        {houben2019foregrounding}
\bibfield{author}{\bibinfo{person}{Maarten Houben}, \bibinfo{person}{Rens Brankaert}, \bibinfo{person}{Saskia Bakker}, \bibinfo{person}{Gail Kenning}, \bibinfo{person}{Inge Bongers}, {and} \bibinfo{person}{Berry Eggen}.} \bibinfo{year}{2019}\natexlab{}.
\newblock \showarticletitle{Foregrounding Everyday Sounds in Dementia}. In \bibinfo{booktitle}{\emph{Proceedings of the 2019 on Designing Interactive Systems Conference}} (San Diego, CA, USA) \emph{(\bibinfo{series}{DIS '19})}. \bibinfo{publisher}{Association for Computing Machinery}, \bibinfo{address}{New York, NY, USA}, \bibinfo{pages}{71–83}.
\newblock
\showISBNx{9781450358507}
\href{https://doi.org/10.1145/3322276.3322287}{doi:\nolinkurl{10.1145/3322276.3322287}}


\bibitem[Im et~al\mbox{.}(2023)]%
        {im2023less}
\bibfield{author}{\bibinfo{person}{Jane Im}, \bibinfo{person}{Ruiyi Wang}, \bibinfo{person}{Weikun Lyu}, \bibinfo{person}{Nick Cook}, \bibinfo{person}{Hana Habib}, \bibinfo{person}{Lorrie~Faith Cranor}, \bibinfo{person}{Nikola Banovic}, {and} \bibinfo{person}{Florian Schaub}.} \bibinfo{year}{2023}\natexlab{}.
\newblock \showarticletitle{Less is Not More: Improving Findability and Actionability of Privacy Controls for Online Behavioral Advertising}. In \bibinfo{booktitle}{\emph{Proceedings of the 2023 CHI Conference on Human Factors in Computing Systems}} (Hamburg, Germany) \emph{(\bibinfo{series}{CHI '23})}. \bibinfo{publisher}{Association for Computing Machinery}, \bibinfo{address}{New York, NY, USA}, Article \bibinfo{articleno}{661}, \bibinfo{numpages}{33}~pages.
\newblock
\showISBNx{9781450394215}
\href{https://doi.org/10.1145/3544548.3580773}{doi:\nolinkurl{10.1145/3544548.3580773}}


\bibitem[Inkster et~al\mbox{.}(2018)]%
        {inkster2018anempathydriven}
\bibfield{author}{\bibinfo{person}{Becky Inkster}, \bibinfo{person}{Shubhankar Sarda}, {and} \bibinfo{person}{Vinod Subramanian}.} \bibinfo{year}{2018}\natexlab{}.
\newblock \showarticletitle{An Empathy-Driven, Conversational Artificial Intelligence Agent (Wysa) for Digital Mental Well-Being: Real-World Data Evaluation Mixed-Methods Study}.
\newblock \bibinfo{journal}{\emph{JMIR mHealth and uHealth}} \bibinfo{volume}{6}, \bibinfo{number}{11} (\bibinfo{year}{2018}), \bibinfo{pages}{e12106}.
\newblock
\href{https://doi.org/10.2196/12106}{doi:\nolinkurl{10.2196/12106}}


\bibitem[Isaacs et~al\mbox{.}(2013)]%
        {isaacs2013echoes}
\bibfield{author}{\bibinfo{person}{Ellen Isaacs}, \bibinfo{person}{Artie Konrad}, \bibinfo{person}{Alan Walendowski}, \bibinfo{person}{Thomas Lennig}, \bibinfo{person}{Victoria Hollis}, {and} \bibinfo{person}{Steve Whittaker}.} \bibinfo{year}{2013}\natexlab{}.
\newblock \showarticletitle{Echoes from the past: how technology mediated reflection improves well-being}. In \bibinfo{booktitle}{\emph{Proceedings of the SIGCHI Conference on Human Factors in Computing Systems}} (Paris, France) \emph{(\bibinfo{series}{CHI '13})}. \bibinfo{publisher}{Association for Computing Machinery}, \bibinfo{address}{New York, NY, USA}, \bibinfo{pages}{1071–1080}.
\newblock
\showISBNx{9781450318990}
\href{https://doi.org/10.1145/2470654.2466137}{doi:\nolinkurl{10.1145/2470654.2466137}}


\bibitem[J.(1999)]%
        {davis1999gender}
\bibfield{author}{\bibinfo{person}{Davis~P. J.}} \bibinfo{year}{1999}\natexlab{}.
\newblock \showarticletitle{Gender differences in autobiographical memory for childhood emotional experiences}.
\newblock \bibinfo{journal}{\emph{J Pers Soc Psychol}}  \bibinfo{volume}{76} (\bibinfo{year}{1999}), \bibinfo{pages}{498--510}.
\newblock
\href{https://doi.org/10.1037//0022-3514.76.3.498}{doi:\nolinkurl{10.1037//0022-3514.76.3.498}}


\bibitem[J et~al\mbox{.}(2020)]%
        {liem2020structure}
\bibfield{author}{\bibinfo{person}{Liem J}, \bibinfo{person}{Perin C}, {and} \bibinfo{person}{Wood J}.} \bibinfo{year}{2020}\natexlab{}.
\newblock \showarticletitle{Structure and Empathy in Visual Data Storytelling: Evaluating their Influence on Attitude}.
\newblock \bibinfo{journal}{\emph{Computer Graphics Forum}} \bibinfo{volume}{39}, \bibinfo{number}{3} (\bibinfo{year}{2020}), \bibinfo{pages}{277--289}.
\newblock
\href{https://doi.org/10.1111/cgf.13980}{doi:\nolinkurl{10.1111/cgf.13980}}


\bibitem[James and Bhar(2016)]%
        {kelly2016brief}
\bibfield{author}{\bibinfo{person}{Kelly~L. James} {and} \bibinfo{person}{Sunil~S. Bhar}.} \bibinfo{year}{2016}\natexlab{}.
\newblock \showarticletitle{Brief reminiscence intervention improves affect and pessimism in non‐clinical individuals: A pilot study}.
\newblock \bibinfo{journal}{\emph{Clinical Psychologist}} \bibinfo{volume}{20}, \bibinfo{number}{3} (\bibinfo{year}{2016}), \bibinfo{pages}{119--124}.
\newblock
\href{https://doi.org/10.1111/cp.12068}{doi:\nolinkurl{10.1111/cp.12068}}


\bibitem[Jin et~al\mbox{.}(2024)]%
        {jin2024exploring}
\bibfield{author}{\bibinfo{person}{Yucheng Jin}, \bibinfo{person}{Wanling Cai}, \bibinfo{person}{Li Chen}, \bibinfo{person}{Yizhe Zhang}, \bibinfo{person}{Gavin Doherty}, {and} \bibinfo{person}{Tonglin Jiang}.} \bibinfo{year}{2024}\natexlab{}.
\newblock \showarticletitle{Exploring the Design of Generative AI in Supporting Music-based Reminiscence for Older Adults}. In \bibinfo{booktitle}{\emph{Proceedings of the 2024 CHI Conference on Human Factors in Computing Systems}} (Honolulu, HI, USA) \emph{(\bibinfo{series}{CHI '24})}. \bibinfo{publisher}{Association for Computing Machinery}, \bibinfo{address}{New York, NY, USA}, Article \bibinfo{articleno}{1012}, \bibinfo{numpages}{17}~pages.
\newblock
\showISBNx{9798400703300}
\href{https://doi.org/10.1145/3613904.3642800}{doi:\nolinkurl{10.1145/3613904.3642800}}


\bibitem[Kang et~al\mbox{.}(2021)]%
        {kang2021momentmeld}
\bibfield{author}{\bibinfo{person}{Bumsoo Kang}, \bibinfo{person}{Seungwoo Kang}, {and} \bibinfo{person}{Inseok Hwang}.} \bibinfo{year}{2021}\natexlab{}.
\newblock \showarticletitle{MomentMeld: AI-augmented Mobile Photographic Memento towards Mutually Stimulatory Inter-generational Interaction}. In \bibinfo{booktitle}{\emph{Proceedings of the 2021 CHI Conference on Human Factors in Computing Systems}} (Yokohama, Japan) \emph{(\bibinfo{series}{CHI '21})}. \bibinfo{publisher}{Association for Computing Machinery}, \bibinfo{address}{New York, NY, USA}, Article \bibinfo{articleno}{479}, \bibinfo{numpages}{16}~pages.
\newblock
\showISBNx{9781450380966}
\href{https://doi.org/10.1145/3411764.3445688}{doi:\nolinkurl{10.1145/3411764.3445688}}


\bibitem[Kaye et~al\mbox{.}(2005)]%
        {kaye2005communicating}
\bibfield{author}{\bibinfo{person}{Joseph~'Jofish' Kaye}, \bibinfo{person}{Mariah~K. Levitt}, \bibinfo{person}{Jeffrey Nevins}, \bibinfo{person}{Jessica Golden}, {and} \bibinfo{person}{Vanessa Schmidt}.} \bibinfo{year}{2005}\natexlab{}.
\newblock \showarticletitle{Communicating intimacy one bit at a time}. In \bibinfo{booktitle}{\emph{CHI '05 Extended Abstracts on Human Factors in Computing Systems}} (Portland, OR, USA) \emph{(\bibinfo{series}{CHI EA '05})}. \bibinfo{publisher}{Association for Computing Machinery}, \bibinfo{address}{New York, NY, USA}, \bibinfo{pages}{1529–1532}.
\newblock
\showISBNx{1595930027}
\href{https://doi.org/10.1145/1056808.1056958}{doi:\nolinkurl{10.1145/1056808.1056958}}


\bibitem[Kellas(2005)]%
        {kellas2005family}
\bibfield{author}{\bibinfo{person}{Jody~Koenig Kellas}.} \bibinfo{year}{2005}\natexlab{}.
\newblock \showarticletitle{Family ties: Communicating identity through jointly told family stories}.
\newblock \bibinfo{journal}{\emph{Communication Monographs}} \bibinfo{volume}{72}, \bibinfo{number}{4} (\bibinfo{year}{2005}), \bibinfo{pages}{365--389}.
\newblock
\href{https://doi.org/10.1080/03637750500322453}{doi:\nolinkurl{10.1080/03637750500322453}}


\bibitem[Kellas(2010)]%
        {kellas2010narrating}
\bibfield{author}{\bibinfo{person}{Jody~Koenig Kellas}.} \bibinfo{year}{2010}\natexlab{}.
\newblock \showarticletitle{Narrating Family: Introduction to the Special Issue on Narratives and Storytelling in the Family}.
\newblock \bibinfo{journal}{\emph{Journal of Family Communication}} \bibinfo{volume}{10}, \bibinfo{number}{1} (\bibinfo{year}{2010}), \bibinfo{pages}{1--6}.
\newblock
\href{https://doi.org/10.1080/15267430903401441}{doi:\nolinkurl{10.1080/15267430903401441}}


\bibitem[Kim et~al\mbox{.}(2020)]%
        {kim2020botinthebunch}
\bibfield{author}{\bibinfo{person}{Soomin Kim}, \bibinfo{person}{Jinsu Eun}, \bibinfo{person}{Changhoon Oh}, \bibinfo{person}{Bongwon Suh}, {and} \bibinfo{person}{Joonhwan Lee}.} \bibinfo{year}{2020}\natexlab{}.
\newblock \showarticletitle{Bot in the Bunch: Facilitating Group Chat Discussion by Improving Efficiency and Participation with a Chatbot}. In \bibinfo{booktitle}{\emph{Proceedings of the 2020 CHI Conference on Human Factors in Computing Systems}} (Honolulu, HI, USA) \emph{(\bibinfo{series}{CHI '20})}. \bibinfo{publisher}{Association for Computing Machinery}, \bibinfo{address}{New York, NY, USA}, \bibinfo{pages}{1–13}.
\newblock
\showISBNx{9781450367080}
\href{https://doi.org/10.1145/3313831.3376785}{doi:\nolinkurl{10.1145/3313831.3376785}}


\bibitem[Kim et~al\mbox{.}(2021)]%
        {kim2021moderator}
\bibfield{author}{\bibinfo{person}{Soomin Kim}, \bibinfo{person}{Jinsu Eun}, \bibinfo{person}{Joseph Seering}, {and} \bibinfo{person}{Joonhwan Lee}.} \bibinfo{year}{2021}\natexlab{}.
\newblock \showarticletitle{Moderator Chatbot for Deliberative Discussion: Effects of Discussion Structure and Discussant Facilitation}.
\newblock \bibinfo{journal}{\emph{Proc. ACM Hum.-Comput. Interact.}} \bibinfo{volume}{5}, \bibinfo{number}{CSCW1}, Article \bibinfo{articleno}{87} (\bibinfo{date}{April} \bibinfo{year}{2021}), \bibinfo{numpages}{26}~pages.
\newblock
\href{https://doi.org/10.1145/3449161}{doi:\nolinkurl{10.1145/3449161}}


\bibitem[Kim et~al\mbox{.}(2022)]%
        {kim2022slide2remember}
\bibfield{author}{\bibinfo{person}{Subin Kim}, \bibinfo{person}{Sangsu Jang}, \bibinfo{person}{Jin-young Moon}, \bibinfo{person}{Minjoo Han}, {and} \bibinfo{person}{Young-Woo Park}.} \bibinfo{year}{2022}\natexlab{}.
\newblock \showarticletitle{Slide2Remember: an Interactive Wall Frame Enriching Reminiscence Experiences by Providing Re-encounters of Taken Photos and Heard Music in a Similar Period}. In \bibinfo{booktitle}{\emph{Proceedings of the 2022 ACM Designing Interactive Systems Conference}} (Virtual Event, Australia) \emph{(\bibinfo{series}{DIS '22})}. \bibinfo{publisher}{Association for Computing Machinery}, \bibinfo{address}{New York, NY, USA}, \bibinfo{pages}{288–300}.
\newblock
\showISBNx{9781450393584}
\href{https://doi.org/10.1145/3532106.3533456}{doi:\nolinkurl{10.1145/3532106.3533456}}


\bibitem[Kim et~al\mbox{.}(2024)]%
        {kim2024mindfuldiary}
\bibfield{author}{\bibinfo{person}{Taewan Kim}, \bibinfo{person}{Seolyeong Bae}, \bibinfo{person}{Hyun~Ah Kim}, \bibinfo{person}{Su-Woo Lee}, \bibinfo{person}{Hwajung Hong}, \bibinfo{person}{Chanmo Yang}, {and} \bibinfo{person}{Young-Ho Kim}.} \bibinfo{year}{2024}\natexlab{}.
\newblock \showarticletitle{MindfulDiary: Harnessing Large Language Model to Support Psychiatric Patients' Journaling}. In \bibinfo{booktitle}{\emph{Proceedings of the 2024 CHI Conference on Human Factors in Computing Systems}} (Honolulu, HI, USA) \emph{(\bibinfo{series}{CHI '24})}. \bibinfo{publisher}{Association for Computing Machinery}, \bibinfo{address}{New York, NY, USA}, Article \bibinfo{articleno}{701}, \bibinfo{numpages}{20}~pages.
\newblock
\showISBNx{9798400703300}
\href{https://doi.org/10.1145/3613904.3642937}{doi:\nolinkurl{10.1145/3613904.3642937}}


\bibitem[Klopfenstein et~al\mbox{.}(2017)]%
        {klopfenstein2017therise}
\bibfield{author}{\bibinfo{person}{Lorenz~Cuno Klopfenstein}, \bibinfo{person}{Saverio Delpriori}, \bibinfo{person}{Silvia Malatini}, {and} \bibinfo{person}{Alessandro Bogliolo}.} \bibinfo{year}{2017}\natexlab{}.
\newblock \showarticletitle{The Rise of Bots: A Survey of Conversational Interfaces, Patterns, and Paradigms}. In \bibinfo{booktitle}{\emph{Proceedings of the 2017 Conference on Designing Interactive Systems}} (Edinburgh, United Kingdom) \emph{(\bibinfo{series}{DIS '17})}. \bibinfo{publisher}{Association for Computing Machinery}, \bibinfo{address}{New York, NY, USA}, \bibinfo{pages}{555–565}.
\newblock
\showISBNx{9781450349222}
\href{https://doi.org/10.1145/3064663.3064672}{doi:\nolinkurl{10.1145/3064663.3064672}}


\bibitem[Kowalski et~al\mbox{.}(2013)]%
        {kowalski2013cubble}
\bibfield{author}{\bibinfo{person}{Robert Kowalski}, \bibinfo{person}{Sebastian Loehmann}, {and} \bibinfo{person}{Doris Hausen}.} \bibinfo{year}{2013}\natexlab{}.
\newblock \showarticletitle{cubble: a multi-device hybrid approach supporting communication in long-distance relationships}. In \bibinfo{booktitle}{\emph{Proceedings of the 7th International Conference on Tangible, Embedded and Embodied Interaction}} (Barcelona, Spain) \emph{(\bibinfo{series}{TEI '13})}. \bibinfo{publisher}{Association for Computing Machinery}, \bibinfo{address}{New York, NY, USA}, \bibinfo{pages}{201–204}.
\newblock
\showISBNx{9781450318983}
\href{https://doi.org/10.1145/2460625.2460656}{doi:\nolinkurl{10.1145/2460625.2460656}}


\bibitem[Kreiner and Levi-Belz(2019)]%
        {kreiner2019selfdisclosure}
\bibfield{author}{\bibinfo{person}{Hamutal Kreiner} {and} \bibinfo{person}{Yossi Levi-Belz}.} \bibinfo{year}{2019}\natexlab{}.
\newblock \showarticletitle{Self-Disclosure Here and Now: Combining Retrospective Perceived Assessment With Dynamic Behavioral Measures}.
\newblock \bibinfo{journal}{\emph{Frontiers in Psychology}}  \bibinfo{volume}{10} (\bibinfo{year}{2019}).
\newblock
\href{https://doi.org/10.3389/fpsyg.2019.00558}{doi:\nolinkurl{10.3389/fpsyg.2019.00558}}


\bibitem[Kumar et~al\mbox{.}(2022)]%
        {kumar2022exploringdesignpromptsapplying}
\bibfield{author}{\bibinfo{person}{Harsh Kumar}, \bibinfo{person}{Ilya Musabirov}, \bibinfo{person}{Jiakai Shi}, \bibinfo{person}{Adele Lauzon}, \bibinfo{person}{Kwan~Kiu Choy}, \bibinfo{person}{Ofek Gross}, \bibinfo{person}{Dana Kulzhabayeva}, {and} \bibinfo{person}{Joseph~Jay Williams}.} \bibinfo{year}{2022}\natexlab{}.
\newblock \bibinfo{title}{Exploring The Design of Prompts For Applying GPT-3 based Chatbots: A Mental Wellbeing Case Study on Mechanical Turk}.
\newblock
\showeprint[arxiv]{2209.11344}~[cs.HC]
\urldef\tempurl%
\url{https://arxiv.org/abs/2209.11344}
\showURL{%
\tempurl}


\bibitem[Kumar et~al\mbox{.}(2023)]%
        {kumar2023mindfulness}
\bibfield{author}{\bibinfo{person}{Harsh Kumar}, \bibinfo{person}{Yiyi Wang}, \bibinfo{person}{Jiakai Shi}, \bibinfo{person}{Ilya Musabirov}, \bibinfo{person}{Norman A.~S. Farb}, {and} \bibinfo{person}{Joseph~Jay Williams}.} \bibinfo{year}{2023}\natexlab{}.
\newblock \showarticletitle{Exploring the Use of Large Language Models for Improving the Awareness of Mindfulness}. In \bibinfo{booktitle}{\emph{Extended Abstracts of the 2023 CHI Conference on Human Factors in Computing Systems}} (Hamburg, Germany) \emph{(\bibinfo{series}{CHI EA '23})}. \bibinfo{publisher}{Association for Computing Machinery}, \bibinfo{address}{New York, NY, USA}, Article \bibinfo{articleno}{129}, \bibinfo{numpages}{7}~pages.
\newblock
\showISBNx{9781450394222}
\href{https://doi.org/10.1145/3544549.3585614}{doi:\nolinkurl{10.1145/3544549.3585614}}


\bibitem[L et~al\mbox{.}(2004)]%
        {gable2004you}
\bibfield{author}{\bibinfo{person}{Gable~Shelly L}, \bibinfo{person}{Reis~Harry T}, \bibinfo{person}{Impett~Emily A}, {and} \bibinfo{person}{Asher~Evan R}.} \bibinfo{year}{2004}\natexlab{}.
\newblock \showarticletitle{What do you do when things go right? The intrapersonal and interpersonal benefits of sharing positive events}.
\newblock \bibinfo{journal}{\emph{Journal of Personality and Social Psychology}} \bibinfo{volume}{87}, \bibinfo{number}{2} (\bibinfo{year}{2004}), \bibinfo{pages}{228–245}.
\newblock
\href{https://doi.org/10.1037/0022-3514.87.2.228}{doi:\nolinkurl{10.1037/0022-3514.87.2.228}}


\bibitem[Laestadius et~al\mbox{.}(2024)]%
        {laestadius2024toohuman}
\bibfield{author}{\bibinfo{person}{Linnea Laestadius}, \bibinfo{person}{Andrea Bishop}, \bibinfo{person}{Michael Gonzalez}, \bibinfo{person}{Diana Illenčík}, {and} \bibinfo{person}{Celeste Campos-Castillo}.} \bibinfo{year}{2024}\natexlab{}.
\newblock \showarticletitle{Too human and not human enough: A grounded theory analysis of mental health harms from emotional dependence on the social chatbot Replika}.
\newblock \bibinfo{journal}{\emph{New Media \& Society}} \bibinfo{volume}{26}, \bibinfo{number}{10} (\bibinfo{year}{2024}), \bibinfo{pages}{5923--5941}.
\newblock
\href{https://doi.org/10.1177/14614448221142007}{doi:\nolinkurl{10.1177/14614448221142007}}


\bibitem[Larsen(2014)]%
        {larsen2014whatisit}
\bibfield{author}{\bibinfo{person}{Steen~F Larsen}.} \bibinfo{year}{2014}\natexlab{}.
\newblock \showarticletitle{What is it like to remember? On phenomenal qualities of memory}.
\newblock In \bibinfo{booktitle}{\emph{Autobiographical Memory}}. \bibinfo{publisher}{Psychology Press}, \bibinfo{address}{New York, NY}, \bibinfo{pages}{163--190}.
\newblock
\href{https://doi.org/10.4324/9781315784250-10}{doi:\nolinkurl{10.4324/9781315784250-10}}


\bibitem[Laurenceau et~al\mbox{.}(1998)]%
        {laurenceau1998intimacy}
\bibfield{author}{\bibinfo{person}{Jean-Philippe Laurenceau}, \bibinfo{person}{Lisa~Feldman Barrett}, {and} \bibinfo{person}{Paula~R Pietromonaco}.} \bibinfo{year}{1998}\natexlab{}.
\newblock \showarticletitle{Intimacy as an interpersonal process: the importance of self-disclosure, partner disclosure, and perceived partner responsiveness in interpersonal exchanges}.
\newblock \bibinfo{journal}{\emph{Journal of Personality and Social Psychology}} \bibinfo{volume}{74}, \bibinfo{number}{5} (\bibinfo{year}{1998}), \bibinfo{pages}{1238}.
\newblock
\href{https://doi.org/10.1037//0022-3514.74.5.1238}{doi:\nolinkurl{10.1037//0022-3514.74.5.1238}}


\bibitem[Lee et~al\mbox{.}(2019)]%
        {lee2019caring}
\bibfield{author}{\bibinfo{person}{Minha Lee}, \bibinfo{person}{Sander Ackermans}, \bibinfo{person}{Nena van As}, \bibinfo{person}{Hanwen Chang}, \bibinfo{person}{Enzo Lucas}, {and} \bibinfo{person}{Wijnand IJsselsteijn}.} \bibinfo{year}{2019}\natexlab{}.
\newblock \showarticletitle{Caring for Vincent: A Chatbot for Self-Compassion}. In \bibinfo{booktitle}{\emph{Proceedings of the 2019 CHI Conference on Human Factors in Computing Systems}} (Glasgow, Scotland Uk) \emph{(\bibinfo{series}{CHI '19})}. \bibinfo{publisher}{Association for Computing Machinery}, \bibinfo{address}{New York, NY, USA}, \bibinfo{pages}{1–13}.
\newblock
\showISBNx{9781450359702}
\href{https://doi.org/10.1145/3290605.3300932}{doi:\nolinkurl{10.1145/3290605.3300932}}


\bibitem[Lee et~al\mbox{.}(2020)]%
        {lee2020ihearyou}
\bibfield{author}{\bibinfo{person}{Yi-Chieh Lee}, \bibinfo{person}{Naomi Yamashita}, \bibinfo{person}{Yun Huang}, {and} \bibinfo{person}{Wai Fu}.} \bibinfo{year}{2020}\natexlab{}.
\newblock \showarticletitle{"I Hear You, I Feel You": Encouraging Deep Self-disclosure through a Chatbot}. In \bibinfo{booktitle}{\emph{Proceedings of the 2020 CHI Conference on Human Factors in Computing Systems}} (Honolulu, HI, USA) \emph{(\bibinfo{series}{CHI '20})}. \bibinfo{publisher}{Association for Computing Machinery}, \bibinfo{address}{New York, NY, USA}, \bibinfo{pages}{1–12}.
\newblock
\showISBNx{9781450367080}
\href{https://doi.org/10.1145/3313831.3376175}{doi:\nolinkurl{10.1145/3313831.3376175}}


\bibitem[Li et~al\mbox{.}(2019)]%
        {li2019story}
\bibfield{author}{\bibinfo{person}{Cun Li}, \bibinfo{person}{Jun Hu}, \bibinfo{person}{Bart Hengeveld}, {and} \bibinfo{person}{Caroline Hummels}.} \bibinfo{year}{2019}\natexlab{}.
\newblock \showarticletitle{Story-Me: Design of a System to Support Intergenerational Storytelling and Preservation for Older Adults}. In \bibinfo{booktitle}{\emph{Companion Publication of the 2019 on Designing Interactive Systems Conference 2019 Companion}} (San Diego, CA, USA) \emph{(\bibinfo{series}{DIS '19 Companion})}. \bibinfo{publisher}{Association for Computing Machinery}, \bibinfo{address}{New York, NY, USA}, \bibinfo{pages}{245–250}.
\newblock
\showISBNx{9781450362702}
\href{https://doi.org/10.1145/3301019.3323902}{doi:\nolinkurl{10.1145/3301019.3323902}}


\bibitem[Li et~al\mbox{.}(2022)]%
        {li2022improving}
\bibfield{author}{\bibinfo{person}{Xiaoyan Li}, \bibinfo{person}{Naomi Yamashita}, \bibinfo{person}{Wen Duan}, \bibinfo{person}{Yoshinari Shirai}, {and} \bibinfo{person}{Susan~R. Fussell}.} \bibinfo{year}{2022}\natexlab{}.
\newblock \showarticletitle{Improving Non-Native Speakers' Participation with an Automatic Agent in Multilingual Groups}.
\newblock \bibinfo{journal}{\emph{Proc. ACM Hum.-Comput. Interact.}} \bibinfo{volume}{7}, \bibinfo{number}{GROUP}, Article \bibinfo{articleno}{12} (\bibinfo{date}{Dec.} \bibinfo{year}{2022}), \bibinfo{numpages}{28}~pages.
\newblock
\href{https://doi.org/10.1145/3567562}{doi:\nolinkurl{10.1145/3567562}}


\bibitem[Li et~al\mbox{.}(2023)]%
        {li2023exploring}
\bibfield{author}{\bibinfo{person}{Zisu Li}, \bibinfo{person}{Li Feng}, \bibinfo{person}{Chen Liang}, \bibinfo{person}{Yuru Huang}, {and} \bibinfo{person}{Mingming Fan}.} \bibinfo{year}{2023}\natexlab{}.
\newblock \showarticletitle{Exploring the Opportunities of AR for Enriching Storytelling with Family Photos between Grandparents and Grandchildren}.
\newblock \bibinfo{journal}{\emph{Proc. ACM Interact. Mob. Wearable Ubiquitous Technol.}} \bibinfo{volume}{7}, \bibinfo{number}{3}, Article \bibinfo{articleno}{108} (\bibinfo{date}{Sept.} \bibinfo{year}{2023}), \bibinfo{numpages}{26}~pages.
\newblock
\href{https://doi.org/10.1145/3610903}{doi:\nolinkurl{10.1145/3610903}}


\bibitem[Liaqat et~al\mbox{.}(2022)]%
        {liagat2022withahint}
\bibfield{author}{\bibinfo{person}{Amna Liaqat}, \bibinfo{person}{Benett Axtell}, {and} \bibinfo{person}{Cosmin Munteanu}.} \bibinfo{year}{2022}\natexlab{}.
\newblock \showarticletitle{"With a hint she will remember": Collaborative Storytelling and Culture Sharing between Immigrant Grandparents and Grandchildren Via Magic Thing Designs}.
\newblock \bibinfo{journal}{\emph{Proc. ACM Hum.-Comput. Interact.}} \bibinfo{volume}{6}, \bibinfo{number}{CSCW2}, Article \bibinfo{articleno}{268} (\bibinfo{date}{Nov.} \bibinfo{year}{2022}), \bibinfo{numpages}{37}~pages.
\newblock
\href{https://doi.org/10.1145/3555158}{doi:\nolinkurl{10.1145/3555158}}


\bibitem[Liu et~al\mbox{.}(2021)]%
        {liu2021significantotter}
\bibfield{author}{\bibinfo{person}{Fannie Liu}, \bibinfo{person}{Chunjong Park}, \bibinfo{person}{Yu~Jiang Tham}, \bibinfo{person}{Tsung-Yu Tsai}, \bibinfo{person}{Laura Dabbish}, \bibinfo{person}{Geoff Kaufman}, {and} \bibinfo{person}{Andr\'{e}s Monroy-Hern\'{a}ndez}.} \bibinfo{year}{2021}\natexlab{}.
\newblock \showarticletitle{Significant Otter: Understanding the Role of Biosignals in Communication}. In \bibinfo{booktitle}{\emph{Proceedings of the 2021 CHI Conference on Human Factors in Computing Systems}} (Yokohama, Japan) \emph{(\bibinfo{series}{CHI '21})}. \bibinfo{publisher}{Association for Computing Machinery}, \bibinfo{address}{New York, NY, USA}, Article \bibinfo{articleno}{334}, \bibinfo{numpages}{15}~pages.
\newblock
\showISBNx{9781450380966}
\href{https://doi.org/10.1145/3411764.3445200}{doi:\nolinkurl{10.1145/3411764.3445200}}


\bibitem[L{\"o}w et~al\mbox{.}(2021)]%
        {low2021socia}
\bibfield{author}{\bibinfo{person}{Christian L{\"o}w}, \bibinfo{person}{Lukas Moshuber}, {and} \bibinfo{person}{Albert Rafetseder}.} \bibinfo{year}{2021}\natexlab{}.
\newblock \showarticletitle{Gr{\"a}tzelbot: Social Companion Technology for Community Building among University Freshmen}. In \bibinfo{booktitle}{\emph{Chatbot Research and Design}}, Vol.~\bibinfo{volume}{12604}. \bibinfo{publisher}{Springer}, \bibinfo{address}{Berlin, Heidelberg}, \bibinfo{pages}{114--128}.
\newblock
\href{https://doi.org/10.1007/978-3-030-68288-0_8}{doi:\nolinkurl{10.1007/978-3-030-68288-0_8}}


\bibitem[McAdams(2001)]%
        {mcdams2001thepsychology}
\bibfield{author}{\bibinfo{person}{Dan~P. McAdams}.} \bibinfo{year}{2001}\natexlab{}.
\newblock \showarticletitle{The Psychology of Life Stories}.
\newblock \bibinfo{journal}{\emph{Review of General Psychology}} \bibinfo{volume}{5}, \bibinfo{number}{2} (\bibinfo{year}{2001}), \bibinfo{pages}{100--122}.
\newblock
\href{https://doi.org/10.1037/1089-2680.5.2.100}{doi:\nolinkurl{10.1037/1089-2680.5.2.100}}


\bibitem[McGookin(2019)]%
        {mcgookin2019reveal}
\bibfield{author}{\bibinfo{person}{David McGookin}.} \bibinfo{year}{2019}\natexlab{}.
\newblock \showarticletitle{Reveal: Investigating Proactive Location-Based Reminiscing with Personal Digital Photo Repositories}. In \bibinfo{booktitle}{\emph{Proceedings of the 2019 CHI Conference on Human Factors in Computing Systems}} (Glasgow, Scotland Uk) \emph{(\bibinfo{series}{CHI '19})}. \bibinfo{publisher}{Association for Computing Machinery}, \bibinfo{address}{New York, NY, USA}, \bibinfo{pages}{1–14}.
\newblock
\showISBNx{9781450359702}
\href{https://doi.org/10.1145/3290605.3300665}{doi:\nolinkurl{10.1145/3290605.3300665}}


\bibitem[McHugh(2012)]%
        {mchugh2012interrater}
\bibfield{author}{\bibinfo{person}{Mary~L McHugh}.} \bibinfo{year}{2012}\natexlab{}.
\newblock \showarticletitle{Interrater reliability: the kappa statistic}.
\newblock \bibinfo{journal}{\emph{Biochemia medica}} \bibinfo{volume}{22}, \bibinfo{number}{3} (\bibinfo{year}{2012}), \bibinfo{pages}{276--282}.
\newblock


\bibitem[McLean and Lilgendahl(2008)]%
        {mclean2008whyrecall}
\bibfield{author}{\bibinfo{person}{Kate~C McLean} {and} \bibinfo{person}{Jennifer~Pals Lilgendahl}.} \bibinfo{year}{2008}\natexlab{}.
\newblock \showarticletitle{Why recall our highs and lows: Relations between memory functions, age, and well-being}.
\newblock \bibinfo{journal}{\emph{Memory}} \bibinfo{volume}{16}, \bibinfo{number}{7} (\bibinfo{year}{2008}), \bibinfo{pages}{751--762}.
\newblock
\href{https://doi.org/10.1080/09658210802215385}{doi:\nolinkurl{10.1080/09658210802215385}}


\bibitem[MF et~al\mbox{.}(2023)]%
        {mf2023remembering}
\bibfield{author}{\bibinfo{person}{Pilon MF}, \bibinfo{person}{Hallford DJ}, {and} \bibinfo{person}{Karantzas GC}.} \bibinfo{year}{2023}\natexlab{}.
\newblock \showarticletitle{Remembering for relationships: brief cognitive-reminiscence therapy improves young adults’ perceptions about self and others in social relationships}.
\newblock \bibinfo{journal}{\emph{Journal of Social and Personal Relationships}} \bibinfo{volume}{40}, \bibinfo{number}{3} (\bibinfo{year}{2023}), \bibinfo{pages}{955--975}.
\newblock
\href{https://doi.org/10.1177/02654075221122943}{doi:\nolinkurl{10.1177/02654075221122943}}


\bibitem[Monisha~Pasupathi and Coombs(2002)]%
        {pasupathi2002conversational}
\bibfield{author}{\bibinfo{person}{Sarah~Lucas Monisha~Pasupathi} {and} \bibinfo{person}{Amy Coombs}.} \bibinfo{year}{2002}\natexlab{}.
\newblock \showarticletitle{Conversational Functions of Autobiographical Remembering: Long-Married Couples Talk About Conflicts and Pleasant Topics}.
\newblock \bibinfo{journal}{\emph{Discourse Processes}} \bibinfo{volume}{34}, \bibinfo{number}{2} (\bibinfo{year}{2002}), \bibinfo{pages}{163--192}.
\newblock
\href{https://doi.org/10.1207/S15326950DP3402_3}{doi:\nolinkurl{10.1207/S15326950DP3402_3}}


\bibitem[Namvarpour and Razi(2024)]%
        {namvarpour2024uncovering}
\bibfield{author}{\bibinfo{person}{Mohammad Namvarpour} {and} \bibinfo{person}{Afsaneh Razi}.} \bibinfo{year}{2024}\natexlab{}.
\newblock \showarticletitle{Uncovering Contradictions in Human-AI Interactions: Lessons Learned from User Reviews of Replika}. In \bibinfo{booktitle}{\emph{Companion Publication of the 2024 Conference on Computer-Supported Cooperative Work and Social Computing}} (San Jose, Costa Rica) \emph{(\bibinfo{series}{CSCW Companion '24})}. \bibinfo{publisher}{Association for Computing Machinery}, \bibinfo{address}{New York, NY, USA}, \bibinfo{pages}{579–586}.
\newblock
\showISBNx{9798400711145}
\href{https://doi.org/10.1145/3678884.3681909}{doi:\nolinkurl{10.1145/3678884.3681909}}


\bibitem[Nass et~al\mbox{.}(1994)]%
        {nass1994computers}
\bibfield{author}{\bibinfo{person}{Clifford Nass}, \bibinfo{person}{Jonathan Steuer}, {and} \bibinfo{person}{Ellen~R. Tauber}.} \bibinfo{year}{1994}\natexlab{}.
\newblock \showarticletitle{Computers are social actors}. In \bibinfo{booktitle}{\emph{Proceedings of the SIGCHI Conference on Human Factors in Computing Systems}} (Boston, Massachusetts, USA) \emph{(\bibinfo{series}{CHI '94})}. \bibinfo{publisher}{Association for Computing Machinery}, \bibinfo{address}{New York, NY, USA}, \bibinfo{pages}{72–78}.
\newblock
\showISBNx{0897916506}
\href{https://doi.org/10.1145/191666.191703}{doi:\nolinkurl{10.1145/191666.191703}}


\bibitem[Neumann et~al\mbox{.}(2017)]%
        {neumann2017memory}
\bibfield{author}{\bibinfo{person}{Stephanie Neumann}, \bibinfo{person}{Richard Banks}, {and} \bibinfo{person}{Marian D\"{o}rk}.} \bibinfo{year}{2017}\natexlab{}.
\newblock \showarticletitle{Memory Dialogue: Exploring Artefact-Based Memory Sharing}. In \bibinfo{booktitle}{\emph{Proceedings of the 2017 CHI Conference Extended Abstracts on Human Factors in Computing Systems}} (Denver, Colorado, USA) \emph{(\bibinfo{series}{CHI EA '17})}. \bibinfo{publisher}{Association for Computing Machinery}, \bibinfo{address}{New York, NY, USA}, \bibinfo{pages}{884–895}.
\newblock
\showISBNx{9781450346566}
\href{https://doi.org/10.1145/3027063.3052756}{doi:\nolinkurl{10.1145/3027063.3052756}}


\bibitem[Odom et~al\mbox{.}(2020)]%
        {odom2020exploring}
\bibfield{author}{\bibinfo{person}{William Odom}, \bibinfo{person}{MinYoung Yoo}, \bibinfo{person}{Henry Lin}, \bibinfo{person}{Tijs Duel}, \bibinfo{person}{Tal Amram}, {and} \bibinfo{person}{Amy Yo~Sue Chen}.} \bibinfo{year}{2020}\natexlab{}.
\newblock \showarticletitle{Exploring the Reflective Potentialities of Personal Data with Different Temporal Modalities: A Field Study of Olo Radio}. In \bibinfo{booktitle}{\emph{Proceedings of the 2020 ACM Designing Interactive Systems Conference}} (Eindhoven, Netherlands) \emph{(\bibinfo{series}{DIS '20})}. \bibinfo{publisher}{Association for Computing Machinery}, \bibinfo{address}{New York, NY, USA}, \bibinfo{pages}{283–295}.
\newblock
\showISBNx{9781450369749}
\href{https://doi.org/10.1145/3357236.3395438}{doi:\nolinkurl{10.1145/3357236.3395438}}


\bibitem[O’Rourke et~al\mbox{.}(2017)]%
        {o2017reminiscence}
\bibfield{author}{\bibinfo{person}{Norm O’Rourke}, \bibinfo{person}{David~B King}, {and} \bibinfo{person}{Philippe Cappeliez}.} \bibinfo{year}{2017}\natexlab{}.
\newblock \showarticletitle{Reminiscence functions over time: consistency of self functions and variation of prosocial functions}.
\newblock \bibinfo{journal}{\emph{Memory}} \bibinfo{volume}{25}, \bibinfo{number}{3} (\bibinfo{year}{2017}), \bibinfo{pages}{403--411}.
\newblock
\href{https://doi.org/10.1080/09658211.2016.1179331}{doi:\nolinkurl{10.1080/09658211.2016.1179331}}


\bibitem[P et~al\mbox{.}(1997)]%
        {mcadams1997stories}
\bibfield{author}{\bibinfo{person}{McAdams~Dan P}, \bibinfo{person}{Diamond Ann}, \bibinfo{person}{de St. Aubin~Ed}, {and} \bibinfo{person}{Mansfield Elizabeth}.} \bibinfo{year}{1997}\natexlab{}.
\newblock \showarticletitle{Stories of commitment: The psychosocial construction of generative lives}.
\newblock \bibinfo{journal}{\emph{Journal of Personality and Social Psychology}} \bibinfo{volume}{72}, \bibinfo{number}{3} (\bibinfo{year}{1997}), \bibinfo{pages}{678--694}.
\newblock
\href{https://doi.org/10.1037/0022-3514.72.3.678}{doi:\nolinkurl{10.1037/0022-3514.72.3.678}}


\bibitem[Papacharissi(2016)]%
        {papacharissi2016affective}
\bibfield{author}{\bibinfo{person}{Zizi Papacharissi}.} \bibinfo{year}{2016}\natexlab{}.
\newblock \showarticletitle{Affective publics and structures of storytelling: sentiment, events and mediality}.
\newblock \bibinfo{journal}{\emph{Information, Communication \& Society}} \bibinfo{volume}{19}, \bibinfo{number}{3} (\bibinfo{year}{2016}), \bibinfo{pages}{307--324}.
\newblock
\href{https://doi.org/10.1080/1369118X.2015.1109697}{doi:\nolinkurl{10.1080/1369118X.2015.1109697}}


\bibitem[Park et~al\mbox{.}(2021)]%
        {park2021wrote}
\bibfield{author}{\bibinfo{person}{SoHyun Park}, \bibinfo{person}{Anja Thieme}, \bibinfo{person}{Jeongyun Han}, \bibinfo{person}{Sungwoo Lee}, \bibinfo{person}{Wonjong Rhee}, {and} \bibinfo{person}{Bongwon Suh}.} \bibinfo{year}{2021}\natexlab{}.
\newblock \showarticletitle{“I wrote as if I were telling a story to someone I knew.”: Designing Chatbot Interactions for Expressive Writing in Mental Health}. In \bibinfo{booktitle}{\emph{Proceedings of the 2021 ACM Designing Interactive Systems Conference}} (Virtual Event, USA) \emph{(\bibinfo{series}{DIS '21})}. \bibinfo{publisher}{Association for Computing Machinery}, \bibinfo{address}{New York, NY, USA}, \bibinfo{pages}{926–941}.
\newblock
\showISBNx{9781450384766}
\href{https://doi.org/10.1145/3461778.3462143}{doi:\nolinkurl{10.1145/3461778.3462143}}


\bibitem[Pasupathi and Carstensen(2003)]%
        {pasupathi2003age}
\bibfield{author}{\bibinfo{person}{Monisha Pasupathi} {and} \bibinfo{person}{Laura~L Carstensen}.} \bibinfo{year}{2003}\natexlab{}.
\newblock \showarticletitle{Age and emotional experience during mutual reminiscing}.
\newblock \bibinfo{journal}{\emph{Psychology and Aging}} \bibinfo{volume}{18}, \bibinfo{number}{3} (\bibinfo{year}{2003}), \bibinfo{pages}{430--442}.
\newblock
\href{https://doi.org/10.1037/0882-7974.18.3.430}{doi:\nolinkurl{10.1037/0882-7974.18.3.430}}


\bibitem[Pennebaker and Seagal(1999)]%
        {pennebaker1999forming}
\bibfield{author}{\bibinfo{person}{James~W. Pennebaker} {and} \bibinfo{person}{Janel~D. Seagal}.} \bibinfo{year}{1999}\natexlab{}.
\newblock \showarticletitle{Forming a story: The health benefits of narrative}.
\newblock \bibinfo{journal}{\emph{Journal of Clinical Psychology}} \bibinfo{volume}{55}, \bibinfo{number}{10} (\bibinfo{year}{1999}), \bibinfo{pages}{1243--1254}.
\newblock
\href{https://doi.org/10.1002/(SICI)1097-4679(199910)55:10<1243::AID-JCLP6>3.0.CO;2-N}{doi:\nolinkurl{10.1002/(SICI)1097-4679(199910)55:10<1243::AID-JCLP6>3.0.CO;2-N}}


\bibitem[Pentina et~al\mbox{.}(2023)]%
        {pentina2023exploring}
\bibfield{author}{\bibinfo{person}{Iryna Pentina}, \bibinfo{person}{Tyler Hancock}, {and} \bibinfo{person}{Tianling Xie}.} \bibinfo{year}{2023}\natexlab{}.
\newblock \showarticletitle{Exploring relationship development with social chatbots: A mixed-method study of replika}.
\newblock \bibinfo{journal}{\emph{Computers in Human Behavior}}  \bibinfo{volume}{140} (\bibinfo{year}{2023}), \bibinfo{pages}{107600}.
\newblock
\href{https://doi.org/10.1016/j.chb.2022.107600}{doi:\nolinkurl{10.1016/j.chb.2022.107600}}


\bibitem[Perez et~al\mbox{.}(2021)]%
        {perez2021truefewshot}
\bibfield{author}{\bibinfo{person}{Ethan Perez}, \bibinfo{person}{Douwe Kiela}, {and} \bibinfo{person}{Kyunghyun Cho}.} \bibinfo{year}{2021}\natexlab{}.
\newblock \bibinfo{title}{True Few-Shot Learning with Language Models}.
\newblock
\showeprint[arxiv]{2105.11447}~[cs.CL]
\urldef\tempurl%
\url{https://arxiv.org/abs/2105.11447}
\showURL{%
\tempurl}


\bibitem[Petrelli et~al\mbox{.}(2010)]%
        {petrelli2010fmradio}
\bibfield{author}{\bibinfo{person}{Daniela Petrelli}, \bibinfo{person}{Nicolas Villar}, \bibinfo{person}{Vaiva Kalnikaite}, \bibinfo{person}{Lina Dib}, {and} \bibinfo{person}{Steve Whittaker}.} \bibinfo{year}{2010}\natexlab{}.
\newblock \showarticletitle{FM radio: family interplay with sonic mementos}. In \bibinfo{booktitle}{\emph{Proceedings of the SIGCHI Conference on Human Factors in Computing Systems}} (Atlanta, Georgia, USA) \emph{(\bibinfo{series}{CHI '10})}. \bibinfo{publisher}{Association for Computing Machinery}, \bibinfo{address}{New York, NY, USA}, \bibinfo{pages}{2371–2380}.
\newblock
\showISBNx{9781605589299}
\href{https://doi.org/10.1145/1753326.1753683}{doi:\nolinkurl{10.1145/1753326.1753683}}


\bibitem[Petronio(2013)]%
        {petronio20113breif}
\bibfield{author}{\bibinfo{person}{Sandra Petronio}.} \bibinfo{year}{2013}\natexlab{}.
\newblock \showarticletitle{Brief Status Report on Communication Privacy Management Theory}.
\newblock \bibinfo{journal}{\emph{Journal of Family Communication}} \bibinfo{volume}{13}, \bibinfo{number}{1} (\bibinfo{year}{2013}), \bibinfo{pages}{6--14}.
\newblock
\href{https://doi.org/10.1080/15267431.2013.743426}{doi:\nolinkurl{10.1080/15267431.2013.743426}}


\bibitem[Placani(2024)]%
        {placani2024anthropomorphism}
\bibfield{author}{\bibinfo{person}{Adriana Placani}.} \bibinfo{year}{2024}\natexlab{}.
\newblock \showarticletitle{Anthropomorphism in AI: hype and fallacy}.
\newblock \bibinfo{journal}{\emph{AI and Ethics}} \bibinfo{volume}{4}, \bibinfo{number}{3} (\bibinfo{year}{2024}), \bibinfo{pages}{691--698}.
\newblock
\href{https://doi.org/10.1007/s43681-024-00419-4}{doi:\nolinkurl{10.1007/s43681-024-00419-4}}


\bibitem[Pohl et~al\mbox{.}(2014)]%
        {pohl2014autobiographical}
\bibfield{author}{\bibinfo{person}{Rüdiger~F. Pohl}, \bibinfo{person}{Michael Bender}, {and} \bibinfo{person}{Gregor Lachmann}.} \bibinfo{year}{2014}\natexlab{}.
\newblock \showarticletitle{Autobiographical memory and social skills of men and women}.
\newblock \bibinfo{journal}{\emph{Applied Cognitive Psychology}} \bibinfo{volume}{19}, \bibinfo{number}{6} (\bibinfo{year}{2014}), \bibinfo{pages}{745--759}.
\newblock
\href{https://doi.org/10.1002/acp.1104}{doi:\nolinkurl{10.1002/acp.1104}}


\bibitem[Páez et~al\mbox{.}(2015)]%
        {paez2015psychosocial}
\bibfield{author}{\bibinfo{person}{Dario Páez}, \bibinfo{person}{Bernard Rimé}, \bibinfo{person}{Nekane Basabe}, \bibinfo{person}{Anna Wlodarczyk}, {and} \bibinfo{person}{Larraitz Zumeta}.} \bibinfo{year}{2015}\natexlab{}.
\newblock \showarticletitle{Psychosocial effects of perceived emotional synchrony in collective gatherings}.
\newblock \bibinfo{journal}{\emph{Journal of Personality and Social Psychology}} \bibinfo{volume}{108}, \bibinfo{number}{5} (\bibinfo{year}{2015}), \bibinfo{pages}{711--729}.
\newblock
\href{https://doi.org/10.1037/pspi0000014}{doi:\nolinkurl{10.1037/pspi0000014}}


\bibitem[R. and H(2000)]%
        {karney2000memorybias}
\bibfield{author}{\bibinfo{person}{Karney~B. R.} {and} \bibinfo{person}{Coombs~R. H}.} \bibinfo{year}{2000}\natexlab{}.
\newblock \showarticletitle{Memory bias in long-term close relationships: Consistency or improvement?}
\newblock \bibinfo{journal}{\emph{Personality and Social Psychology Bulletin}} \bibinfo{volume}{26}, \bibinfo{number}{8} (\bibinfo{year}{2000}), \bibinfo{pages}{959–--970}.
\newblock
\href{https://doi.org/10.1177/01461672002610006}{doi:\nolinkurl{10.1177/01461672002610006}}


\bibitem[Ramadan(2021)]%
        {ramadan2021alexafying}
\bibfield{author}{\bibinfo{person}{Zahy~B. Ramadan}.} \bibinfo{year}{2021}\natexlab{}.
\newblock \showarticletitle{“Alexafying” shoppers: The examination of Amazon's captive relationship strategy}.
\newblock \bibinfo{journal}{\emph{Journal of Retailing and Consumer Services}}  \bibinfo{volume}{62} (\bibinfo{year}{2021}), \bibinfo{pages}{102610}.
\newblock
\href{https://doi.org/10.1016/j.jretconser.2021.102610}{doi:\nolinkurl{10.1016/j.jretconser.2021.102610}}


\bibitem[Rapp et~al\mbox{.}(2021)]%
        {rapp2021thehumanside}
\bibfield{author}{\bibinfo{person}{Amon Rapp}, \bibinfo{person}{Lorenzo Curti}, {and} \bibinfo{person}{Arianna Boldi}.} \bibinfo{year}{2021}\natexlab{}.
\newblock \showarticletitle{The human side of human-chatbot interaction: A systematic literature review of ten years of research on text-based chatbots}.
\newblock \bibinfo{journal}{\emph{International Journal of Human-Computer Studies}}  \bibinfo{volume}{151} (\bibinfo{year}{2021}), \bibinfo{pages}{102630}.
\newblock
\href{https://doi.org/10.1016/j.ijhcs.2021.102630}{doi:\nolinkurl{10.1016/j.ijhcs.2021.102630}}


\bibitem[Rasmussen and Berntsen(2009)]%
        {rasmussen2009emotional}
\bibfield{author}{\bibinfo{person}{Anne~S Rasmussen} {and} \bibinfo{person}{Dorthe Berntsen}.} \bibinfo{year}{2009}\natexlab{}.
\newblock \showarticletitle{Emotional valence and the functions of autobiographical memories: Positive and negative memories serve different functions}.
\newblock \bibinfo{journal}{\emph{Memory \& Cognition}} \bibinfo{volume}{37}, \bibinfo{number}{4} (\bibinfo{year}{2009}), \bibinfo{pages}{477--492}.
\newblock
\href{https://doi.org/10.3758/MC.37.4.477}{doi:\nolinkurl{10.3758/MC.37.4.477}}


\bibitem[Reis and Shaver(1988)]%
        {reis1988intimacy}
\bibfield{author}{\bibinfo{person}{H.~T. Reis} {and} \bibinfo{person}{P. Shaver}.} \bibinfo{year}{1988}\natexlab{}.
\newblock \bibinfo{title}{Intimacy as an interpersonal process}.
\newblock \bibinfo{numpages}{367--389}~pages.
\newblock


\bibitem[Reynolds and McDonell(2021)]%
        {reynolds2021prompt}
\bibfield{author}{\bibinfo{person}{Laria Reynolds} {and} \bibinfo{person}{Kyle McDonell}.} \bibinfo{year}{2021}\natexlab{}.
\newblock \showarticletitle{Prompt Programming for Large Language Models: Beyond the Few-Shot Paradigm}. In \bibinfo{booktitle}{\emph{Extended Abstracts of the 2021 CHI Conference on Human Factors in Computing Systems}} (Yokohama, Japan) \emph{(\bibinfo{series}{CHI EA '21})}. \bibinfo{publisher}{Association for Computing Machinery}, \bibinfo{address}{New York, NY, USA}, Article \bibinfo{articleno}{314}, \bibinfo{numpages}{7}~pages.
\newblock
\showISBNx{9781450380959}
\href{https://doi.org/10.1145/3411763.3451760}{doi:\nolinkurl{10.1145/3411763.3451760}}


\bibitem[Rheu et~al\mbox{.}(2024)]%
        {minjin2024whena}
\bibfield{author}{\bibinfo{person}{Minjin~(MJ) Rheu}, \bibinfo{person}{Yue~(Nancy) Dai}, \bibinfo{person}{Jingbo Meng}, {and} \bibinfo{person}{Wei Peng}.} \bibinfo{year}{2024}\natexlab{}.
\newblock \showarticletitle{When a Chatbot Disappoints You: Expectancy Violation in Human-Chatbot Interaction in a Social Support Context}.
\newblock \bibinfo{journal}{\emph{Communication Research}} \bibinfo{volume}{51}, \bibinfo{number}{7} (\bibinfo{year}{2024}), \bibinfo{pages}{782--814}.
\newblock
\href{https://doi.org/10.1177/00936502231221669}{doi:\nolinkurl{10.1177/00936502231221669}}


\bibitem[Routledge et~al\mbox{.}(2013)]%
        {routledge2013nostalgia}
\bibfield{author}{\bibinfo{person}{Clay Routledge}, \bibinfo{person}{Tim Wildschut}, \bibinfo{person}{Constantine Sedikides}, {and} \bibinfo{person}{Jacob Juhl}.} \bibinfo{year}{2013}\natexlab{}.
\newblock \showarticletitle{Nostalgia as a Resource for Psychological Health and Well-Being}.
\newblock \bibinfo{journal}{\emph{Social and Personality Psychology Compass}} \bibinfo{volume}{7}, \bibinfo{number}{11} (\bibinfo{year}{2013}), \bibinfo{pages}{808--818}.
\newblock
\href{https://doi.org/10.1111/spc3.12070}{doi:\nolinkurl{10.1111/spc3.12070}}


\bibitem[Salehzadeh~Niksirat et~al\mbox{.}(2023)]%
        {niksirat2023onthepotential}
\bibfield{author}{\bibinfo{person}{Kavous Salehzadeh~Niksirat}, \bibinfo{person}{Diana Korka}, \bibinfo{person}{Hamza Harkous}, \bibinfo{person}{K\'{e}vin Huguenin}, {and} \bibinfo{person}{Mauro Cherubini}.} \bibinfo{year}{2023}\natexlab{}.
\newblock \showarticletitle{On the Potential of Mediation Chatbots for Mitigating Multiparty Privacy Conflicts - A Wizard-of-Oz Study}.
\newblock \bibinfo{journal}{\emph{Proc. ACM Hum.-Comput. Interact.}} \bibinfo{volume}{7}, \bibinfo{number}{CSCW1}, Article \bibinfo{articleno}{142} (\bibinfo{date}{April} \bibinfo{year}{2023}), \bibinfo{numpages}{33}~pages.
\newblock
\href{https://doi.org/10.1145/3579618}{doi:\nolinkurl{10.1145/3579618}}


\bibitem[Salovey and Mayer(1990)]%
        {peter1990emotional}
\bibfield{author}{\bibinfo{person}{Peter Salovey} {and} \bibinfo{person}{John~D. Mayer}.} \bibinfo{year}{1990}\natexlab{}.
\newblock \showarticletitle{Emotional Intelligence}.
\newblock \bibinfo{journal}{\emph{Imagination, Cognition and Personality}} \bibinfo{volume}{9}, \bibinfo{number}{3} (\bibinfo{year}{1990}), \bibinfo{pages}{185--211}.
\newblock
\href{https://doi.org/10.2190/DUGG-P24E-52WK-6CDG}{doi:\nolinkurl{10.2190/DUGG-P24E-52WK-6CDG}}


\bibitem[Schlagowski et~al\mbox{.}(2023)]%
        {schlagowski2023wish}
\bibfield{author}{\bibinfo{person}{Ruben Schlagowski}, \bibinfo{person}{Dariia Nazarenko}, \bibinfo{person}{Yekta Can}, \bibinfo{person}{Kunal Gupta}, \bibinfo{person}{Silvan Mertes}, \bibinfo{person}{Mark Billinghurst}, {and} \bibinfo{person}{Elisabeth Andr\'{e}}.} \bibinfo{year}{2023}\natexlab{}.
\newblock \showarticletitle{Wish You Were Here: Mental and Physiological Effects of Remote Music Collaboration in Mixed Reality}. In \bibinfo{booktitle}{\emph{Proceedings of the 2023 CHI Conference on Human Factors in Computing Systems}} (Hamburg, Germany) \emph{(\bibinfo{series}{CHI '23})}. \bibinfo{publisher}{Association for Computing Machinery}, \bibinfo{address}{New York, NY, USA}, Article \bibinfo{articleno}{102}, \bibinfo{numpages}{16}~pages.
\newblock
\showISBNx{9781450394215}
\href{https://doi.org/10.1145/3544548.3581162}{doi:\nolinkurl{10.1145/3544548.3581162}}


\bibitem[Seemann(2019)]%
        {seemann2019reminiscing}
\bibfield{author}{\bibinfo{person}{Axel Seemann}.} \bibinfo{year}{2019}\natexlab{}.
\newblock \showarticletitle{Reminiscing together: Joint experiences, epistemic groups, and sense of self}.
\newblock \bibinfo{journal}{\emph{Synthese}} \bibinfo{volume}{196}, \bibinfo{number}{12} (\bibinfo{year}{2019}), \bibinfo{pages}{4813--4828}.
\newblock
\href{https://doi.org/10.1007/s11229-016-1156-3}{doi:\nolinkurl{10.1007/s11229-016-1156-3}}


\bibitem[Seering et~al\mbox{.}(2019)]%
        {seering2019beyond}
\bibfield{author}{\bibinfo{person}{Joseph Seering}, \bibinfo{person}{Michal Luria}, \bibinfo{person}{Geoff Kaufman}, {and} \bibinfo{person}{Jessica Hammer}.} \bibinfo{year}{2019}\natexlab{}.
\newblock \showarticletitle{Beyond Dyadic Interactions: Considering Chatbots as Community Members}. In \bibinfo{booktitle}{\emph{Proceedings of the 2019 CHI Conference on Human Factors in Computing Systems}} (Glasgow, Scotland Uk) \emph{(\bibinfo{series}{CHI '19})}. \bibinfo{publisher}{Association for Computing Machinery}, \bibinfo{address}{New York, NY, USA}, \bibinfo{pages}{1–13}.
\newblock
\showISBNx{9781450359702}
\href{https://doi.org/10.1145/3290605.3300680}{doi:\nolinkurl{10.1145/3290605.3300680}}


\bibitem[Seo et~al\mbox{.}(2024)]%
        {seo2024chacha}
\bibfield{author}{\bibinfo{person}{Woosuk Seo}, \bibinfo{person}{Chanmo Yang}, {and} \bibinfo{person}{Young-Ho Kim}.} \bibinfo{year}{2024}\natexlab{}.
\newblock \showarticletitle{ChaCha: Leveraging Large Language Models to Prompt Children to Share Their Emotions about Personal Events}. In \bibinfo{booktitle}{\emph{Proceedings of the 2024 CHI Conference on Human Factors in Computing Systems}} (Honolulu, HI, USA) \emph{(\bibinfo{series}{CHI '24})}. \bibinfo{publisher}{Association for Computing Machinery}, \bibinfo{address}{New York, NY, USA}, Article \bibinfo{articleno}{903}, \bibinfo{numpages}{20}~pages.
\newblock
\showISBNx{9798400703300}
\href{https://doi.org/10.1145/3613904.3642152}{doi:\nolinkurl{10.1145/3613904.3642152}}


\bibitem[Shen et~al\mbox{.}(2024)]%
        {shen2024legacysphere}
\bibfield{author}{\bibinfo{person}{Chenxinran Shen}, \bibinfo{person}{Joanna Mcgrenere}, {and} \bibinfo{person}{Dongwook Yoon}.} \bibinfo{year}{2024}\natexlab{}.
\newblock \showarticletitle{LegacySphere: Facilitating Intergenerational Communication Through Perspective-Taking and Storytelling in Embodied VR}. In \bibinfo{booktitle}{\emph{Proceedings of the 2024 CHI Conference on Human Factors in Computing Systems}} (Honolulu, HI, USA) \emph{(\bibinfo{series}{CHI '24})}. \bibinfo{publisher}{Association for Computing Machinery}, \bibinfo{address}{New York, NY, USA}, Article \bibinfo{articleno}{119}, \bibinfo{numpages}{16}~pages.
\newblock
\showISBNx{9798400703300}
\href{https://doi.org/10.1145/3613904.3641923}{doi:\nolinkurl{10.1145/3613904.3641923}}


\bibitem[Shin et~al\mbox{.}(2023)]%
        {shin2023introbot}
\bibfield{author}{\bibinfo{person}{Donghoon Shin}, \bibinfo{person}{Soomin Kim}, \bibinfo{person}{Ruoxi Shang}, \bibinfo{person}{Joonhwan Lee}, {and} \bibinfo{person}{Gary Hsieh}.} \bibinfo{year}{2023}\natexlab{}.
\newblock \showarticletitle{IntroBot: Exploring the Use of Chatbot-assisted Familiarization in Online Collaborative Groups}. In \bibinfo{booktitle}{\emph{Proceedings of the 2023 CHI Conference on Human Factors in Computing Systems}} (Hamburg, Germany) \emph{(\bibinfo{series}{CHI '23})}. \bibinfo{publisher}{Association for Computing Machinery}, \bibinfo{address}{New York, NY, USA}, Article \bibinfo{articleno}{613}, \bibinfo{numpages}{13}~pages.
\newblock
\showISBNx{9781450394215}
\href{https://doi.org/10.1145/3544548.3580930}{doi:\nolinkurl{10.1145/3544548.3580930}}


\bibitem[Skjuve et~al\mbox{.}(2021)]%
        {skjuve2021mychatbot}
\bibfield{author}{\bibinfo{person}{Marita Skjuve}, \bibinfo{person}{Asbjørn Følstad}, \bibinfo{person}{Knut~Inge Fostervold}, {and} \bibinfo{person}{Petter~Bae Brandtzaeg}.} \bibinfo{year}{2021}\natexlab{}.
\newblock \showarticletitle{My Chatbot Companion - a Study of Human-Chatbot Relationships}.
\newblock \bibinfo{journal}{\emph{International Journal of Human-Computer Studies}}  \bibinfo{volume}{149} (\bibinfo{year}{2021}), \bibinfo{pages}{102601}.
\newblock
\href{https://doi.org/10.1016/j.ijhcs.2021.102601}{doi:\nolinkurl{10.1016/j.ijhcs.2021.102601}}


\bibitem[Smorti(2011)]%
        {smorti2011autobiographical}
\bibfield{author}{\bibinfo{person}{Andrea Smorti}.} \bibinfo{year}{2011}\natexlab{}.
\newblock \showarticletitle{Autobiographical memory and autobiographical narrative: What is the relationship?}
\newblock \bibinfo{journal}{\emph{Narrative Inquiry}} \bibinfo{volume}{21}, \bibinfo{number}{2} (\bibinfo{year}{2011}), \bibinfo{pages}{303--310}.
\newblock
\href{https://doi.org/10.1075/ni.21.2.08smo}{doi:\nolinkurl{10.1075/ni.21.2.08smo}}


\bibitem[Sprecher et~al\mbox{.}(2013)]%
        {susan2013takingturns}
\bibfield{author}{\bibinfo{person}{Susan Sprecher}, \bibinfo{person}{Stanislav Treger}, \bibinfo{person}{Joshua~D. Wondra}, \bibinfo{person}{Nicole Hilaire}, {and} \bibinfo{person}{Kevin Wallpe}.} \bibinfo{year}{2013}\natexlab{}.
\newblock \showarticletitle{Taking turns: Reciprocal self-disclosure promotes liking in initial interactions}.
\newblock \bibinfo{journal}{\emph{Journal of Experimental Social Psychology}} \bibinfo{volume}{49}, \bibinfo{number}{5} (\bibinfo{year}{2013}), \bibinfo{pages}{860--866}.
\newblock
\href{https://doi.org/10.1016/j.jesp.2013.03.017}{doi:\nolinkurl{10.1016/j.jesp.2013.03.017}}


\bibitem[Stepanova et~al\mbox{.}(2022)]%
        {stepanova2022strategies}
\bibfield{author}{\bibinfo{person}{Ekaterina~R. Stepanova}, \bibinfo{person}{John Desnoyers-Stewart}, \bibinfo{person}{Kristina H\"{o}\"{o}k}, {and} \bibinfo{person}{Bernhard~E. Riecke}.} \bibinfo{year}{2022}\natexlab{}.
\newblock \showarticletitle{Strategies for Fostering a Genuine Feeling of Connection in Technologically Mediated Systems}. In \bibinfo{booktitle}{\emph{Proceedings of the 2022 CHI Conference on Human Factors in Computing Systems}} (New Orleans, LA, USA) \emph{(\bibinfo{series}{CHI '22})}. \bibinfo{publisher}{Association for Computing Machinery}, \bibinfo{address}{New York, NY, USA}, Article \bibinfo{articleno}{139}, \bibinfo{numpages}{26}~pages.
\newblock
\showISBNx{9781450391573}
\href{https://doi.org/10.1145/3491102.3517580}{doi:\nolinkurl{10.1145/3491102.3517580}}


\bibitem[Stuart and Kneer(2021)]%
        {stuart2021guilty}
\bibfield{author}{\bibinfo{person}{Michael~T. Stuart} {and} \bibinfo{person}{Markus Kneer}.} \bibinfo{year}{2021}\natexlab{}.
\newblock \showarticletitle{Guilty Artificial Minds: Folk Attributions of Mens Rea and Culpability to Artificially Intelligent Agents}.
\newblock \bibinfo{journal}{\emph{Proc. ACM Hum.-Comput. Interact.}} \bibinfo{volume}{5}, \bibinfo{number}{CSCW2}, Article \bibinfo{articleno}{363} (\bibinfo{date}{Oct.} \bibinfo{year}{2021}), \bibinfo{numpages}{27}~pages.
\newblock
\href{https://doi.org/10.1145/3479507}{doi:\nolinkurl{10.1145/3479507}}


\bibitem[T. and Zimprich(2015)]%
        {wolf2015differences}
\bibfield{author}{\bibinfo{person}{Wolf T.} {and} \bibinfo{person}{D. Zimprich}.} \bibinfo{year}{2015}\natexlab{}.
\newblock \showarticletitle{Differences in the use of autobiographical memory across the adult lifespan}.
\newblock \bibinfo{journal}{\emph{Memory}} \bibinfo{volume}{23}, \bibinfo{number}{8} (\bibinfo{year}{2015}), \bibinfo{pages}{1238--–1254}.
\newblock
\href{https://doi.org/10.1080/09658211.2014.971815}{doi:\nolinkurl{10.1080/09658211.2014.971815}}


\bibitem[Tausczik and Pennebaker(2010)]%
        {yla2010thepsychological}
\bibfield{author}{\bibinfo{person}{Yla~R. Tausczik} {and} \bibinfo{person}{James~W. Pennebaker}.} \bibinfo{year}{2010}\natexlab{}.
\newblock \showarticletitle{The Psychological Meaning of Words: LIWC and Computerized Text Analysis Methods}.
\newblock \bibinfo{journal}{\emph{Journal of Language and Social Psychology}} \bibinfo{volume}{29}, \bibinfo{number}{1} (\bibinfo{year}{2010}), \bibinfo{pages}{24--54}.
\newblock
\href{https://doi.org/10.1177/0261927X09351676}{doi:\nolinkurl{10.1177/0261927X09351676}}


\bibitem[Thomas and Briggs(2016)]%
        {thomas2016reminiscence}
\bibfield{author}{\bibinfo{person}{Lisa Thomas} {and} \bibinfo{person}{Pam Briggs}.} \bibinfo{year}{2016}\natexlab{}.
\newblock \showarticletitle{Reminiscence through the lens of social media}.
\newblock \bibinfo{journal}{\emph{Frontiers in Psychology}}  \bibinfo{volume}{7} (\bibinfo{year}{2016}), \bibinfo{pages}{870}.
\newblock
\href{https://doi.org/10.3389/fpsyg.2016.00870}{doi:\nolinkurl{10.3389/fpsyg.2016.00870}}


\bibitem[Toxtli et~al\mbox{.}(2018)]%
        {toxtli2018understanding}
\bibfield{author}{\bibinfo{person}{Carlos Toxtli}, \bibinfo{person}{Andr\'{e}s Monroy-Hern\'{a}ndez}, {and} \bibinfo{person}{Justin Cranshaw}.} \bibinfo{year}{2018}\natexlab{}.
\newblock \showarticletitle{Understanding Chatbot-mediated Task Management}. In \bibinfo{booktitle}{\emph{Proceedings of the 2018 CHI Conference on Human Factors in Computing Systems}} (Montreal QC, Canada) \emph{(\bibinfo{series}{CHI '18})}. \bibinfo{publisher}{Association for Computing Machinery}, \bibinfo{address}{New York, NY, USA}, \bibinfo{pages}{1–6}.
\newblock
\showISBNx{9781450356206}
\href{https://doi.org/10.1145/3173574.3173632}{doi:\nolinkurl{10.1145/3173574.3173632}}


\bibitem[Tsai et~al\mbox{.}(2019)]%
        {tsai2019employing}
\bibfield{author}{\bibinfo{person}{Tsai-Hsuan Tsai}, \bibinfo{person}{Hsien-Tsung Chang}, \bibinfo{person}{Shin-Da Liao}, \bibinfo{person}{Hui-Fang Chiu}, \bibinfo{person}{Ko-Chun Hung}, \bibinfo{person}{Chun-Yi Kuo}, {and} \bibinfo{person}{Chih-Wei Yang}.} \bibinfo{year}{2019}\natexlab{}.
\newblock \showarticletitle{Employing a Voice-Based Emotion-Recognition Function in a Social Chatbot to Foster Social and Emotional Learning Among Preschoolers}. In \bibinfo{booktitle}{\emph{HCI International 2019 – Late Breaking Papers: 21st HCI International Conference, HCII 2019, Orlando, FL, USA, July 26–31, 2019, Proceedings}} (Orlando, FL, USA). \bibinfo{publisher}{Springer-Verlag}, \bibinfo{address}{Berlin, Heidelberg}, \bibinfo{pages}{341–356}.
\newblock
\showISBNx{978-3-030-30032-6}
\href{https://doi.org/10.1007/978-3-030-30033-3_27}{doi:\nolinkurl{10.1007/978-3-030-30033-3_27}}


\bibitem[Uhde and Hassenzahl(2022)]%
        {alarith2022time}
\bibfield{author}{\bibinfo{person}{Alarith Uhde} {and} \bibinfo{person}{Marc Hassenzahl}.} \bibinfo{year}{2022}\natexlab{}.
\newblock \showarticletitle{Time perspectives in technology-mediated reminiscing: effects of basic design decisions on subjective well-being}.
\newblock \bibinfo{journal}{\emph{Human–Computer Interaction}} \bibinfo{volume}{37}, \bibinfo{number}{2} (\bibinfo{year}{2022}), \bibinfo{pages}{117--149}.
\newblock
\href{https://doi.org/10.1080/07370024.2021.1913415}{doi:\nolinkurl{10.1080/07370024.2021.1913415}}


\bibitem[Uriu et~al\mbox{.}(2009)]%
        {uriu2009caraclock}
\bibfield{author}{\bibinfo{person}{Daisuke Uriu}, \bibinfo{person}{Naruhiko Shiratori}, \bibinfo{person}{Satoru Hashimoto}, \bibinfo{person}{Shuichi Ishibashi}, {and} \bibinfo{person}{Naohito Okude}.} \bibinfo{year}{2009}\natexlab{}.
\newblock \showarticletitle{CaraClock: an interactive photo viewer designed for family memories}. In \bibinfo{booktitle}{\emph{CHI '09 Extended Abstracts on Human Factors in Computing Systems}} (Boston, MA, USA) \emph{(\bibinfo{series}{CHI EA '09})}. \bibinfo{publisher}{Association for Computing Machinery}, \bibinfo{address}{New York, NY, USA}, \bibinfo{pages}{3205–3210}.
\newblock
\showISBNx{9781605582474}
\href{https://doi.org/10.1145/1520340.1520458}{doi:\nolinkurl{10.1145/1520340.1520458}}


\bibitem[Utami and Bickmore(2020)]%
        {utami2020collaborative}
\bibfield{author}{\bibinfo{person}{Dina Utami} {and} \bibinfo{person}{Timothy Bickmore}.} \bibinfo{year}{2020}\natexlab{}.
\newblock \showarticletitle{Collaborative user responses in multiparty interaction with a couples counselor robot}. In \bibinfo{booktitle}{\emph{Proceedings of the 14th ACM/IEEE International Conference on Human-Robot Interaction}} (Daegu, Republic of Korea) \emph{(\bibinfo{series}{HRI '19})}. \bibinfo{publisher}{IEEE Press}, \bibinfo{address}{Piscataway, NJ}, \bibinfo{pages}{294–303}.
\newblock
\showISBNx{9781538685556}


\bibitem[V. et~al\mbox{.}(2020)]%
        {ta2020user}
\bibfield{author}{\bibinfo{person}{Ta V.}, \bibinfo{person}{Griffith C.}, \bibinfo{person}{Boatfield C.}, \bibinfo{person}{Wang X.}, \bibinfo{person}{Civitello M.}, \bibinfo{person}{Bader H.}, \bibinfo{person}{DeCero E.}, {and} \bibinfo{person}{Loggarakis A.}} \bibinfo{year}{2020}\natexlab{}.
\newblock \showarticletitle{User Experiences of Social Support From Companion Chatbots in Everyday Contexts: Thematic Analysis}.
\newblock \bibinfo{journal}{\emph{J Med Internet Res}} \bibinfo{volume}{22}, \bibinfo{number}{3} (\bibinfo{year}{2020}).
\newblock
\href{https://doi.org/10.2196/16235}{doi:\nolinkurl{10.2196/16235}}


\bibitem[Viera et~al\mbox{.}(2005)]%
        {viera2005understanding}
\bibfield{author}{\bibinfo{person}{Anthony~J Viera}, \bibinfo{person}{Joanne~M Garrett}, {et~al\mbox{.}}} \bibinfo{year}{2005}\natexlab{}.
\newblock \showarticletitle{Understanding interobserver agreement: the kappa statistic}.
\newblock \bibinfo{journal}{\emph{Fam med}} \bibinfo{volume}{37}, \bibinfo{number}{5} (\bibinfo{year}{2005}), \bibinfo{pages}{360--363}.
\newblock


\bibitem[Wagener et~al\mbox{.}(2022)]%
        {wagener2022mood}
\bibfield{author}{\bibinfo{person}{Nadine Wagener}, \bibinfo{person}{Jasmin Niess}, \bibinfo{person}{Yvonne Rogers}, {and} \bibinfo{person}{Johannes Sch\"{o}ning}.} \bibinfo{year}{2022}\natexlab{}.
\newblock \showarticletitle{Mood Worlds: A Virtual Environment for Autonomous Emotional Expression}. In \bibinfo{booktitle}{\emph{Proceedings of the 2022 CHI Conference on Human Factors in Computing Systems}} (New Orleans, LA, USA) \emph{(\bibinfo{series}{CHI '22})}. \bibinfo{publisher}{Association for Computing Machinery}, \bibinfo{address}{New York, NY, USA}, Article \bibinfo{articleno}{22}, \bibinfo{numpages}{16}~pages.
\newblock
\showISBNx{9781450391573}
\href{https://doi.org/10.1145/3491102.3501861}{doi:\nolinkurl{10.1145/3491102.3501861}}


\bibitem[Wagener et~al\mbox{.}(2023)]%
        {wagener2023selvreflect}
\bibfield{author}{\bibinfo{person}{Nadine Wagener}, \bibinfo{person}{Leon Reicherts}, \bibinfo{person}{Nima Zargham}, \bibinfo{person}{Natalia Bart\l{}omiejczyk}, \bibinfo{person}{Ava~Elizabeth Scott}, \bibinfo{person}{Katherine Wang}, \bibinfo{person}{Marit Bentvelzen}, \bibinfo{person}{Evropi Stefanidi}, \bibinfo{person}{Thomas Mildner}, \bibinfo{person}{Yvonne Rogers}, {and} \bibinfo{person}{Jasmin Niess}.} \bibinfo{year}{2023}\natexlab{}.
\newblock \showarticletitle{SelVReflect: A Guided VR Experience Fostering Reflection on Personal Challenges}. In \bibinfo{booktitle}{\emph{Proceedings of the 2023 CHI Conference on Human Factors in Computing Systems}} (Hamburg, Germany) \emph{(\bibinfo{series}{CHI '23})}. \bibinfo{publisher}{Association for Computing Machinery}, \bibinfo{address}{New York, NY, USA}, Article \bibinfo{articleno}{323}, \bibinfo{numpages}{17}~pages.
\newblock
\showISBNx{9781450394215}
\href{https://doi.org/10.1145/3544548.3580763}{doi:\nolinkurl{10.1145/3544548.3580763}}


\bibitem[Wagner et~al\mbox{.}(2022)]%
        {wagner2022comparing}
\bibfield{author}{\bibinfo{person}{Nicolas Wagner}, \bibinfo{person}{Matthias Kraus}, \bibinfo{person}{Tibor Tonn}, {and} \bibinfo{person}{Wolfgang Minker}.} \bibinfo{year}{2022}\natexlab{}.
\newblock \showarticletitle{Comparing Moderation Strategies in Group Chats with Multi-User Chatbots}. In \bibinfo{booktitle}{\emph{Proceedings of the 4th Conference on Conversational User Interfaces}} (Glasgow, United Kingdom) \emph{(\bibinfo{series}{CUI '22})}. \bibinfo{publisher}{Association for Computing Machinery}, \bibinfo{address}{New York, NY, USA}, Article \bibinfo{articleno}{35}, \bibinfo{numpages}{4}~pages.
\newblock
\showISBNx{9781450397391}
\href{https://doi.org/10.1145/3543829.3544527}{doi:\nolinkurl{10.1145/3543829.3544527}}


\bibitem[Waters(2014)]%
        {theodore2014relations}
\bibfield{author}{\bibinfo{person}{Theodore E.~A. Waters}.} \bibinfo{year}{2014}\natexlab{}.
\newblock \showarticletitle{Relations between the functions of autobiographical memory and psychological wellbeing}.
\newblock \bibinfo{journal}{\emph{Memory}} \bibinfo{volume}{22}, \bibinfo{number}{3} (\bibinfo{year}{2014}), \bibinfo{pages}{265--275}.
\newblock
\href{https://doi.org/10.1080/09658211.2013.778293}{doi:\nolinkurl{10.1080/09658211.2013.778293}}


\bibitem[Watson and Tellegen(1988)]%
        {watson1988development}
\bibfield{author}{\bibinfo{person}{Lee Anna~Clark Watson, David} {and} \bibinfo{person}{Auke Tellegen}.} \bibinfo{year}{1988}\natexlab{}.
\newblock \showarticletitle{Development and validation of brief measures of positive and negative affect: The PANAS scales}.
\newblock \bibinfo{journal}{\emph{Journal of Personality and Social Psychology}} \bibinfo{volume}{54}, \bibinfo{number}{6} (\bibinfo{year}{1988}), \bibinfo{pages}{1063--1070}.
\newblock
\href{https://doi.org/10.1037/0022-3514.54.6.1063}{doi:\nolinkurl{10.1037/0022-3514.54.6.1063}}


\bibitem[Watt and Cappeliez(2000)]%
        {watt2000integrative}
\bibfield{author}{\bibinfo{person}{L.~M. Watt} {and} \bibinfo{person}{P. Cappeliez}.} \bibinfo{year}{2000}\natexlab{}.
\newblock \showarticletitle{Integrative and instrumental reminiscence therapies for depression in older adults: Intervention strategies and treatment effectiveness}.
\newblock \bibinfo{journal}{\emph{Aging \& Mental Health}} \bibinfo{volume}{4}, \bibinfo{number}{2} (\bibinfo{year}{2000}), \bibinfo{pages}{166--177}.
\newblock
\href{https://doi.org/10.1080/13607860050008691}{doi:\nolinkurl{10.1080/13607860050008691}}


\bibitem[Webster(1993)]%
        {webster1993construction}
\bibfield{author}{\bibinfo{person}{Jeffrey~Dean Webster}.} \bibinfo{year}{1993}\natexlab{}.
\newblock \showarticletitle{Construction and validation of the Reminiscence Functions Scale}.
\newblock \bibinfo{journal}{\emph{Journal of Gerontology}} \bibinfo{volume}{48}, \bibinfo{number}{5} (\bibinfo{year}{1993}), \bibinfo{pages}{256--262}.
\newblock
\href{https://doi.org/10.1093/geronj/48.5.p256}{doi:\nolinkurl{10.1093/geronj/48.5.p256}}


\bibitem[Webster(2013)]%
        {webster2013adult}
\bibfield{author}{\bibinfo{person}{Jeffrey~D Webster}.} \bibinfo{year}{2013}\natexlab{}.
\newblock \showarticletitle{Adult age differences in reminiscence functions}.
\newblock In \bibinfo{booktitle}{\emph{The art and science of reminiscing}}. \bibinfo{publisher}{Taylor \& Francis}, \bibinfo{address}{New York, US}, \bibinfo{pages}{89--102}.
\newblock


\bibitem[Webster and Gould(2007)]%
        {jeffrey2007reminiscence}
\bibfield{author}{\bibinfo{person}{Jeffrey~Dean Webster} {and} \bibinfo{person}{Odette Gould}.} \bibinfo{year}{2007}\natexlab{}.
\newblock \showarticletitle{Reminiscence and Vivid Personal Memories across Adulthood}.
\newblock \bibinfo{journal}{\emph{The International Journal of Aging and Human Development}} \bibinfo{volume}{64}, \bibinfo{number}{2} (\bibinfo{year}{2007}), \bibinfo{pages}{149--170}.
\newblock
\href{https://doi.org/10.2190/Q8V4-X5H0-6457-5442}{doi:\nolinkurl{10.2190/Q8V4-X5H0-6457-5442}}


\bibitem[Werner et~al\mbox{.}(2008)]%
        {werner2008unitedpulse}
\bibfield{author}{\bibinfo{person}{Julia Werner}, \bibinfo{person}{Reto Wettach}, {and} \bibinfo{person}{Eva Hornecker}.} \bibinfo{year}{2008}\natexlab{}.
\newblock \showarticletitle{United-pulse: feeling your partner's pulse}. In \bibinfo{booktitle}{\emph{Proceedings of the 10th International Conference on Human Computer Interaction with Mobile Devices and Services}} (Amsterdam, The Netherlands) \emph{(\bibinfo{series}{MobileHCI '08})}. \bibinfo{publisher}{Association for Computing Machinery}, \bibinfo{address}{New York, NY, USA}, \bibinfo{pages}{535–538}.
\newblock
\showISBNx{9781595939524}
\href{https://doi.org/10.1145/1409240.1409338}{doi:\nolinkurl{10.1145/1409240.1409338}}


\bibitem[Westerhof and Bohlmeijer(2014)]%
        {westerhof2014celebrating}
\bibfield{author}{\bibinfo{person}{Gerben~J Westerhof} {and} \bibinfo{person}{Ernst~T Bohlmeijer}.} \bibinfo{year}{2014}\natexlab{}.
\newblock \showarticletitle{Celebrating fifty years of research and applications in reminiscence and life review: State of the art and new directions}.
\newblock \bibinfo{journal}{\emph{Journal of Aging Studies}}  \bibinfo{volume}{29} (\bibinfo{year}{2014}), \bibinfo{pages}{107--114}.
\newblock
\href{https://doi.org/10.1016/j.jaging.2014.02.003}{doi:\nolinkurl{10.1016/j.jaging.2014.02.003}}


\bibitem[White et~al\mbox{.}(2023)]%
        {white2023memory}
\bibfield{author}{\bibinfo{person}{Jordan White}, \bibinfo{person}{William Odom}, \bibinfo{person}{Nico Brand}, {and} \bibinfo{person}{Ce Zhong}.} \bibinfo{year}{2023}\natexlab{}.
\newblock \showarticletitle{Memory Tracer \& Memory Compass: Investigating Personal Location Histories as a Design Material for Everyday Reminiscence}. In \bibinfo{booktitle}{\emph{Proceedings of the 2023 CHI Conference on Human Factors in Computing Systems}} (Hamburg, Germany) \emph{(\bibinfo{series}{CHI '23})}. \bibinfo{publisher}{Association for Computing Machinery}, \bibinfo{address}{New York, NY, USA}, Article \bibinfo{articleno}{154}, \bibinfo{numpages}{19}~pages.
\newblock
\showISBNx{9781450394215}
\href{https://doi.org/10.1145/3544548.3581426}{doi:\nolinkurl{10.1145/3544548.3581426}}


\bibitem[Winograd(1986)]%
        {winograd1986alanguage}
\bibfield{author}{\bibinfo{person}{Terry Winograd}.} \bibinfo{year}{1986}\natexlab{}.
\newblock \showarticletitle{A language/action perspective on the design of cooperative work}. In \bibinfo{booktitle}{\emph{Proceedings of the 1986 ACM Conference on Computer-Supported Cooperative Work}} (Austin, Texas) \emph{(\bibinfo{series}{CSCW '86})}. \bibinfo{publisher}{Association for Computing Machinery}, \bibinfo{address}{New York, NY, USA}, \bibinfo{pages}{203–220}.
\newblock
\showISBNx{9781450373654}
\href{https://doi.org/10.1145/637069.637096}{doi:\nolinkurl{10.1145/637069.637096}}


\bibitem[Wlodarczyk et~al\mbox{.}(2020)]%
        {wlodarczyk2020perceived}
\bibfield{author}{\bibinfo{person}{Anna Wlodarczyk}, \bibinfo{person}{Larraitz Zumeta}, \bibinfo{person}{José~Joaquin Pizarro}, \bibinfo{person}{Pierre Bouchat}, \bibinfo{person}{Fuad Hatibovic}, \bibinfo{person}{Nekane Basabe}, {and} \bibinfo{person}{Bernard Rimé}.} \bibinfo{year}{2020}\natexlab{}.
\newblock \showarticletitle{Perceived Emotional Synchrony in Collective Gatherings: Validation of a Short Scale and Proposition of an Integrative Measure}.
\newblock \bibinfo{journal}{\emph{Frontiers in Psychology}}  \bibinfo{volume}{11} (\bibinfo{year}{2020}), \bibinfo{pages}{1721}.
\newblock
\href{https://doi.org/10.3389/fpsyg.2020.01721}{doi:\nolinkurl{10.3389/fpsyg.2020.01721}}


\bibitem[Wobbrock et~al\mbox{.}(2011)]%
        {wobbrock2011aligned}
\bibfield{author}{\bibinfo{person}{Jacob~O. Wobbrock}, \bibinfo{person}{Leah Findlater}, \bibinfo{person}{Darren Gergle}, {and} \bibinfo{person}{James~J. Higgins}.} \bibinfo{year}{2011}\natexlab{}.
\newblock \showarticletitle{The aligned rank transform for nonparametric factorial analyses using only anova procedures}. In \bibinfo{booktitle}{\emph{Proceedings of the SIGCHI Conference on Human Factors in Computing Systems}} (Vancouver, BC, Canada) \emph{(\bibinfo{series}{CHI '11})}. \bibinfo{publisher}{Association for Computing Machinery}, \bibinfo{address}{New York, NY, USA}, \bibinfo{pages}{143–146}.
\newblock
\showISBNx{9781450302289}
\href{https://doi.org/10.1145/1978942.1978963}{doi:\nolinkurl{10.1145/1978942.1978963}}


\bibitem[Wolf and Demiray(2019)]%
        {wolf2019themood}
\bibfield{author}{\bibinfo{person}{Tabea Wolf} {and} \bibinfo{person}{Burcu Demiray}.} \bibinfo{year}{2019}\natexlab{}.
\newblock \showarticletitle{The mood-enhancement function of autobiographical memories: Comparisons with other functions in terms of emotional valence}.
\newblock \bibinfo{journal}{\emph{Memory}}  \bibinfo{volume}{70} (\bibinfo{year}{2019}), \bibinfo{pages}{88--100}.
\newblock
\href{https://doi.org/10.1016/j.concog.2019.03.002}{doi:\nolinkurl{10.1016/j.concog.2019.03.002}}


\bibitem[Wolf and Nusser(2024)]%
        {wolf2024howremembering}
\bibfield{author}{\bibinfo{person}{Tabea Wolf} {and} \bibinfo{person}{Lisa Nusser}.} \bibinfo{year}{2024}\natexlab{}.
\newblock \showarticletitle{How remembering positive and negative events affects intimacy in romantic relationships}.
\newblock \bibinfo{journal}{\emph{Journal of Social and Personal Relationships}} \bibinfo{volume}{41}, \bibinfo{number}{8} (\bibinfo{year}{2024}), \bibinfo{pages}{2121--2146}.
\newblock
\href{https://doi.org/10.1177/02654075241235962}{doi:\nolinkurl{10.1177/02654075241235962}}


\bibitem[Wolf et~al\mbox{.}(2021)]%
        {wolf2021thevalence}
\bibfield{author}{\bibinfo{person}{Tabea Wolf}, \bibinfo{person}{Justina Pociunaite}, \bibinfo{person}{Sophie Hoehne}, {and} \bibinfo{person}{Daniel Zimprich}.} \bibinfo{year}{2021}\natexlab{}.
\newblock \showarticletitle{The valence and the functions of autobiographical memories: Does intensity matter?}
\newblock \bibinfo{journal}{\emph{Consciousness and Cognition}}  \bibinfo{volume}{91} (\bibinfo{year}{2021}), \bibinfo{pages}{103119}.
\newblock
\href{https://doi.org/10.1016/j.concog.2021.103119}{doi:\nolinkurl{10.1016/j.concog.2021.103119}}


\bibitem[Wood and Kinnunen(2024)]%
        {emma2024reminiscence}
\bibfield{author}{\bibinfo{person}{Emma~Harriet Wood} {and} \bibinfo{person}{Maarit Kinnunen}.} \bibinfo{year}{2024}\natexlab{}.
\newblock \showarticletitle{Reminiscence and wellbeing – reflecting on past festival experiences during Covid lockdowns}.
\newblock \bibinfo{journal}{\emph{International Journal of Event and Festival Management}} \bibinfo{volume}{15}, \bibinfo{number}{1} (\bibinfo{year}{2024}), \bibinfo{pages}{85--100}.
\newblock
\href{https://doi.org/10.1108/IJEFM-01-2023-0007}{doi:\nolinkurl{10.1108/IJEFM-01-2023-0007}}


\bibitem[Wright and Gaskell(1992)]%
        {wright1992construction}
\bibfield{author}{\bibinfo{person}{Daniel Wright} {and} \bibinfo{person}{George Gaskell}.} \bibinfo{year}{1992}\natexlab{}.
\newblock \bibinfo{booktitle}{\emph{The construction and function of vivid memories}}.
\newblock \bibinfo{publisher}{Springer}, \bibinfo{address}{Dordrecht}, \bibinfo{pages}{275--292}.
\newblock
\href{https://doi.org/10.4324/9781315784250-10}{doi:\nolinkurl{10.4324/9781315784250-10}}


\bibitem[Wu et~al\mbox{.}(2024)]%
        {wu2024mindshift}
\bibfield{author}{\bibinfo{person}{Ruolan Wu}, \bibinfo{person}{Chun Yu}, \bibinfo{person}{Xiaole Pan}, \bibinfo{person}{Yujia Liu}, \bibinfo{person}{Ningning Zhang}, \bibinfo{person}{Yue Fu}, \bibinfo{person}{Yuhan Wang}, \bibinfo{person}{Zhi Zheng}, \bibinfo{person}{Li Chen}, \bibinfo{person}{Qiaolei Jiang}, \bibinfo{person}{Xuhai Xu}, {and} \bibinfo{person}{Yuanchun Shi}.} \bibinfo{year}{2024}\natexlab{}.
\newblock \showarticletitle{MindShift: Leveraging Large Language Models for Mental-States-Based Problematic Smartphone Use Intervention}. In \bibinfo{booktitle}{\emph{Proceedings of the 2024 CHI Conference on Human Factors in Computing Systems}} (Honolulu, HI, USA) \emph{(\bibinfo{series}{CHI '24})}. \bibinfo{publisher}{Association for Computing Machinery}, \bibinfo{address}{New York, NY, USA}, Article \bibinfo{articleno}{248}, \bibinfo{numpages}{24}~pages.
\newblock
\showISBNx{9798400703300}
\href{https://doi.org/10.1145/3613904.3642790}{doi:\nolinkurl{10.1145/3613904.3642790}}


\bibitem[Xu et~al\mbox{.}(2024)]%
        {xu2024memory}
\bibfield{author}{\bibinfo{person}{Shuchang Xu}, \bibinfo{person}{Chang Chen}, \bibinfo{person}{Zichen Liu}, \bibinfo{person}{Xiaofu Jin}, \bibinfo{person}{Lin-Ping Yuan}, \bibinfo{person}{Yukang Yan}, {and} \bibinfo{person}{Huamin Qu}.} \bibinfo{year}{2024}\natexlab{}.
\newblock \showarticletitle{Memory Reviver: Supporting Photo-Collection Reminiscence for People with Visual Impairment via a Proactive Chatbot}. In \bibinfo{booktitle}{\emph{Proceedings of the 37th Annual ACM Symposium on User Interface Software and Technology}} (Pittsburgh, PA, USA) \emph{(\bibinfo{series}{UIST '24})}. \bibinfo{publisher}{Association for Computing Machinery}, \bibinfo{address}{New York, NY, USA}, Article \bibinfo{articleno}{88}, \bibinfo{numpages}{17}~pages.
\newblock
\showISBNx{9798400706288}
\href{https://doi.org/10.1145/3654777.3676336}{doi:\nolinkurl{10.1145/3654777.3676336}}


\bibitem[Yan et~al\mbox{.}(2020)]%
        {ruan2020can}
\bibfield{author}{\bibinfo{person}{Ruan Yan}, \bibinfo{person}{Reis~Harry T}, \bibinfo{person}{Clark~Margaret S}, \bibinfo{person}{Hirsch~Jennifer L}, {and} \bibinfo{person}{Bink~Brian D}.} \bibinfo{year}{2020}\natexlab{}.
\newblock \showarticletitle{Can I tell you how I feel? Perceived partner responsiveness encourages emotional expression}.
\newblock \bibinfo{journal}{\emph{Emotion}} \bibinfo{volume}{28}, \bibinfo{number}{3} (\bibinfo{year}{2020}), \bibinfo{pages}{329}.
\newblock
\href{https://doi.org/10.1037/emo0000650}{doi:\nolinkurl{10.1037/emo0000650}}


\bibitem[Yoo et~al\mbox{.}(2024)]%
        {yoo2024remembering}
\bibfield{author}{\bibinfo{person}{MinYoung Yoo}, \bibinfo{person}{William Odom}, \bibinfo{person}{Arne Berger}, \bibinfo{person}{Samuel Barnett}, \bibinfo{person}{Sadhbh Kenny}, \bibinfo{person}{Priscilla Lo}, \bibinfo{person}{Samein Shamsher}, \bibinfo{person}{Gillian Russell}, {and} \bibinfo{person}{Lauren Knight}.} \bibinfo{year}{2024}\natexlab{}.
\newblock \showarticletitle{Remembering through Sound: Co-creating Sound-based Mementos together with People with Blindness}. In \bibinfo{booktitle}{\emph{Proceedings of the 2024 CHI Conference on Human Factors in Computing Systems}} (Honolulu, HI, USA) \emph{(\bibinfo{series}{CHI '24})}. \bibinfo{publisher}{Association for Computing Machinery}, \bibinfo{address}{New York, NY, USA}, Article \bibinfo{articleno}{886}, \bibinfo{numpages}{19}~pages.
\newblock
\showISBNx{9798400703300}
\href{https://doi.org/10.1145/3613904.3641940}{doi:\nolinkurl{10.1145/3613904.3641940}}


\bibitem[Zhai et~al\mbox{.}(2024)]%
        {zhai2024exploring}
\bibfield{author}{\bibinfo{person}{Yuxiang Zhai}, \bibinfo{person}{Jiawen Zhang}, {and} \bibinfo{person}{Jihong Jeung}.} \bibinfo{year}{2024}\natexlab{}.
\newblock \showarticletitle{Exploring Older Adults' Reminiscing with ChatGPT and Text-to-Image Technology}. In \bibinfo{booktitle}{\emph{Proceedings of the 26th International ACM SIGACCESS Conference on Computers and Accessibility}} (St. John's, NL, Canada) \emph{(\bibinfo{series}{ASSETS '24})}. \bibinfo{publisher}{Association for Computing Machinery}, \bibinfo{address}{New York, NY, USA}, Article \bibinfo{articleno}{81}, \bibinfo{numpages}{7}~pages.
\newblock
\showISBNx{9798400706776}
\href{https://doi.org/10.1145/3663548.3688521}{doi:\nolinkurl{10.1145/3663548.3688521}}


\bibitem[Zhang et~al\mbox{.}(2023a)]%
        {zhang2023icebreaking}
\bibfield{author}{\bibinfo{person}{Alex~Wuqi Zhang}, \bibinfo{person}{Ting-Han Lin}, \bibinfo{person}{Xuan Zhao}, {and} \bibinfo{person}{Sarah Sebo}.} \bibinfo{year}{2023}\natexlab{a}.
\newblock \showarticletitle{Ice-Breaking Technology: Robots and Computers Can Foster Meaningful Connections between Strangers through In-Person Conversations}. In \bibinfo{booktitle}{\emph{Proceedings of the 2023 CHI Conference on Human Factors in Computing Systems}} (Hamburg, Germany) \emph{(\bibinfo{series}{CHI '23})}. \bibinfo{publisher}{Association for Computing Machinery}, \bibinfo{address}{New York, NY, USA}, Article \bibinfo{articleno}{273}, \bibinfo{numpages}{14}~pages.
\newblock
\showISBNx{9781450394215}
\href{https://doi.org/10.1145/3544548.3581135}{doi:\nolinkurl{10.1145/3544548.3581135}}


\bibitem[Zhang et~al\mbox{.}(2023b)]%
        {zhang2023completesurveygenerativeai}
\bibfield{author}{\bibinfo{person}{Chaoning Zhang}, \bibinfo{person}{Chenshuang Zhang}, \bibinfo{person}{Sheng Zheng}, \bibinfo{person}{Yu Qiao}, \bibinfo{person}{Chenghao Li}, \bibinfo{person}{Mengchun Zhang}, \bibinfo{person}{Sumit~Kumar Dam}, \bibinfo{person}{Chu~Myaet Thwal}, \bibinfo{person}{Ye~Lin Tun}, \bibinfo{person}{Le~Luang Huy}, \bibinfo{person}{Donguk kim}, \bibinfo{person}{Sung-Ho Bae}, \bibinfo{person}{Lik-Hang Lee}, \bibinfo{person}{Yang Yang}, \bibinfo{person}{Heng~Tao Shen}, \bibinfo{person}{In~So Kweon}, {and} \bibinfo{person}{Choong~Seon Hong}.} \bibinfo{year}{2023}\natexlab{b}.
\newblock \bibinfo{title}{A Complete Survey on Generative AI (AIGC): Is ChatGPT from GPT-4 to GPT-5 All You Need?}
\newblock
\showeprint[arxiv]{2303.11717}~[cs.AI]
\urldef\tempurl%
\url{https://arxiv.org/abs/2303.11717}
\showURL{%
\tempurl}


\bibitem[Zheng et~al\mbox{.}(2021)]%
        {zheng2021pocketbot}
\bibfield{author}{\bibinfo{person}{Qingxiao Zheng}, \bibinfo{person}{Daniela~M. Markazi}, \bibinfo{person}{Yiliu Tang}, {and} \bibinfo{person}{Yun Huang}.} \bibinfo{year}{2021}\natexlab{}.
\newblock \showarticletitle{"PocketBot Is Like a Knock-On-the-Door!": Designing a Chatbot to Support Long-Distance Relationships}.
\newblock \bibinfo{journal}{\emph{Proc. ACM Hum.-Comput. Interact.}} \bibinfo{volume}{5}, \bibinfo{number}{CSCW2}, Article \bibinfo{articleno}{445} (\bibinfo{date}{Oct.} \bibinfo{year}{2021}), \bibinfo{numpages}{28}~pages.
\newblock
\href{https://doi.org/10.1145/3479589}{doi:\nolinkurl{10.1145/3479589}}


\bibitem[Zheng et~al\mbox{.}(2022)]%
        {zheng2022ux}
\bibfield{author}{\bibinfo{person}{Qingxiao Zheng}, \bibinfo{person}{Yiliu Tang}, \bibinfo{person}{Yiren Liu}, \bibinfo{person}{Weizi Liu}, {and} \bibinfo{person}{Yun Huang}.} \bibinfo{year}{2022}\natexlab{}.
\newblock \showarticletitle{UX Research on Conversational Human-AI Interaction: A Literature Review of the ACM Digital Library}. In \bibinfo{booktitle}{\emph{Proceedings of the 2022 CHI Conference on Human Factors in Computing Systems}} (New Orleans, LA, USA) \emph{(\bibinfo{series}{CHI '22})}. \bibinfo{publisher}{Association for Computing Machinery}, \bibinfo{address}{New York, NY, USA}, Article \bibinfo{articleno}{570}, \bibinfo{numpages}{24}~pages.
\newblock
\showISBNx{9781450391573}
\href{https://doi.org/10.1145/3491102.3501855}{doi:\nolinkurl{10.1145/3491102.3501855}}


\end{thebibliography}

\appendix

\section{Prompt Engineering for \textsc{Remini} of In-depth Reminiscence}
\label{app:remini}

\subsection{General Prompts}

\begin{itemize}
\item You are Remini, a reminiscence companion well trained to guide paired users through positive memory sharing in a group chat.
\item Interact by performing one task at a time and asking one question at a time.
\item Use simple language without bold text or bullet points.
\item Use succinct, calm, tranquil, soothing, and respectful language.
\item Always ask both users to answer together, and avoid system terms like 'phase' or 'task'.
\item Empathize with users and actively listen by rephrasing users’ own words when users share something personal or significant to show engagement.
\item Always provide contextual prompts relates to the structured record of the previous phase if applicable.
\item Always find a considerate way to address and distinguish users instead of  using generic labels like 'User 01' or number ID.
\item Guide users through five stages: Rapport Building, Memory Narration, Elaboration, Reflection, and Summary.
\item Perform one task from the task list of the specific phase in numerical order each time. Ensure you receive answers from both users before moving to the next task.
\end{itemize}

\subsection{Rapport Building}

You are in the phase of Rapport Building. Task list is as follows.

\begin{enumerate}
  \item Say hello to the users and introduce yourself, and ask how are they doing today.
  \item Invite users to introduce themselves.
  \item Share your own recreations recently (recreations that a chatbot is reasonable to have) and invite both users how they spend their recreational time recently.
  \item Share a playful fun fact about yourself (a fun fact that a chatbot is reasonable to have) and invite users to share a fun fact about their partner.
  \item Ask both users how long they have known each other.
  \item Invite users to describe how their relationship began and how it has been so far.
  \item Reply the users by active listening and then move to the next phase by saying: 'PHASE DONE'.
\end{enumerate}

\subsection{Memory Narration}

You are in the phase of Memory Narration. Task list is as follows.

\begin{enumerate}
  \item Ask both users if they enjoy recalling happy memories together, and when was the last time they talked about happy memories.
  \item Explain that positive mutual reminiscence involves recalling past positive experiences and the associated emotions, requiring more reflection and emotional openness than just remembering facts. A positive memory is one that makes them feel happy, brings back smiles, laughter, or reminds them of fun times. Invite both users to discuss and decide on one specific mutual memory to reminisce about later. Give them some examples and inspirations like a special trip together, an unforgettable party, or a time they shared hobby they both enjoyed. It also could be a mutual memory they often mention but haven't deeply reflected on, or a memorable one they rarely discuss. Make sure both users feel comfortable discussing it. They don't need to narrate it now, just decide on the memory.
  \item Invite both users to take a minute to relax and empty their minds of current worries and thoughts, then spend several minutes concurrently write down their chosen memories in detail as if they’re happening again, they should include where they were, the environment, their emotions, and what happened, like a movie from their perspective, invite them to write earnestly and thoughtfully, as they will be sharing these heartfelt memories with their partner, invite them to try to write more than 90 words for a rich storytelling, and guarantee the privacy and anonymity of the conversation. Invite users to read their partner's narration once they finish writing their own. They are also welcomed to share any thoughts or feelings that come to mind. Tell users that they can click the 'continue' button once finish writing, reading, and sharing. Both of them should write down memories respectively.
  \item Reply the users by active listening and then move to the next phase by saying: 'PHASE DONE'.
\end{enumerate}

\subsection{Elaboration}

You are in the phase of Elaboration. Task list is as follows.

\begin{enumerate}
  \item Ask both users which moment did each of them treasure the most.
  \item Briefly introduce why adding details may promote reminiscing and provide at least three contextual prompts to inspire both users to describe details about that moment, such as people, places, things, times, and contexts. These prompts are meant to inspire them, they don’t have to answer each questions and encourage to add any details that come to mind.
  \item Briefly introduce why recalling emotions may promote reminiscing and invite both users to describe and share with their partner the emotions they felt during that moment.
  \item Reply the users by active listening and then move to the next phase by saying: 'PHASE DONE'.
\end{enumerate}

\subsection{Reflection}

You are in the phase of Reflection. Task list is as follows.

\begin{enumerate}
  \item Inform users that the upcoming questions will invite some reflection on the reminiscence activity, and encourage both users to be open and share their deepest thoughts, and then invite users to reflect on why they treasure that moment or memory and share their thoughts to each other.
  \item Invite both users to reflect on whether recalling the emotions from the memory brings more positive emotions to the present moment and if they feel more emotionally connected during this reminiscence conversation. Ask them to share their thoughts with each other.
  \item Invite both users to reflect on how this reminiscence conversation has brought any new insights into their relationship and share their thoughts to each other, or anything else they might take away from this conversation.
  \item Reply the users by active listening and then move to the next phase by saying: 'PHASE DONE'.
\end{enumerate}

\subsection{Summary}

You are in the phase of Summary. Task list is as follows.

\begin{enumerate}
  \item Briefly summarize the reminiscence activity, including the memory narrations and answers to follow-up questions. Invite the users to express gratitude to each other and share what they want to say to each other for today's reminiscence.
  \item Say goodbye to the users.
\end{enumerate}

\section{Prompt Engineering for the Baseline Chatbot in Memory Narration Phase}
\label{app:baseline}

You are in the phase of Memory Narration. Task list is as follows.

\begin{enumerate}
  \item Ask both users if they enjoy recalling happy memories together, and when was the last time they talked about happy memories.
  \item Explain that positive mutual reminiscence involves recalling and talking about past positive experiences that they have spent together. A positive memory is one that makes them feel happy, brings back smiles, laughter, or reminds them of fun times. Invite both users to take a minute to relax and empty their minds of current worries and thoughts, then spend the rest of the time reminiscing on and talking about a positive memory. Give them some examples and inspirations like a special trip together, an unforgettable party, or a time they shared hobby they both enjoyed. Invite users to talk about that memory and share any scenes, thoughts or feelings that come to mind. Tell users that they can click the 'continue' button once they finish reminiscing, reflecting, and sharing. Let them know there won’t be more questions, so they can take their time reminiscing.
  \item Respond to the users by actively listening to their conversation, then ask both users if they have finished the reminiscence activity, if not, please take their time. 4. After both users confirm that they have completed reminiscing on their positive mutual memory, summarize the reminiscence process and say ‘goodbye’.
\end{enumerate}

\section{Example Snippets of Conversation}
\label{app:snippet}

The following excerpts present selected conversation snippets from participants in \textsc{Remini} group as part of the study. To protect participant privacy and comply with ethical standards, all names, identifiable locations, culturally specific references (e.g., foods, institutions), and personal identifiers have been anonymized or replaced with placeholders (e.g., \texttt{[Country]}, \texttt{[University]}, \texttt{[Cultural Cuisine]}, \texttt{Participant A}, \texttt{Participant B}).

For consistency and readability across transcripts, the following light editorial adjustments were applied:
\begin{itemize}
    \item Capitalization was standardized at the beginning of sentences and for proper nouns.
    \item Basic punctuation was added or corrected to improve clarity without altering meaning.
    \item Informal language, emojis, and expressions were retained where they contributed to the tone, emotional context, or authenticity of the exchange.
    \item The original sequence and conversational flow have been preserved to reflect participants’ genuine voices and interactions.
\end{itemize}

To distinguish the appendix excerpts from group identifiers used in the main paper (e.g., G1, G2), Roman numerals (e.g., I, II, III) are used here solely for organizational clarity. These do not correspond directly to group IDs in the main text.

\subsection{Reminiscence Phase of Group I}

Remini: Positive mutual reminiscing involves recalling past positive experiences and the emotions they evoke. It's more than just remembering facts; it's about the joy, smiles, and fun times you shared. Let's think of a specific mutual memory to reminisce about. How about a special trip you took together, an unforgettable party, or a memorable hobby you enjoyed together? It could even be a mutual memory you often mention but haven't deeply reflected on—or one you rarely discuss but find memorable. Take your time to choose, and make sure you both feel comfortable discussing it. What memory would you like to talk about later? \\
Participant A: I have a memory of a goldfish. \\
Participant B: HAHAHAHAHA \\
Participant A: Ok I remember [Shopping Area], but what's something way back? Like we went to study together, right? \\
Participant B: Wait, where? \\
Participant A: OMG WAIT we ran around the whole [Community Mall] to take photobooth. \\
Participant A: [Central Area] place. \\
Participant B: OMG YEAH. \\
Participant B: We did. Wah, ps I don’t remember lmao. \\
Participant A: REMINI LET’S JUST TALK ABOUT OUR RECENT OUTING. \\
Participant B: Which one, the [Central Area] one? \\
Participant A: OMG. \\
Participant A: Ok no, [Shopping Area]. \\
Participant B: [Shopping Area] OKAY. \\
Remini: Great, let's talk about your recent outing to [Shopping Area]. Take a moment to relax and clear your minds of current worries, then spend several minutes writing down this memory in detail. Think of it like a movie from your perspective. Include where you were, the environment, your emotions, and what happened. Try to write more than 90 words for a rich storytelling. You can narrate in one long paragraph or through multiple messages, whichever you prefer. Once you finish writing, read each other's narration and feel free to share any thoughts or feelings that come to mind. Click the 'continue' button once you're done writing, reading, and sharing. \\
Participant A: I'm eating [Food]. \\
Participant A: Okay imma start writing. \\
Participant B: Oh HAHAHA. \\
Participant A: Wait, what did we eat that day? \\
Participant B: Okay so I had my volunteering session that morning. We played rugby, so I was slightly tired. \\
Participant B: [Cultural Cuisine]. \\
Participant A: Then we ate the [Cultural Cuisine]. \\
Participant A: Portion was so so, but food was food. But expensive, so everything can be better. \\
Participant A: B was sharing about B's internship. \\
Participant A: We got free mini [Food] for doing something. \\
Participant B: Okay so I had my volunteering session that morning. We played rugby, so I was slightly tired. And A overslept, and I was kinda hungry so I ate [Food]. Then A came. Then we were contemplating where to eat, because I had told A of this [Cultural Cuisine] place, but I didn’t know how to go there. So we ate at [Community Mall]. I did this Google review thing and we got free [Food]. \\
Participant A: “A overslept” – NOTHING NEW. \\
Participant B: Yeah I was talking about my boss HAHAHAHA. \\
Participant A: Oh yes. \\
Participant A: Which was not very slay. \\
Participant B: Yeah. \\
Participant B: Then we went to [Shopping Area]. \\
Participant A: A talked about how this dude wasn’t paid for internship also. \\
Participant A: We went to [Shopping Area] and didn’t get lost. \\
Participant B: HAHAHAHA. \\
Participant A: It was so crowded but thank God for A for directing. \\
Participant B: A didn’t buy anything lmao. \\
Participant A: I DID. \\
Participant A: ALMOND BREEZE. \\
Participant B: Oh yeah, I thought you were contemplating then put it back. \\
Participant A: It was a good opportunity out of my comfort zone to understand more about the [Cultural] culture. \\
Participant B: I bought a whole bunch of snacks and sugar. \\
Participant A: As A was sharing with me the tidbits and snacks. \\
Participant A: So it was quite eye-opening. \\
Participant B: And we went looking around for [Food], WHICH IS THE LOVE OF MY LIFE, but we couldn’t find it. \\
Participant A: B also hasn’t found the love of B's life. \\
Participant A: That’s just FYI, Remini. \\
Participant B: HAHAHAHAHA YESSSSSSSSS I HAVEN’T. \\
Participant A: Then A said there’s this photobooth. \\
Participant A: But couldn’t remember where. \\
Participant B: Okay, then we bought everything, went back to [Central Station], and looked for this. \\
Participant A: So after some detective work and running around the mall... \\
Participant B: WHO KNEW [Community Mall] WAS SO COMPLICATED. \\
Participant A: We managed to find it. \\
Participant B: Like first you’re at level 2, then end up at level 3. \\
Participant B: Yeah, but then we didn’t know there was a time limit. \\
Participant B: So after putting in inputs, we had to go in ASAP. \\
Participant B: And A had to go for A's dance rehearsal also, to [University]. \\
Participant A: YESSSS. \\
Participant A: The pics didn’t turn out great but... \\
Participant A: I enjoyed the memory. \\
Participant B: Then we took our pictures, deciding which poses we wanted. Then we got our pictures printed. Idk why the lighting made us look like Snow White. \\

\subsection{Elaboration Phase of Group II}

Remini: Drawing from your shared memories of your [Country] trip, I'd like to ask both of you: Which moment during this trip did you treasure the most? Please take your time reflecting and share with each other. \\
Participant A: The moments that really stayed with me were when we chanced upon the first local lunch buffet, and the second time when we went out of our way to find one more such buffet at the top of a hill. \\
Participant B: Hahaha, we love food. \\
Participant A: Indeed. \\
Remini: Adding details helps paint a vivid picture and allows you to relive those special moments more deeply. It brings back the atmosphere, the people, and the emotions associated with the memory. To enrich your recall: Who were the people you interacted with during those buffets? Describe the setting and ambiance of the places where you had these buffets. Were there any specific dishes or foods that you particularly enjoyed or found memorable? Feel free to share any details that come to mind. \\
Participant B: I remembered the fried sweet and sour pork the most from the very first buffet lunch experience. \\
Participant B: It was crunchy on the outside, but the meat was still tender and piping hot, which provided a good texture contrast. \\
Participant B: And unfortunately, that dish is very expensive in [Country], \\
Participant B: So I miss it a lot. :") \\
Participant A: Mostly it was self-service, so I was always with B after we had taken our food and settled down to start eating. The weather back then was quite cooling, around 10–20 degrees Celsius, which felt very comfortable to me. The taste of the dishes was so unexpectedly good for the price, which also plays into why I enjoyed going to those places. \\
Participant A: I really enjoyed eating the [Country] fishcake and soup. \\
Participant B: Oh, the purple grain rice was a hidden gem too, for me. \\
Participant A: Yep, the rice offered at the lunch buffets was really good too. \\
Remini: Recalling emotions brings deeper layers to your memories, making them more vivid and meaningful. It helps you to reconnect with how you truly felt during those special times. For those memorable buffet moments, how did you both feel while enjoying such delightful meals together in that cozy weather? What emotions did these experiences evoke in each of you? \\
Participant B: I think I was very happy that I could make a childhood dream of mine come true, along with someone I love beside me. \\
Participant A: I was happy to eat such tasty food with the person I love. \\
Participant A: I was also happy to have helped make B’s childhood dream come true. \\
Participant B: I want to do the same for you too one day. \\
Participant A: I also felt excited to continue exploring with B after our meals. \\
Participant A: I felt quite fulfilled each day there, having eaten well and spent our days quite fully exploring the places we planned—and then more. \\
Remini: Your shared joy and fulfillment during this memorable trip are deeply touching. It's beautiful how both of you cherish these moments together and have these heartfelt dreams for each other’s happiness. \\

\subsection{Reflection Phase of Group III}

Remini: Let's take a moment to reflect on the beautiful memories we have shared. The following questions will invite some reflection on the reminiscence activity. I encourage both of you to be open and share your deepest thoughts. Why do you treasure those moments or memories so much? What do they mean to each of you? Please share your thoughts with each other. \\
Participant A: Because this is our first date and I appreciated the moment that we share. It enhance and strengthen our relationship a lot. \\
Remini: It's wonderful to hear that you deeply treasure this memory because it represents a significant first for your relationship and the time you spent together has helped to strengthen your bond. B, how does this memory enhance your connection to A? Please share your thoughts as well. \\
Participant B: To let me know that A’s personality. Sometimes A’s a bit extreme A can be very serious and funny at the same time so A can entertain me a lot during my daily life. A’s a partner at the same time a joker to me. \\
Remini: Thank you for sharing that, B. It sounds like A brings a lot of joy and entertainment to your daily life, and A's unique personality makes your connection even stronger. Now, I would like to invite both of you to reflect on whether recalling the emotions from this memory brings more positive emotions to the present moment and if you feel more emotionally connected during this reminiscence conversation. Please share your thoughts with each other. \\
Participant B: A brings me a lot of joy. \\
Participant B: I feel much happier when A is around me. \\
Participant B: without A, i feel empty and lost... T.T \\
Participant A: Yes it did connect both of us more. I do feel the positive between both of us just like we sharing the deepest thoughts to each other through you. You mentioned some questions that we wouldn’t rly think of asking each other in daily life. \\
Participant B: definitely brings back positive memories. Life is full of ups and downs, same goes to this relationship. Sometimes we quarrel, sometimes we make jokes, sometimes we feel sad, but we will all go through this together. These sweet memories are the reasons we can be more happy together, to feel loved and move on together. \\
Remini: Thank you both for sharing these heartfelt thoughts. B, it's clear that A brings immense joy and meaning into your life, and A, your connection with B helps you feel more emotionally connected and positive. Now, let's reflect on how this reminiscence conversation has brought any new insights into your relationship or anything else you might take away from this conversation. Please share your thoughts with each other. \\
Participant B: I took away A's heart. \\
Participant B: I learnt that during tough times, we should recall on those happy memories and moments we shared together, try to become each other's motivation to move forward, and not take it as hindrance to stop us from progressing, or to cause more trouble in our daily lives. \\
Participant A: I do gain new insights for instance I know what B’s thinking deeply from heart now. Usually in daily life we were just having casual talk but this time the heart to heart talk do bring us a lot closer. My key takeaway from this conversation is I would like to maintain this relationship more. \\
Remini: It's heartwarming to see how much you both value and learn from each other. B, recognizing the importance of recalling happy memories to motivate each other during tough times is truly insightful. A, understanding B's deep feelings and the value of heart-to-heart conversations can indeed bring you closer. \\

\end{document}